\documentclass[12pt]{article}
\usepackage{amssymb,amsmath}
\usepackage{graphicx,epsfig}
\usepackage{caption}
\usepackage{subcaption}
\usepackage{enumerate}
\usepackage{epstopdf}
\usepackage{xcolor}
\usepackage{booktabs}
\usepackage{float}
\usepackage{rotating}
\usepackage{tikz}
\usepackage[section]{placeins}
\usepackage{textcomp}
\usepackage[normalem]{ulem}
\usepackage{derivative}
\usepackage{amsmath}

\usepackage{dsfont}
\usepackage{amsthm}
\usepackage{enumitem}
\usepackage{comment}
\usepackage{tabularx}
\usepackage{hyperref}
\usepackage{upgreek}
\usepackage[margin=3cm]{geometry}
\usepackage[english]{babel}
\usepackage[numbers, square]{natbib} 
\usepackage{algpseudocode}
\usepackage[linesnumbered,lined,boxed,commentsnumbered,ruled,vlined]{algorithm2e}
\RestyleAlgo{ruled}
\SetKwComment{Comment}{/* }{ */}
\hypersetup{
    colorlinks=true,
    linkcolor=cyan,
    filecolor=cyan,      
    urlcolor=cyan,
    citecolor = cyan
    }
\graphicspath{{Rips sequences/}}
 \numberwithin{equation}{section}

\newtheorem{t1}{Theorem}[section]

\newtheorem{l1}{Lemma}[section]
\newtheorem{r1}{Remark}[section]

\newtheorem{R1}{Result}[section]
\newtheorem{A1}{Assumption}[section]
\newcommand{\HRule}{\rule{\linewidth}{0.8mm}}
\begin{document}
\title{
\HRule\quad
\LARGE\bfseries
 M-Estimation in Censored Regression Model under Endogeneity
\HRule
}
\author{
	Swati Shukla$^1$, Subhra Sankar Dhar$^2$, Shalabh$^3$\\
	Department of Statistics and Data Science,\\ Indian Institute of Technology Kanpur,
	 Kanpur 208016, India\\
	 Emails: $^1$shuklaswati1602@gmail.com, $^2$subhra@iitk.ac.in,  $^3$ shalab@iitk.ac.in}
\maketitle
\begin{center}
    \textbf{Abstract} 
\end{center} 
This paper develops a general M-estimation framework for censored regression models under endogeneity. A two-stage estimation is proposed to address the issue of endogeneity in the first stage and the issue of censoring in linear regression models in the second stage. The issue of endogeneity is addressed via instrumental variables using the control function approach. The issue of censoring in linear regression models is addressed via M-estimation using a suitably defined class of loss functions yielding a class of estimators. Thus, the proposed general M-estimation framework nests several existing estimators as special cases for the respective loss functions. Moreover, this provides flexibility of accommodating a broad range of robust loss functions that have not previously been studied in the context of censored regression. We establish consistency and asymptotic normality for the proposed class of estimators. An empirical study based on simulated datasets is conducted to illustrate general M-estimation for various choices of loss functions and to examine the finite-sample performance of various resulting estimators.   


\section{Introduction}
This paper proposes a general $M$-estimation framework for the semiparametric estimation of linear regression models in the presence of exogenous censoring and endogeneity. Censoring and endogeneity are common features of linear regression models and arise naturally in many empirical applications. One such example comes from the standard labor supply model, which assumes that each individual has some desired hours of work, say $y^{*}$, which is subject to the constraints of how many hours an individual wants and does not want to work. Thus, the variable $y^{*}$ is assumed to be the interior solution to a utility maximization problem subject to constraints on the number of working hours and leisure. Further, it is also assumed that the firm sets the lower limit of working hours, say $c$. Therefore, the dependent variable $y$ is the actual number of hours worked, which is denoted as $y=y^{*}$ if $y^{*}>c$, and $y=c$ if $y^{*}< c$. 

\noindent In this model, one is interested in examining the effects of other independent variables, such as non-labour income, education, age, wage, etc., on the dependent variable $y$. Therefore, one can fit a regression model to see the effect of these independent variables on $y$. However, it is usually understood that wages and desired hours of work are jointly determined for each individual; wages influence how many hours a person chooses to work, but hours worked can, in turn, affect the wage rate (e.g., through experience, overtime, or bargaining power). This simultaneity, or reverse causality between wages and hours, implies that wage is correlated with the unobserved error term in the hours equation, giving rise to the problem of endogeneity in the regression model. Now, since the dependent variable $y$ is censored below 0 and there is one endogenous explanatory variable, i.e., wage, in the model, this leads to the problem of estimating the censored regression model with an endogenous explanatory variable.

We propose an estimation procedure based on the \emph{control function approach} (see, e.g., \cite{Roger1984}), which is essentially an instrumental variable estimation technique. The control function estimation framework involves the following two steps. In the first step, endogenous variables are regressed on instrumental variables to find residuals, which are referred to as control variables. In the second stage, we include these control variables as additional explanatory variables to account for endogeneity in the model. This yields a new model that is free from endogeneity with an additional regressor and an appropriately defined error. The model framework, the underlying assumptions, and more details on the control function approach are provided in Section~\ref{The model}.

\subsection{Literature review} 
The semiparametric approaches to addressing endogeneity in censored regression models include those of \cite{das2002}, \cite{blundell2007}, and \cite{chernozhukov2015}, which rely on the symmetry of error distributions or conditional quantile restrictions. Building on these contributions, \cite{HongTamer} and \cite{KHAN2009104} propose moment-inequality approaches that offer robustness but result in estimation and inference procedures that are difficult to implement in practice. In a related vein, \cite{HONORE2004293} derive moment conditions under both full independence and weaker conditional quantile restrictions of error distributions. Subsequently, \cite{CHEN201830} extends this framework by proposing a sequential estimator for censored quantile regression models. A key property common to these approaches is that when the model is correctly specified, estimators based on the independence restriction are typically more efficient than those relying on quantile restrictions. This is shown by the extensive numerical evidence in \cite{HONORE1994241} and \cite{HONORE2004293}. 
\cite{CHEN2020239} introduced moment-based estimators under independence and conditional quantile restrictions under complicated error assumptions that are often difficult to verify in practice.  

In contrast to the existing approaches discussed above, we extend the work of \cite{Lee1992} to develop a general $M$-estimation framework for censored regression in the presence of endogeneity. 
The proposed estimation framework is based on an appropriately defined class of loss functions (see Section~\ref{Two-Stage Estimation via M-Estimation}) yielding a class of estimators indexed by loss functions. As such the proposed framework generalizes the existing quantile based approaches to more general functionals determined by their respective loss functions and provide alternatives to robust and semiparametric estimation of censored regression.   

\subsection{Contribution}
This paper develops an estimation framework for censored regression models with endogenous regressors. We propose a two-step M-estimation procedure based on the control function approach: the first step recovers the endogenous variation, while the second step estimates the structural parameters through a general M-estimation criterion. We adopt the idea of tuning the censoring threshold with the chosen loss function from \cite{Lee1992} and generalize the work from the Huber loss function to an appropriate class of loss functions (see Section~\ref{Two-Stage Estimation via M-Estimation}). The proposed class of loss functions is rich enough to incorporate convex, non-convex, and redescending (see e.g., \cite{SHEVLYAKOV20082906}) loss functions, providing flexibility and alternatives to robust estimation in censored regression. Moreover, the proposed class of estimators nests several existing estimators, such as the censored least absolute deviations (CLAD) estimator (\cite{POWELL1984}) for the absolute loss, when the tuning censoring threshold (see Section~\ref{Two-Stage Estimation via M-Estimation}) is set to be 0.

We establish consistency and asymptotic normality of the proposed general $M$-estimators indexed by their respective loss functions. In addition, we illustrate the general $M$-estimators by choosing loss functions with various properties, such as convex, non-convex, and redescending via a Monte Carlo study, and provide a comparative analysis with some existing estimators. In the course of the empirical study, we also propose an algorithm for data driven choice of tuning parameters involved in the loss functions to define a tuning censoring threshold (see Section~\ref{Two-Stage Estimation via M-Estimation}).

\subsection{ Mathematical Challenges} One of the main challenges in theoretical analysis is that the objective function is not  continuously differentiable in the present context as the regression function is not differentiable. Consequently, the usual approaches based on Taylor expansions of the objective function are not directly applicable here. However, if the population counterpart of the objective function is differentiable, we can adopt the technique discussed in \cite{PakesandPollard}, with appropriate modifications, for a generic Lipschitz-continuous loss function. Alternatively, one may adopt the technique proposed by \cite{huber1967under}, with slight modifications, although the choice between these approaches remains somewhat subjective.

\noindent It is important to note that the method outlined in \cite{PakesandPollard} applies more directly to the estimator developed in \cite{Honoré1992} than to the estimators discussed in this paper. This is because the population moment conditions associated with those panel estimators are not defined for multi-step estimation problems and do not rely on preliminary estimates of certain parameters. To this end, we extend the regularity conditions in \cite{Honoré1992} to a broader class of loss functions in order to establish consistency and asymptotic normality.

\subsection{Organisation of the Paper}
The remainder of the paper is organised as follows. Section \ref{Model description and estimation procedure} introduces the model and describes the estimation procedure. Section \ref{Asymptotic properties} develops the theoretical framework and states the main asymptotic results. Section \ref{Simulated Data Study} presents a simulation study examining the finite-sample performance of the proposed estimator. Section \ref{Conclusion} consists of some concluding remarks. Proofs of all theorems are provided in Appendix (Section \ref{appendix}).

\section{Model Description and Estimation Procedure  }\label{Model description and estimation procedure}
\subsection{The Model} \label{The model} 
Let $Y_1, \ldots, Y_n$ be random variables generated from the following regression model, with the $p + 1 $ regressors denoted by  $\boldsymbol{\Tilde{X}}\in \mathbb{R}^{p}$ and $W \in \mathbb{R}$:
\begin{equation} \label{Eq:2.1}
    Y_i = \max\left(\boldsymbol{\Tilde{x}_i}^\top \boldsymbol{\alpha}_0 + w_i \gamma_{1, 0} + \eta_i, c\right),
\end{equation}
where $\{\boldsymbol{\Tilde{x}_i}, i= 1, \ldots, n\}$ are i.i.d. observations of the random vector $\boldsymbol{\Tilde{X}}$, $(\boldsymbol{\alpha}_0^\top, \gamma_{1, 0})^\top \in \mathbb{R}^{p+2}$ is the vector of unknown true parameters including the intercept parameter, $c$ is some known real number, $\{w_i, i = 1, \dots, n\}$ are observations of the random variables $\{W_i, i = 1, \ldots, n\}$ which are $n$ replications of $W$ generated from the linear regression model
\begin{equation} \label{Eq:2.2}
    W_i = \boldsymbol{z}_i^T \boldsymbol{\delta}_0 + \vartheta_i.
\end{equation}
Here, $\{\boldsymbol{z}_i = (\boldsymbol{\Tilde{x}_i}^\top, z_{1, i})^\top, i =1, \ldots, n\}$ are exogenous regressors, where $z_{1, i}$'s are i.i.d observations of the \emph{instrumental variable} $Z_1 \in \mathbb{R}$, $\boldsymbol{\delta}_0$ is the unknown true parameter in the parameter space $\Delta$, and the structure of the unobserved random errors $\vartheta_i$ are specified according to the control function approach (see \cite{Smith}, \cite{NEWEY19}, \cite{RIVERS1988347}) as:
\begin{equation} \label{Eq:2.3}
    \eta_i = \vartheta_i \gamma_{2, 0} + \varepsilon_i,
\end{equation}
where $\mathbb{E}(\vartheta_i|\boldsymbol{z}_i)=0$, $var(\vartheta_i|\boldsymbol{z}_i) = \sigma^2_\vartheta > 0$, $\mathbb{E}(\varepsilon_i|\boldsymbol{\Tilde{x}}_i, w_i)=0$, and $var(\varepsilon_i|\boldsymbol{\Tilde{x}}_i, w_i) = \sigma^2_\varepsilon > 0$, $\sigma^2_\vartheta$ and $\sigma^2_\varepsilon$ are unknown constants. Moreover, the control approach requires the following assumptions.  
\begin{A1}\label{a:2.1}
$\varepsilon_i$ and $\vartheta_i$ are independent. 
\end{A1}
\begin{A1}\label{a:2.2}
    The joint distribution of $(\eta_i, \vartheta_i)^\top$ is independent of $\boldsymbol{Z}:=(\boldsymbol{\Tilde{X}}^\top, Z_1)^\top$. 
\end{A1}
The above assumptions imply that $\varepsilon_i$ is independent of $\boldsymbol{Z}$. Note that by Equation~\eqref{Eq:2.3}, $W_i$ is endogenous provided that $\gamma_{2, 0} \neq 0$ which governs the degree of endogeneity. The control function approach considers the following model obtained by redefining the error structure according to Equation~\eqref{Eq:2.3},
\begin{equation} \label{Eq:2.4}
    Y_i = \max\left(\boldsymbol{\Tilde{x}_i}^\top \boldsymbol{\alpha}_0 + w_i \gamma_{1, 0} + \vartheta_i \gamma_{2, 0} + \varepsilon_i, c\right).
\end{equation}
The model in Equation~\eqref{Eq:2.1} is called the censored regression model with censoring threshold $c$. Tobit models are a particular case of such models for $c = 0$, and for normally distributed errors. In this paper, we consider $c = 0$ without loss of generality and consider a more general class of error distributions. That is, we consider the following model derived from Equation~\eqref{Eq:2.4}:  
\begin{equation} \label{Eq:2.5}
    Y_i = \max\left(\boldsymbol{x}_i^\top \boldsymbol{\beta}_0 + \varepsilon_i, 0\right),
\end{equation} 
where $\boldsymbol{x}_i \equiv \boldsymbol{x}_i(\boldsymbol{\boldsymbol{\delta}}_0)  = (\boldsymbol{\Tilde{x}}_i^\top, w_i, \vartheta_i(\boldsymbol{\boldsymbol{\delta}_0}))^\top$, $\boldsymbol{\beta}_0 = (\boldsymbol{\alpha}_0^\top, \gamma_{1, 0}, \gamma_{2, 0})^\top$ is the vector of unknown parameters in the parameter space $\mathcal{B}$.

We remark here that one can formulate and implement the proposed estimation framework for multiple endogenous regressors under a routine set of assumptions required for a multivariate regression model. Hence, the restriction to a single endogenous regressor can be considered as an expositional and mathematical simplicity rather than a potential limitation of the proposed estimation framework. 

\subsection{Two-Stage Estimation via M-Estimation}\label{Two-Stage Estimation via M-Estimation}
We adopt a two-stage estimation procedure to estimate $\boldsymbol{\beta}_0$ of the model in Equation~\eqref{Eq:2.5}. Note that one of the regressors $\vartheta_i(\boldsymbol{\delta}_0)$ depends on the unknown parameter $\boldsymbol{\delta}_0$ of the model in Equation~\eqref{Eq:2.2} which we call the \emph{first-stage} model henceforth. Since $\boldsymbol{\delta}_0$ is not known in practice, we estimate it from the first-stage model to observe the regressor $\vartheta_i(\boldsymbol{\delta}_0)$. Subsequently, having an estimate of $\boldsymbol{\delta}_0$, say $\boldsymbol{\hat{\delta}}_n$, we estimate $\boldsymbol{\beta}_0$ from the following model, which we call the \emph{second-stage} model: 
\begin{equation} \label{Eq:2.6}
  Y_i = \max\left(\boldsymbol{\hat{x}}_i^\top \boldsymbol{\beta}_0 + \epsilon_i(\boldsymbol{\hat{\delta}}_n), 0\right), 
\end{equation}
where $\boldsymbol{\hat{x}}_i \equiv \boldsymbol{\hat{x}}_i(\boldsymbol{\hat{\delta}}_n) = (\boldsymbol{\Tilde{x}}_i^\top, w_i, e_{i}(\boldsymbol{\hat{\delta}}_n))^\top$, $\epsilon_i(\boldsymbol{\hat{\delta}}_n) = \varepsilon_i + \gamma_{2, 0}(\vartheta_i - e_{i}(\boldsymbol{\hat{\delta}}_n))$, and $e_i(\boldsymbol{\hat{\delta}}_n):= w_i - \boldsymbol{z}_i^\top \boldsymbol{\hat{\delta}}_n$ are residuals from the first-stage model. In this paper, we consider $\boldsymbol{\hat{\delta}}_n$ as the least-squares estimator (LSE) from the first-stage model to establish consistency and asymptotic normality of the proposed class of $M$-estimators. 

Thus, the second-stage estimation involves estimating $\boldsymbol{\beta}_0$ from the second-stage model using some appropriate techniques. In this paper, we propose the following class of estimators of $\boldsymbol{\beta}_0$ obtained using the M-estimation framework:
\begin{equation} \label{Eq:2.7}
    \boldsymbol{\hat{\beta}}_n(\rho, \boldsymbol{\hat{\delta}}_n):= \left\{ \boldsymbol{\hat{\beta}}_n \in \arg \min_{\boldsymbol{\beta} \in \mathcal{B}} \sum_{i = 1}^n \rho\left(Y_i - \max\left(\boldsymbol{\hat{x}_i}^\top \boldsymbol{\beta}, c_{\rho}\right)\right) : \rho \in \mathcal{L} \right\},
\end{equation}
where $\rho$ is a loss function, $c_{\rho} > 0$ is a tuning parameter depending on $\rho$, and the class of loss functions $\mathcal{L}$ is defined as  
\begin{equation} \label{Eq:2.8}
  \mathcal{L}:= \left\{\rho: \mathbb{R} \to [0, \infty) : \rho(0) = 0,\ \rho(x) = \rho(-x),\ \left|\rho(x)-\rho(y)\right|\leq k_{\rho}|x-y| \right\},
\end{equation}
here $k_{\rho} > 0$ is the Lipschitz constant associated with the loss function $\rho$. This class of loss functions consists of several convex and non-convex loss functions. In particular, it includes convex loss functions such as Huber and non-convex loss functions such as Hampel (\cite{Andrews1972}) and Tukey biweight (\cite{BeatonTukey1974}) loss functions. Thus, the proposed class of estimators includes several existing estimators such as \cite{Lee1992} for semiparametric estimation of censored regression models. We refer to Section~\ref{Loss Functions: Properties} for a detailed discussion on the choice of $c_{\rho}$ and the properties of the aforementioned loss functions.  

\section{Asymptotic Properties} \label{Asymptotic properties}
 This section provides asymptotic properties such as consistency and asymptotic normality of the proposed estimator $\boldsymbol{\hat{\beta}}_n(\rho, \boldsymbol{\hat{\delta}}_n)$ for a fix choice of $\rho \in \mathcal{L}$.

\subsection{Consistency}\label{SC}
In this section, we establish the consistency of the proposed estimator $\boldsymbol{\hat{\beta}}_n(\rho, \boldsymbol{\hat{\delta}}_n)$ to its population counterpart $\boldsymbol{\beta}_{0}(\rho, \boldsymbol{\delta}_0)$. Let $Y$ denote an identical copy of $Y_i$ and $\boldsymbol{x} \equiv\boldsymbol{x}(\boldsymbol{\delta}_0)$ denote an observed value of the random vector $\boldsymbol{X}:=(\boldsymbol{\Tilde{X}}^\top, W, \vartheta(\boldsymbol{\delta}_0))^\top$, define the following population objective function for any $\rho \in \mathcal{L}$ and $\boldsymbol{\beta} \in \mathcal{B}$,
\begin{equation}\label{Eq:3.1}
    Q(\boldsymbol{\beta}, \boldsymbol{\delta}_0):= \mathbb{E} \left[\rho \left(Y - \max\left(\boldsymbol{X}^\top \boldsymbol{\beta}, c_{\rho}\right)\right)\right]. 
\end{equation} 
We require the following assumptions on $\rho \in \mathcal{L}$ and error distribution to define $\boldsymbol{\beta}_{0}(\rho, \boldsymbol{\delta}_0)$. 
\begin{A1}\label{a:3.1}
For each $\rho \in \mathcal{L}$, $\rho$ is differentiable except possibly on a finite set $D_{\rho}$ with derivative $\psi$, where $D_{\rho}$ denotes the set of points where $\rho$ is not differentiable. Moreover, on $D_{\rho}$, the left and right derivatives of $\rho$ denoted by $\psi^{-}$ and $\psi^{+}$, respectively, exists. 
\end{A1}

\begin{A1}\label{a:3.2}
    The conditional distribution of the error $\varepsilon$ (see Equation~\eqref{Eq:2.3}) given $\boldsymbol{x}$ is absolutely continuous with the conditional density $f_{\varepsilon \mid \boldsymbol{x}}(.)$ that is strictly unimodal, symmetric around zero and satisfies $\mathbb{E}\!\left[\psi(\varepsilon)\mid\boldsymbol{x}\right]=0$. Moreover, the first derivative of $f_{\varepsilon \mid \boldsymbol{x}}(.)$ exists, and is integrable. 
\end{A1}
\begin{A1} \label{a:3.3} 
    For any $\xi>0$ and $\boldsymbol{\beta} \in \mathcal{B}$, $\inf_{\|\beta - \beta_0\| > \xi} \big(Q(\boldsymbol{\beta},\boldsymbol{\delta}_{0})-Q(\boldsymbol{\beta}_{0},\boldsymbol{\delta}_{0})\big) > 0.$
\end{A1}
Then under Assumption~\ref{a:3.2} and Assumption~\ref{a:3.3} , $\boldsymbol{\beta}_{0}(\rho, \boldsymbol{\delta}_0)$ is defined as \begin{equation} \label{Eq:3.2}
 \boldsymbol{\beta}_{0}(\rho, \boldsymbol{\delta}_0) := \arg \min_{\boldsymbol{\beta} \in \mathcal{B}} Q(\boldsymbol{\beta}, \boldsymbol{\delta}_0).
\end{equation}
Equivalently, $\boldsymbol{\beta}_{0}(\rho, \boldsymbol{\delta}_0) \equiv \boldsymbol{\beta}_0$ can be defined as a solution to the following first-order moment condition 
\begin{equation}\label{Eq:3.3}
   \mathbb{E}\left[
\mathds{1}\!\left(\boldsymbol{X}^{\top}\boldsymbol{\beta}_{0}>c_{\rho}\right)\psi\!\left(Y-\boldsymbol{X}^{\top}\boldsymbol{\beta}_{0}\right)\boldsymbol{X}
\right]=\boldsymbol{0}. 
\end{equation}

Now, we present the consistency of  $\boldsymbol{\hat{\beta}}_n(\rho, \boldsymbol{\hat{\delta}}_n)$ for any fixed choice of $\rho \in \mathcal{L}$ under the following assumptions on the parameter space $\mathcal{B}$ and regressors $\boldsymbol{X}$ of the second-stage model. 
\begin{A1}\label{a:3.4}
    $\mathcal{B}$ is compact and $\boldsymbol{\beta}_{0}$ is an interior point of $\mathcal{B}$. 
\end{A1}
\begin{A1}\label{a:3.5}
    $\|\boldsymbol{X}\|\leq \mathcal{K}_{0}$ and $\|\boldsymbol{Z}\|\leq \mathcal{K}_{1}$ almost surely, where $\boldsymbol{X}:=(\boldsymbol{\Tilde{X}}^\top, W, \vartheta(\boldsymbol{\delta}_0))^\top$, $ \boldsymbol{Z}:=(\boldsymbol{\Tilde{X}}^\top, Z_1)^\top$, $\mathcal{K}_{0}$ and $\mathcal{K}_{1}$ are some real numbers.
\end{A1}

\begin{t1}\label{th1} Under Assumptions~\ref{a:2.1}-\ref{a:2.2} and Assumptions~\ref{a:3.1}-\ref{a:3.5}, for any fixed $\rho \in \mathcal{L}$, 
$$\boldsymbol{\hat{\beta}}_n(\rho, \boldsymbol{\hat{\delta}}_n) \xrightarrow{a.s.} \boldsymbol{\beta}_{0}(\rho, \boldsymbol{\delta}_0) \text{ provided } \boldsymbol{\hat{\delta}}_n \xrightarrow{a.s.} \boldsymbol{\delta}_0 \text{ as } n \to \infty .$$ 
\end{t1} 
\begin{proof}[$\textbf{Proof}$]
The proof is provided in Appendix \ref{app:B.1}.
\end{proof}

\subsection{Asymptotic Normality} \label{Asymptotic Normality}
This section provides asymptotic distribution of $\boldsymbol{\hat{\beta}}_n(\rho, \boldsymbol{\hat{\delta}}_n)$ for any fixed $\rho \in \mathcal{L}$. First, we define the sample counterpart of the population objective function as
\begin{equation}\label{Eq:3.4}
    Q_{n}(\boldsymbol{\beta}, \boldsymbol{\hat{\delta}}_{n}):= \frac{1}{n}\sum_{i=1}^{n}\rho\left(Y_{i}-  \max \left\{c_{\rho},{\boldsymbol{\hat{x}}}^\top_{i}\boldsymbol{\beta}\right\}\right).
\end{equation}
Note that $Q_{n}(\boldsymbol{\beta}, \boldsymbol{\hat{\delta}}_{n})$ is not differentiable in $\boldsymbol{\beta}$ as $\max \left\{c_{\rho},{\boldsymbol{\hat{x}}}^\top_{i}\boldsymbol{\beta}\right\}$ is not differentiable on $\{\boldsymbol{\beta} \in \mathcal{B}:\boldsymbol{\hat{x}}^\top_{i}\boldsymbol{\beta} = c_{\rho} \}$. Hence, the standard approach of establishing asymptotic normality based on the Taylor expansion of the sample objective function does not directly apply in the present context.

We adopt the technique developed by \cite{pollard1985new} and \cite{PakesandPollard}, which is well-suited to establishing asymptotic normality of estimators for non-smooth regression functions. For instance, \cite{Honoré1992} adopted these techniques for panel data with non-differentiable moment restrictions. However, unlike \cite{Honoré1992}, in the present context, we adopt a two-stage estimation, which requires modifying the framework of \cite{PakesandPollard}, accordingly. In particular, since the first-stage estimation affects the randomness of the second-stage estimation, we derive the joint asymptotic distribution of $(\boldsymbol{\hat{\beta}}_n^\top(\rho, \boldsymbol{\hat{\delta}}_n),\boldsymbol{\hat{\delta}}_{n}^\top)^\top$. This requires their population counterpart $(\boldsymbol{\beta}_0^\top(\rho, \boldsymbol{\delta}_0),\boldsymbol{\delta}_0^\top)^\top \equiv (\boldsymbol{\beta}_0^\top, \boldsymbol{\delta}_0^\top)^\top$ to solve the following joint first-order moment condition
\begin{align}\label{Eq:3.5}
\mathbb{J}(\boldsymbol{\beta},\boldsymbol{\delta}) &=  
          \begin{bmatrix}
          \mathbb{J}_{2}(\boldsymbol{\beta},\boldsymbol{\delta})  \\    \mathbb{J}_{1}(\boldsymbol{\delta})
          \end{bmatrix}  = \boldsymbol{0}, 
  \end{align}
\begin{equation}\label{Eq:3.6}
\begin{split}
  \text{ where }  \mathbb{J}_{2}(\boldsymbol{\beta},\boldsymbol{\delta}) &= \mathbb{E}[\mathds{1}(\boldsymbol{X}^\top(\boldsymbol{\delta})\boldsymbol{\beta}>c_{\rho})\psi(Y-\boldsymbol{X}^\top(\boldsymbol{\delta})\boldsymbol{\beta})\boldsymbol{X}(\boldsymbol{\delta})] \text{ and } \mathbb{J}_{1}(\boldsymbol{\delta})=\mathbb{E}[(W-\boldsymbol{Z}^\top\boldsymbol{\delta})\boldsymbol{Z}].
   \end{split}
\end{equation}
That is, we require $\mathbb{J}(\boldsymbol{\beta}_0, \boldsymbol{\delta}_0) = \boldsymbol{0}$. Note that, in Equation~\eqref{Eq:3.6}, $\boldsymbol{X}(\boldsymbol{\delta})$ is defined by replacing $\boldsymbol{\delta}_0$ in $\boldsymbol{X}(\boldsymbol{\delta}_0)$ (see Equation~\eqref{Eq:2.5}) by any arbitrary point $\boldsymbol{\delta} \in \Delta$. The following assumption, similar to Assumption 6 of \cite{CHEN201830} is sufficient to ensure the differentiability of $Q(\boldsymbol{\beta}, \boldsymbol{\delta}_0)$ at $\boldsymbol{\beta}_0$.  

\begin{A1}\label{a:3.6} There exist constants $0<k_{1}<\infty$ and $ 0< \ell_0 < \infty$ such that, for every $0<\ell\leq \ell_{0},$ we have
$$\mathbb{E}\left[\mathds{1}\left(\left|c_{\rho}-\boldsymbol{X}^\top\boldsymbol{\beta}_0\right|\leq\ell\right)\right]
  \leq k_{1}\ell.$$
 \end{A1}
In addition, we need the following assumptions on $\psi$ and some moment conditions on the regressors $\boldsymbol{Z}$ of the first-stage model.  
  \begin{A1}\label{a:3.7} $\psi(t) > 0$ on $(0, \infty)$, $\psi(t) = -\psi(t)$ for all $t \in \mathbb{R}$, and $\sup_{t \in \mathbb{R}} |\psi(t)|= \phi_{1}$, where $\phi_1 < \infty$ is some real number.       
 \end{A1}

 \begin{A1}\label{a:3.8} The following matrices exist, and $\boldsymbol{\Sigma}_{2, \boldsymbol{\beta}_0}$ and  $\boldsymbol{\Sigma}_{1, \boldsymbol{\delta}_0}$,  are positive definite. 
\begin{equation}\label{Eq:3.7}
    \begin{split}
    \boldsymbol{\Sigma}_{2,\boldsymbol{\beta}_{0}}&=\mathbb{E}[\mathds{1}(\boldsymbol{X}^\top\boldsymbol{\beta}_{0}>c_{\rho})C_{\psi}\boldsymbol{X}\boldsymbol{X}^\top],\;
    \boldsymbol{\Sigma}_{2,\boldsymbol{\delta}_{0}}=\mathbb{E}[\mathds{1}(\boldsymbol{X}^\top\boldsymbol{\beta}_{0}>c_{\rho})C_{\psi}\gamma_{2,0}\boldsymbol{X}\boldsymbol{Z}^\top],\\
     \boldsymbol{\Sigma}_{1,\boldsymbol{\delta}_{0}}&=\mathbb{E}[\boldsymbol{Z}\boldsymbol{Z}^\top],\quad
     \mathbb{D}_{1}=\mathbb{E}[\vartheta^2\boldsymbol{Z}\boldsymbol{Z}^\top],\quad
      \mathbb{D}_{2}=\mathbb{E}[\mathds{1}(\boldsymbol{X}^\top\boldsymbol{\beta}_{0}>c_{\rho})\psi^{2}(\varepsilon)\boldsymbol{X}\boldsymbol{X}^\top].
    \end{split}
    \end{equation}
Here, $ 0 < C_{\psi} \leq \phi_2$, for some real number $\phi_2 > 0$, is defined as 
  \begin{equation} \label{Eq:3.8}
      C_{\psi}:=- \int_{- \infty}^{\infty} \psi(t) f^{\prime}_{\varepsilon|\boldsymbol{x}} (t) dt = \int_{- \infty}^{\infty} \psi^{\prime}(t) f_{\varepsilon|\boldsymbol{x}} (t) dt,
  \end{equation}
 where the second equality holds whenever $\psi$ is differentiable or $\psi^{'}$ exist almost surely. In addition, we assume that $\boldsymbol{\Sigma}_{2,\boldsymbol{\delta}_{0}}$ has full column rank, and $\mathbb{D}_1$ and $\mathbb{D}_2$ have full rank.
\end{A1}
 
\begin{t1}\label{th2} Let $\boldsymbol{\beta}_0$ and $\boldsymbol{\hat{\beta}}_n$ denote $\boldsymbol{\beta}_0(\rho, \boldsymbol{\delta}_0)$ (see Equation~\eqref{Eq:3.2}) and $\boldsymbol{\hat{\beta}}_n(\rho, \boldsymbol{\hat{\delta}}_n)$ (see Equation~\eqref{Eq:2.7}), respectively, for any fixed $\rho \in \mathcal{L}$. Then, under the conditions of Theorem~\ref{th1} and Assumptions~\ref{a:3.6}-\ref{a:3.8}, we have 
$$\sqrt{n}\begin{pmatrix}
          \begin{bmatrix}
           \boldsymbol{\hat{\beta}}_{n}\\
          \boldsymbol{\hat{\delta}}_{n}
          \end{bmatrix} -
          \begin{bmatrix}
           \boldsymbol{\beta}_{0}\\
          \boldsymbol{\delta}_{0}
         \end{bmatrix}
         \end{pmatrix}\xrightarrow{D} \mathbb{N}_{2p+3}(\boldsymbol{0}, \boldsymbol{\Sigma}^{-1} \mathbb{D} (\boldsymbol{\Sigma}^{-1})^\top)$$
where the matrices $\boldsymbol{\Sigma}$ and $\mathbb{D}$ are given by:
\begin{equation}\label{Eq:3.9}
    \boldsymbol{\Sigma}=\begin{bmatrix}
        \boldsymbol{\Sigma}_{2,\boldsymbol{\beta}_{0}} &\boldsymbol{\Sigma}_{2,\boldsymbol{\delta}_{0}}\\
        \boldsymbol{0}&\boldsymbol{\Sigma}_{1,\boldsymbol{\delta}_{0}}
    \end{bmatrix},\; \mathbb{D}=\begin{bmatrix}
        \mathbb{D}_{2} &\boldsymbol{0}\\
        \boldsymbol{0}&\mathbb{D}_{1}
    \end{bmatrix},\end{equation}
with the components defined in Equation~\eqref{Eq:3.7}. 
    
Moreover, an application of partitioned matrix multiplication to this result will yield the conclusions: 
\begin{equation}
    \sqrt{n}(\boldsymbol{\hat{\delta}}_{n}-\boldsymbol{\delta}_{0})\xrightarrow{D} \mathbb{N}_{p+1}(\boldsymbol{0},\boldsymbol{\Sigma}_{1,\boldsymbol{\delta}_{0}}^{-1}\mathbb{D}_{1}(\boldsymbol{\Sigma}_{1,\boldsymbol{\delta}_{0}}^{-1})^\top)
\end{equation}
and 
\begin{equation} \label{Eq:3.10}
    \sqrt{n}( \boldsymbol{\hat{\beta}}_{n}-\boldsymbol{\beta}_{0})\xrightarrow{D} \mathbb{N}_{p+2}(\boldsymbol{0},\boldsymbol{\Sigma}_{2,\boldsymbol{\beta}_{0}}^{-1}\{\mathbb{D}_{2}+\boldsymbol{\Sigma}_{2,\boldsymbol{\delta}_{0}}\boldsymbol{\Omega}_{1}\boldsymbol{\Sigma}_{2,\boldsymbol{\delta}_{0}}^\top\}(\boldsymbol{\Sigma}_{2,\boldsymbol{\beta}_{0}}^{-1})^\top),
\end{equation}
where $\boldsymbol{\Omega}_{1}=\boldsymbol{\Sigma}_{1,\boldsymbol{\delta}_{0}}^{-1}\mathbb{D}_{1}(\boldsymbol{\Sigma}_{1,\boldsymbol{\delta}_{0}}^{-1})^\top.$
\end{t1}
\begin{proof}[$\textbf{Proof}$]
The proof is provided in Appendix \ref{app:B.2}.
\end{proof}

\begin{r1} \label{Remark 3.2}
The expression of $C_{\psi}$ in Assumption~\ref{a:3.8} is derived as follows. Consider $\Sigma_{2, \boldsymbol{\beta}_0}:= \mathbb{E}\left[ \mathds{1} \left( \boldsymbol{X}^\top \boldsymbol{\beta}_0 > c_{\rho} \right)\boldsymbol{X} \Psi(\boldsymbol{\beta_0};\boldsymbol{X}) \right]$, where $\Psi(\boldsymbol{\beta}; \boldsymbol{x})$ is defined as  
\begin{equation}
    \Psi(\boldsymbol{\beta};\boldsymbol{x}):= - \frac{\partial}{\partial \boldsymbol{\boldsymbol{\beta}^\top}} \mathbb{E}\left[ \psi\left( Y - \boldsymbol{x}^\top \boldsymbol{\beta}\right) | \boldsymbol{X} = \boldsymbol{x}\right] = -\frac{\partial}{\partial \boldsymbol{\boldsymbol{\beta}^\top}} \int_{- \infty}^{\infty} \psi\left(u - \boldsymbol{x}^\top \left(\boldsymbol{\beta} - \boldsymbol{\beta_0}\right)\right) f_{\varepsilon | \boldsymbol{x}} (u) du. 
\end{equation}
Now, by the change of variable $u - \boldsymbol{x}^\top (\boldsymbol{\beta} - \boldsymbol{\beta_0}) = t$ in the above integral and using the assumptions \ref{a:3.2} and \ref{a:3.7} to differentiate under the integral sign yields 
\begin{equation} \label{Eq:3.12}
  \Psi(\boldsymbol{\beta};\boldsymbol{x}) = -\int_{- \infty}^{\infty} \psi(t) f_{\varepsilon | \boldsymbol{x}}^\prime \left(t + \boldsymbol{x}^\top \left(\boldsymbol{\beta} - \boldsymbol{\beta_0}\right)  \right) \boldsymbol{x}^\top dt.  
\end{equation}
Hence, by Equation~\eqref{Eq:3.12}, we have $\Psi(\boldsymbol{\beta}_0;\boldsymbol{x}) = C_{\psi} \boldsymbol{x}^\top$. Thus $C_{\psi}$ is defined in Equation~\eqref{Eq:3.8} where the second equality follows from applying integration by parts, whenever $\psi$ is differentiable. 
\end{r1}

\begin{r1} \label{Remark 3.3}  Under the assumptions \ref{a:3.2} and \ref{a:3.7}, $C_{\psi} > 0$, which is essential for $\Sigma_{2, \boldsymbol{\beta}_0}$ to be positive definite to hold Assumption~\ref{a:3.8}. This follows from the fact that both $\psi$ and $f_{\varepsilon | \boldsymbol{x}}^\prime$ have opposite signs in $(-\infty, 0)$ and $(0, \infty)$, since $\psi$ is odd by Assumption~\ref{a:3.7} and $f_{\varepsilon | \boldsymbol{x}}^\prime(t) < 0$ on $(0, \infty)$ by strict unimodality and symmetry of $f_{\varepsilon}|\boldsymbol{x}$ around 0 (see Assumption~\ref{a:3.2}). This implies that the function $\psi(t) f_{\varepsilon | \boldsymbol{x}}^\prime(t)$ is even which simplifies $C_{\psi}$ as
$$C_\psi = - 2\int_{0}^{\infty} \psi(t) f_{\varepsilon | \boldsymbol{x}}^\prime(t) dt .$$
Now, further observe that $\psi(t) > 0$ on $(0, \infty)$ by Assumption~\ref{a:3.7} and $f_{\varepsilon | \boldsymbol{x}}^\prime(t) < 0$ on $(0, \infty)$. This implies that $C_{\psi} > 0$.
\end{r1}

\begin{r1} \label{Remark 3.4}
The matrix $\boldsymbol{\Sigma}_{2,\boldsymbol{\delta}_{0}}\boldsymbol{\Omega}_{1}\boldsymbol{\Sigma}_{2,\boldsymbol{\delta}_{0}}^\top$ in Equation~\eqref{Eq:3.10} represents the variability introduced in the second-stage estimation by the first-stage estimation. 
In particular, when no endogenous regressors are present, that is $\gamma_{2, 0} = 0$, then $\Sigma_{2, \boldsymbol{\delta}_0} = \boldsymbol{0}$. In this case, the asymptotic covariance matrix coincides with that of the standard M-estimation for a censored regression model involving only exogenous regressors. 

Moreover, the derived expression for the asymptotic covariance matrix in Equation~\eqref{Eq:3.10} clearly shows that the presence of an endogenous regressor increases the variability of the second-stage estimator by the amount $\boldsymbol{\Sigma}_{2,\boldsymbol{\beta}_{0}}^{-1}\{\boldsymbol{\Sigma}_{2,\boldsymbol{\delta}_{0}}\boldsymbol{\Omega}_{1}\boldsymbol{\Sigma}_{2,\boldsymbol{\delta}_{0}}^\top\}(\boldsymbol{\Sigma}_{2,\boldsymbol{\beta}_{0}}^{-1})^\top$, which is a positive semidefinite matrix. In particular, the additional term  $\boldsymbol{\Sigma}_{2,\boldsymbol{\beta}_{0}}^{-1}\{\boldsymbol{\Sigma}_{2,\boldsymbol{\delta}_{0}}\boldsymbol{\Omega}_{1}\boldsymbol{\Sigma}_{2,\boldsymbol{\delta}_{0}}^\top\}(\boldsymbol{\Sigma}_{2,\boldsymbol{\beta}_{0}}^{-1})^\top$ captures the effect of endogeneity. Consequently, ignoring this adjustment term when $\gamma_{2, 0} \neq 0$, may lead to misleading large-sample inference. For instance, the derived asymptotic confidence intervals may be narrower compared to the case when the adjustment term is included in the asymptotic covariance matrix.
\end{r1}

\subsection{Consistent Estimation of  \texorpdfstring{$\boldsymbol{\Sigma}$}{} and \texorpdfstring{$\mathbb{D}$}{}}
In this subsection, we provide consistent estimators of $\boldsymbol{\Sigma}$ and $\mathbb{D}$ that can be used to facilitate large-sample inference such as hypothesis testing and confidence intervals for $\boldsymbol{\beta}_0(\rho, \boldsymbol{\delta}_0)$. 

Let $\hat{f}_{\varepsilon|\boldsymbol{x}}(.)$ and $\hat{f}^\prime_{\varepsilon|\boldsymbol{x}}(.)$ denote estimates of $f_{\varepsilon|\boldsymbol{x}}(.)$ and $f^\prime_{\varepsilon|\boldsymbol{x}}(.)$, respectively. Then, we define the estimate $\hat{C}_{\psi}$ of $C_{\psi}$ as
\begin{equation} \label{Eq:3.14}
   \hat{C}_{\psi}:= - \int_{- \infty}^{\infty} \psi(t) \hat{f}^\prime_{\varepsilon|\boldsymbol{x}}(t) dt \text{ or when $\psi^\prime$ exists then } \hat{C}_{\psi}:= \int_{- \infty}^{\infty} \psi^{\prime}(t) \hat{f}_{\varepsilon|\boldsymbol{x}}(t) dt. 
\end{equation}
We estimate matrices $\boldsymbol{\Sigma}$ and $\mathbb{D}$ as
 \begin{equation}\label{Eq:3.15}
 \hat{\boldsymbol{\Sigma}}=
\begin{bmatrix}\hat{\boldsymbol{\Sigma}}_{2,\boldsymbol{\beta}}&\hat{\boldsymbol{\Sigma}}_{2,\boldsymbol{\delta}}\\
     \boldsymbol{0}&\hat{\boldsymbol{\Sigma}}_{1,\boldsymbol{\delta}}\\
 \end{bmatrix},\;
 \hat{\mathbb{D}}=
\begin{bmatrix}\hat{\mathbb{D}}_{2}&\boldsymbol{0}\\
     \boldsymbol{0}&\hat{\mathbb{D}}_{1}\\
 \end{bmatrix},
 \end{equation}    
where the individual components are defined as follows:
\begin{equation}\label{Eq:3.16}
    \begin{split}
    \hat{\boldsymbol{\Sigma}}_{2,\boldsymbol{\beta}}&=\frac{1}{n}\sum\limits_{i=1}^{n}\mathds{1}(\hat{\boldsymbol{x}}_{i}^\top \boldsymbol{\hat{\beta}}_{n}>c_{\rho}) \hat{C}_{\psi} \hat{\boldsymbol{x}}_{i}\hat{\boldsymbol{x}}^\top_{i},\quad
     \hat{\boldsymbol{\Sigma}}_{2,\boldsymbol{\delta}}=\frac{1}{n}\sum\limits_{i=1}^{n}\mathds{1}(\hat{\boldsymbol{x}}_{i}^\top \boldsymbol{\hat{\beta}}_{n}>c_{\rho})\hat{C}_{\psi}\hat{\gamma}_{2, 0}\hat{\boldsymbol{x}}_{i}\boldsymbol{z}_{i}^\top,\\
     \hat{\boldsymbol{\Sigma}}_{1,\boldsymbol{\delta}}&=\frac{1}{n}\sum\limits_{i=1}^{n}\boldsymbol{z}_{i}\boldsymbol{z}_{i}^\top, \quad
     \hat{\mathbb{D}}_{1}=\frac{1}{n}\sum\limits_{i=1}^{n}e_{i}^2\boldsymbol{z}_{i}\boldsymbol{z}_{i}^\top, \quad
     \hat{\mathbb{D}}_{2} =\frac{1}{n}\sum\limits_{i=1}^{n}\mathds{1}(\hat{\boldsymbol{x}}_{i}^\top \boldsymbol{\hat{\beta}}_{n}>c_{\rho})\psi^{2}(\hat{\varepsilon}_{i})\hat{\boldsymbol{x}}_{i}\hat{\boldsymbol{x}}_{i}^\top.
    \end{split}  
 \end{equation}
Here, $\hat{\varepsilon}_{i}=Y_{i}-\hat{\boldsymbol{x}}^\top_{i} \boldsymbol{\hat{\beta}}_{n}(\rho, \boldsymbol{\hat{\delta}}_n)$, $e_{i}= W_{i} - \boldsymbol{z}^\top_{i}\boldsymbol{\hat{\delta}}_{n}$ and $\hat{\gamma}_{2,0}:= v_{p+3}^\top \boldsymbol{\hat{\beta}}_{n}(\rho, \boldsymbol{\hat{\delta}}_n)$ is an estimator of $\gamma_{2,0}$, where $v_{p+3} = (\boldsymbol{0}^\top, 1)^\top \in \mathbb{R}^{p+3}$ and $\boldsymbol{0}$ is a null vector in $\mathbb{R}^{p+2}$.   

\begin{t1}\label{th3}
Under the conditions of Theorem \ref{th2},  $\Hat{\boldsymbol{\Sigma}}$ and $\Hat{\mathbb{D}}$ are consistent. That is, $$\hat{\boldsymbol{\Sigma}}-\boldsymbol{\Sigma}=o_{P}(1) \text{ and } \hat{\mathbb{D}}-\mathbb{D}=o_{P}(1) \text{ provided } \hat{C}_{\psi} - C_{\psi} = o_P(1).$$
\end{t1}
\begin{proof}[$\textbf{Proof}$]
    The proof is provided in Appendix \ref{app:B.3}.
\end{proof}

\section{Empirical Study}\label{Simulated Data Study}
This section presents an empirical study based on simulated datasets to illustrate the proposed class of estimators for the model defined in Equation~\eqref{Eq:2.6}. We consider Huber (\cite{Huber1964}), Hampel (\cite{Andrews1972}), and Tukey biweight (\cite{BeatonTukey1974}) loss functions, yielding three different estimators derived from the proposed class. We examine the finite-sample performance of these estimators in terms of mean bias, median bias, mean squared error (MSE), and median absolute error (MAE) and provide a comparative analysis with some existing estimators. In the following subsections, we first examine the required properties of the chosen loss functions and subsequently present some simulation studies.

\subsection{Loss Functions: Properties} \label{Loss Functions: Properties}
We illustrate the properties of the following loss functions for the censored regression model $Y_i = \max (\boldsymbol{x}_i^\top \boldsymbol{\beta}_0 + \varepsilon_i, 0)$ for exogenous regressors under the standard assumptions on the error $\varepsilon_i, \ i =1, \ldots, n$. In what follows, $\mathds{1}(.)$ and sign(.) denote the indicator function and the sign function, respectively. 

\subsubsection{Huber Loss Function} \label{Huber Loss Function}
The Huber loss function (\cite{Huber1964}) is defined as  
\begin{equation}
    \rho(t) := 0.5t^2 \ \mathds{1}\left(|t| < \lambda\right) + \left(\lambda|t| - 0.5 \lambda^2\right) \mathds{1}(|t| \geq \lambda), 
\end{equation}
where $\lambda > 0$ is a tuning parameter. In the context of the censored regression model, this loss function is adopted in \cite{Lee1992} and the resulting estimator is called the \emph{winsorized mean estimator} (WME) for censored regression. In this case $c_{\rho} = \lambda$ in Equation~\eqref{Eq:2.7}. 

The WME is obtained by minimizing $\sum_{i = 1}^n \rho(Y_i - \max(\boldsymbol{x}_i^\top \boldsymbol{\beta}, \lambda))$ with respect to $\boldsymbol{\beta} \in \mathcal{B}$, under the assumption that the true value $\boldsymbol{\beta}_0$ solves: 
\begin{equation} \label{Eq:4.2}
    \mathbb{E}\left[\psi\left(Y_i - \max\left( \boldsymbol{x}_i^\top \boldsymbol{\beta}, \lambda \right)\right)\right] = 0,
\end{equation}
where $\psi$ is the first derivative of $\rho$ defined as
\begin{equation}
    \psi(t):= t \ \mathds{1}\left(|t| < \lambda \right) + \lambda \ \text{sign}(t) \ \mathds{1}(|t| > \lambda).
\end{equation}
The condition in Equation~\eqref{Eq:4.2} is equivalent to $\mathbb{E}\left(\psi(\varepsilon_i)|\boldsymbol{x}_i\right) = 0$. It is also straightforward to verify that the Huber loss satisfies the required properties of the class $\mathcal{L}$ (see Equation~\eqref{Eq:2.8}), Assumption~\ref{a:3.1}, and Assumption~\ref{a:3.7}. Moreover, as $ \lambda \to 0$, the Huber loss reduces to the absolute loss function, yielding the censored least absolute deviation estimator (CLAD) proposed in \cite{POWELL1984}. 

\subsubsection{Hampel Loss Function}\label{Hampel Loss Function}
The Hampel loss function (\cite{Andrews1972}) is defined as
\begin{equation}
    \rho(t):= 0.5t^2 \mathds{1}\left(|t| \leq \lambda_1\right) +  \rho_1(t) \mathds{1} \left( \lambda_1 < |t| \leq \lambda_2\right) + \rho_2(t) \mathds{1}\left( \lambda_2 < |t| \leq \lambda_3\right) + \rho_3(t) \mathds{1}\left( |t| > \lambda_3\right),
\end{equation}
where $\rho_1(t) := \lambda_1 |t| - 0.5\lambda_1^2, \ \rho_2(t):= \lambda_1 (\lambda_2 - 0.5 \lambda_1) + (\lambda_3 - \lambda_2)0.5 \lambda_1\left[1- (\lambda_3 - \lambda_2)^{-2} (\lambda_3 - |t|)^2\right]$, $\rho_3(t):= 0.5\lambda_1 (\lambda_2  + \lambda_3 - \lambda_1 )$ and $\lambda_1, \ \lambda_2, \text{ and } \lambda_3$ are tuning parameters such that $0< \lambda_1 < \lambda_2 < \lambda_3$. 

In the context of censored regression models, this loss function has not been used in the literature. In this paper, we use this loss function in the proposed framework of general $M$-estimation defined in Equation~\eqref{Eq:2.7} with $c_{\rho} = \lambda_3$ under the assumption that $\boldsymbol{\beta}_0$ solves:
\begin{equation}
    \mathbb{E}\left[\psi\left(Y_i - \max\left( \boldsymbol{x}_i^\top \boldsymbol{\beta}, \lambda_3 \right)\right)\right] = 0,
\end{equation}
where $\psi$ is the first derivative of $\rho$ defined as 
\begin{equation}
    \psi(t):= t \mathds{1}(|t| \leq \lambda_1 ) + \lambda_1 \operatorname{sign}(t) \mathds{1}(\lambda_1 < |t| \leq \lambda_2 ) + \lambda_1 \operatorname{sign}(t)  \frac{(\lambda_3 - |t|)}{\lambda_3 - \lambda_2} \mathds{1}(\lambda_2 < |t| \leq \lambda_3 ).    
\end{equation}
The population moment condition $\mathbb{E}(\psi(\varepsilon)|\boldsymbol{x}) = 0$ holds as $\psi$ is odd and distribution of $\varepsilon$ is symmetric around 0 (see Assumption~\ref{a:3.2}). Moreover, other required conditions of the class $\mathcal{L}$, Assumption~\ref{a:3.1}, and Assumption~\ref{a:3.7} are also satisfied. 

The Hampel loss function is non-convex with a redescending score function $\psi$. That is $\psi$ has the property that $\lim_{|t| \to \infty}\psi (t) = 0$. The resulting estimators corresponding to such loss functions are called \emph{redescending M-estimators} (see, e.g., \cite{SHEVLYAKOV20082906}). Intuitively, this property ensures that observations with extremely large residuals have diminishing influence on the estimating equation, thereby enhancing robustness against outliers.

\noindent It behaves like the squared-error loss for small residuals, assigns a constant influence to moderate residuals, and progressively reduces the influence of large residuals until it becomes zero. In particular, observations with residuals exceeding the threshold $\lambda_3$ receive zero weight. Consequently, do not affect the estimation process, regardless of the magnitude of the residual. Owing to this redescending property, the Hampel loss function is more resistant to extreme outliers than the Huber loss function. The estimator obtained by minimizing the objective function based on this loss is referred to as the Hampel M-estimator (HME) for censored regression.

\subsubsection{Tukey Biweight Loss Function}\label{Tukey Biweight Loss Function}
The Tukey biweight (bisquare) loss function (\cite{BeatonTukey1974}) is defined as
\begin{equation}
    \rho(t):= \frac{\lambda^2}{6} \left[1 - \left(1- \lambda^{-2} t^2\right)^3\right] \mathds{1} \left(|t| \leq \lambda\right) + \frac{\lambda^2}{6} \mathds{1} \left(|t| > \lambda\right),
\end{equation}
where $\lambda > 0$ is a tuning parameter. In the context of censored regression, this loss function has also not been used so far. In this paper, we use this loss function in Equation~\eqref{Eq:2.7} with $c_{\rho} = \lambda$ under the population moment condition $\mathbb{E}(\psi(\varepsilon)|\boldsymbol{x}) = 0$, where $\psi$ is defined as
\begin{equation}\label{Eq:4.8}
\psi(t) = t \left(1 - \lambda^{-2} t^2\right)^2 \mathds{1}(|t| \leq \lambda). 
\end{equation}
It is straightforward to verify that the population moment condition holds and the loss function admits the desired properties of the class $\mathcal{L}$, Assumption~\ref{a:3.1}, and Assumption~\ref{a:3.7}. 

This loss function possesses the same properties as the Hampel loss function, such as non-convexity and the redescending score function. However, this loss function has an edge over the Hampel loss function, as it contains only one tuning parameter. Moreover, this loss function also has an edge over both the Huber and the Hampel loss functions since it is twice differentiable. Consequently, this loss function is preferred when the aim is to construct confidence intervals or to perform hypothesis testing for the parameters. This is because the consistent estimation of $\Sigma$ (see Equation~\eqref{Eq:3.9}, \eqref{Eq:3.14}, and \eqref{Eq:3.15}) requires consistent estimation of the error density $f_{\varepsilon|\boldsymbol{x}}$ rather than its derivative $f^\prime_{\varepsilon|\boldsymbol{x}}$, which is difficult in practice. The estimator obtained by minimizing the objective function based on this loss is referred to as the Tukey biweight M-estimator (TME) for censored regression. 

\subsection{Monte Carlo Study}
This section presents a Monte Carlo study to examine the finite-sample performance of $\boldsymbol{\hat{\beta}}_{n}(\rho, \boldsymbol{\hat{\delta}}_n)$ (see Equation~\eqref{Eq:2.7}) for the Huber, the Hampel, and the Tukey biweight loss functions. We generate a random sample $Y_1,\ldots,Y_n $ of size $n$ as:
\begin{equation}
    Y_i = \max\left(\beta_0 + x_{i}\beta_1 + w_i\beta_2 + \vartheta_i \beta_3 + \varepsilon_i, 0\right),
\end{equation}
where $\{x_{i}\}$ are observed values of the exogenous regressor $X\sim\mathbb{N}(0,1)$ and $\{w_i\}$ are observed values of the endogenous regressor generated as
\begin{equation}
    W_i = \delta_0 + x_{i} \delta_1  + z_{i} \delta_2 + \vartheta_i. 
\end{equation} 
Here, $\{z_i\}$ are observed values of the instrumental variable $Z \sim \mathbb{U}(0,1)$ and $\{\vartheta_i\}$ are i.i.d. random variables from $\mathbb{N}(0,1)$. The model errors $\{\varepsilon_i\}$ are i.i.d.\ random variables generated from a
variety of symmetric distributions, including the standard normal distribution
$\mathbb{N}(0,1)$, the Student's $t$-distribution with 3 degrees of freedom ($t_3$),
the standard Cauchy distribution $\mathbb{C}(0,1)$, and two contaminated
distributions. The first is a contaminated Laplace distribution with contamination
proportion $0.2$, that is,
$\varepsilon_i \sim 0.8\,\mathbb{DE}(0,1)+0.2\,\mathbb{DE}(0,5)$, where
$\mathbb{DE}(0,b)$ denotes the Laplace distribution with location $0$ and scale $b$.
The second is a contaminated Student's $t$-distribution with contamination proportion
$0.1$, that is, $\varepsilon_i \sim 0.9\,t_{1}+0.1\,t_{2}$, where $t_n$
denotes the Student's $t$-distribution with $n$ degrees of freedom.
We initialize the true parameters $ (\beta_0, \beta_1, \beta_2)^\top = (\delta_0, \delta_1, \delta_2)^\top = (1, 2, 3)^\top$ and $\beta_{3}=0.5$.  
 
We consider existing estimators, such as symmetric trimmed censored least squares (SCLS) \cite{SCLS} and the censored regression quantile
estimator (CRQ) \cite{POWELL1986143} with the quantile level $\tau=0.75$, as benchmarks for comparison with the estimators derived from the proposed class. 
\subsubsection{Data-driven tuning-parameter selection}
Now, in the following algorithm, we provide a data-driven choice of the tuning parameters involved in the loss functions discussed in Section~\ref{Loss Functions: Properties}. We propose to define tuning parameters as some appropriate quantiles of the distribution of the random variable $Y_i - \boldsymbol{\hat{x}}_i^\top \boldsymbol{\beta}$ for some suitable choice of $\boldsymbol{\beta} \in \mathcal{B}$. Here,   $\boldsymbol{\hat{x}}_i := (1, x_i, w_i, \vartheta_i(\boldsymbol{\hat{\delta}}_n))^\top$, $\boldsymbol{\hat{\delta}}_n$ is LSE from the first stage, and $\boldsymbol{\beta}:= (\beta_0, \beta_1, \beta_2, \beta_3)^\top$. The proposed algorithm specifies $\boldsymbol{\beta}$ as the censored absolute deviation (CLAD) estimator $\boldsymbol{\hat{\beta}}_n$. We use the fact that $Y_i - \boldsymbol{\hat{x}}_i^\top \boldsymbol{\hat{\beta}}_n$ can be written as a right censored random variable $T_i:= \min(-\varepsilon_i, \boldsymbol{\hat{x}}_i^\top \boldsymbol{\hat{\beta}}_n)$ with the random censoring threshold $\boldsymbol{\hat{x}}_i^\top \boldsymbol{\hat{\beta}}_n$. We tacitly assume that random censoring thresholds $\{\boldsymbol{\hat{x}}_i^\top \boldsymbol{\hat{\beta}}_n\}$ and $\{\varepsilon_i\}$ are independent and estimate the distribution of $T_i$ using Kaplan-Meier estimator (\cite{KaplanMeier1958}) from the observed data $\{r_i\}, r_i := Y_i - \boldsymbol{\hat{x}}_i^\top \boldsymbol{\hat{\beta}}_n$. Then, we define tuning parameters as some appropriate quantiles of this estimated distribution. In particular, the Huber tuning constant (see Section~\ref{Huber Loss Function}) is set to the 0.65th quantile, while the Tukey tuning constant (see Section~\ref{Tukey Biweight Loss Function}) is set to the 0.85th quantile. The three Hampel tuning constants $\lambda_1, \lambda_2$, and $\lambda_3$ (see Section~\ref{Hampel Loss Function}) are chosen as the 0.65th, 0.75th, and 0.85th quantiles, respectively.

 \begin{algorithm}[htbp]
\caption{Data-driven selection of loss function tuning parameters}
\label{alg:tuning}
\begin{algorithmic}[1]

\Require Left-censored sample $\{(Y_i,x_{i},w_{i},z_i)\}_{i=1}^{n}$ with censoring threshold $0$.
\Ensure Tuning parameters for the Huber, Hampel, and Tukey loss functions.

\State Estimate LSE $\boldsymbol{\hat{\delta}}_n$ from the first- stage model.

\State Define $\boldsymbol{\hat{x}}_i := (1, x_i, w_i, \vartheta_i(\boldsymbol{\hat{\delta}}_n ))^\top$, $\vartheta_i(\boldsymbol{\hat{\delta}}_n):= w_i - \Tilde{Z}_i^\top\boldsymbol{\hat{\delta}}_n$, where $\Tilde{Z}_i:= (1, x_i, z_i)^\top$. 

\State Estimate CLAD $\boldsymbol{\hat{\beta}}_n$ from the second-stage model.

\State Define $r_i:= Y_i - \boldsymbol{\hat{x}}_i^\top \boldsymbol{\hat{\beta}}_n, \ i = 1, \ldots, n $.

\State Estimate distribution $\hat{F}_n(.)$ of $\{r_i\}$ using Kaplan-Meier estimator.  

\State Define $Q_{\tau}:= \inf\{t: \hat{F}_n(t) \geq \tau \}, \ \tau \in (0, 1) $.

\State Set
\[
\lambda_{\mathrm{Huber}}=Q_{0.65}, \qquad
(\lambda_1,\lambda_2,\lambda_3)_{\mathrm{Hampel}}=(Q_{0.65},Q_{0.75},Q_{0.85}), \qquad
\lambda_{\mathrm{Tukey}}=Q_{0.85}.
\]

\State \Return $\lambda_{\mathrm{Huber}}$, $(\lambda_1,\lambda_2,\lambda_3)_{\mathrm{Hampel}}$, and $\lambda_{\mathrm{Tukey}}$.

\end{algorithmic}
\end{algorithm}

We use \texttt{optim} function in \textsf{R} to estimate $\beta_{0}$, $\beta_{1}$, $\beta_{2}$, and $\beta_{3}$ using the Nelder--Mead simplex algorithm. To improve numerical stability and convergence, the initial values supplied to the optimization algorithm are taken as the censored least absolute deviations (CLAD) estimates obtained from the Tobit model under the control function approach. The simulation procedure is repeated $r=2000$ times for each sample size $n\in\{100, 200,400,500,600,800,1000\}$.

The performance of the estimators is evaluated using the empirical bias, empirical mean squared error (MSE), mean absolute error (MAE), and median absolute error (MedAE) of the parameter estimates $\hat{\beta}_{0}$, $\hat{\beta}_{1}$, $\hat{\beta}_{2}$, and $\hat{\beta}_{3}$, defined as
$$\mathrm{Bias}(\hat{\beta}_{j})
=\frac{1}{r}\sum_{k=1}^{r}
\left(\hat{\beta}_{jk}-\beta_{j}\right),
\qquad j=0,1,2,3,$$
$$\mathrm{MSE}(\hat{\beta}_{j})
=\frac{1}{r}\sum_{k=1}^{r}
\left(\hat{\beta}_{jk}-\beta_{j}\right)^2,
\qquad j=0,1,2,3,$$
$$\mathrm{MAE}(\hat{\beta}_{j})
=\frac{1}{r}\sum_{k=1}^{r}
\left|\hat{\beta}_{jk}-\beta_{j}\right|,
\qquad j=0,1,2,3,$$
and
$$\mathrm{MedAE}(\hat{\beta}_{j})
=
\operatorname{median}
\left\{
\left|\hat{\beta}_{jk}-\beta_{j}\right|
\right\}_{k=1}^{r},
\qquad j=0,1,2,3.$$

Here, $\hat{\beta}_{jk}$ denotes the estimate of the parameter $\beta_{j}$ obtained from the $k$th Monte Carlo replication, where $k=1,2,\ldots,r$. 

The overall censoring percentage (the percentage of censored observations) ranges from approximately $18\%$ to $20\%$ across the considered simulation settings. The simulation results in Tables \ref{table:1}–\ref{table:5} evaluate the finite-sample performance of the CRQ, SCLS, WME, HME, and TME estimators. It is evident from the results that across all five error distributions, the Mean Bias, MSE, MAE, and MedAE of each estimator decrease steadily as the sample size increases, suggesting improved precision in larger samples and supporting the consistency of the estimators. The empirical mean biases remain small and tend towards zero as $n$ grows, so the differences among the estimators arise mainly from their variability rather than from bias. The only notable exception is SCLS at $n=100$ under the standard Cauchy distribution $\mathbb{C}(0,1)$ (Table \ref{table:2}) and the contaminated mixture $0.9\,t_1+0.1\,t_2$ (Table \ref{table:5}), where the bias is also non-negligible. Now, in the following, we summarize the results in Tables \ref{table:1}–\ref{table:5}. 

Table \ref{table:1} presents the results for the model error $\varepsilon \sim \mathbb{N}(0,1)$. Here, the censoring percentage remains 18\% (approximately) across all the sample sizes. In this case, the SCLS estimator outperforms all other estimators exhibiting the lowest MSE, MAE, and MedAE for all three regression coefficients across all the sample sizes. Next, the WME outperforms CRQ, HME and TME and is close to SCLS as its performance measures are marginally larger. The CRQ estimator also performs well, typically outperforming HME and TME in terms of MSE, MAE, and MedAE. In contrast, HME and TME tend to produce larger values of these measures, indicating relatively weaker performance under normally distributed errors. The mean biases of all estimators are comparatively small, and the MSE, MAE, and MedAE decreases as the sample size increases. Thus, the results indicate that, under the standard normal error distribution, SCLS provides the most accurate estimates among the methods examined, while WME serves as a competitive alternative. The consistently lower MSE, MAE, and MedAE values for SCLS and WME reflect their favorable finite-sample performance. In contrast, HME and TME generally show less favorable performance which may be influenced by the choice of initial values and tuning parameters.

Table \ref{table:2} presents the results for the model error $\varepsilon \sim \mathbb{C}(0,1)$. Here, the censoring percentage remains 20\% (approximately) across all the sample sizes. While SCLS shows the best overall performance under the standard normal error distribution, its performance deteriorates considerably under the standard Cauchy error distribution. In particular, SCLS produces an extremely large MSE for all three regression coefficients at $n=100$ and continues to display relatively larger MSE, MAE, and MedAE values across all the sample sizes. This pattern reflects its sensitivity to heavy-tailed errors. In contrast, the CRQ, WME, HME, and TME estimators generally perform better, with their MSE, MAE, and MedAE values decreasing as the sample size increases. Among these estimators, TME generally attains the smallest MSE, MAE, and MedAE values for all three regression coefficients at larger sample sizes, especially for $n=400$ and above. HME performs competitively and yields results close to those of TME, whereas WME offers a substantial improvement over SCLS. The CRQ estimator performs reasonably well, particularly at smaller sample sizes, but is generally outperformed by the proposed M-estimators as the sample size increases. 

Table \ref{table:3} presents the results for the model error $\varepsilon \sim t_3$. Here, the censoring percentage remains 18\% (approximately) across all the sample sizes. In this case, the WME outperforms all the other estimators exhibiting the smallest MSE, MAE, and MedAE for all three regression coefficients across all the sample sizes. The SCLS and CRQ estimators also perform reasonably well, however, their MSE, MAE, and MedAE values are generally higher than those of WME. The HME and TME estimators show Moderate performance, with their relative performance varying across coefficients and sample sizes. Furthermore, the MSE, MAE, and MedAE values decrease as the sample size increases for all estimators, indicating improved performance with larger samples. The mean biases also remain relatively small and generally approach to zero as the sample size increases. Overall, these findings suggest that WME provides the most favorable performance under the heavy-tailed \(t_3\) error distribution.

Tables~\ref{table:4} presents the results for the model error $\varepsilon \sim 0.8\,\mathbb{DE}(0,1)+0.2\,\mathbb{DE}(0,5)$. Here, the censoring percentage remains 19\% (approximately) across all the sample sizes. In this case, all the estimators exhibit generally good performance. The TME generally yields the smallest MSE, MAE, and MedAE across regression coefficients and sample sizes, while WME and HME exhibit comparable performance. In contrast, CRQ yields comparatively larger values, while SCLS performs the least favorably. The MSE, MAE, and MedAE generally decrease as the sample size increases, while the mean biases remain relatively small for WME, HME, and TME. 

Table~\ref{table:5} presents the results for the model error $ \varepsilon \sim0.9\,t_{1}+0.1\,t_{2}$. Here, the censoring percentage remains 20\% (approximately) across all the sample sizes. In this case, the differences between the estimators become more apparent compared to the results in Tables~\ref{table:4}. The SCLS estimator exhibits exceptionally large MSE values at \(n=100\) and continues to perform poorly for larger sample sizes. At \(n=100\), WME yields the smallest MSE for all three regression coefficients, whereas TME generally performs best for \(n\geq 200\), followed closely by HME and WME. The CRQ estimator performs better than SCLS but generally remains less favourable than the WME, HME, and TME. In this case also, the MSE, MAE, and MedAE generally decrease as the sample size increases, while the mean biases remain relatively small for WME, HME, and TME. 

\begin{table}[H]
\centering
\small
\caption{Results for the model error $\varepsilon \sim \mathbb{N}(0,1)$.}
\label{table:1}
\resizebox{\textwidth}{!}{%
\begin{tabular}{|c|c|c|c|c|c|c|c|c|c|c|c|c|}
\hline
\textbf{Estimators}
& \multicolumn{12}{c|}{\textbf{$\mathbb{N}(0,1)$, $n=100$, C.P.=18\%}} \\
\cline{2-13}
& \multicolumn{4}{c|}{$\hat{\beta}_{1}$}
& \multicolumn{4}{c|}{$\hat{\beta}_{2}$}
& \multicolumn{4}{c|}{$\hat{\beta}_{3}$} \\
\cline{2-13}
& \textbf{Mean Bias} & \textbf{MSE} & \textbf{MAE} & \textbf{MedAE}
& \textbf{Mean Bias} & \textbf{MSE} & \textbf{MAE} & \textbf{MedAE}
& \textbf{Mean Bias} & \textbf{MSE} & \textbf{MAE} & \textbf{MedAE} \\
\hline
\textbf{CRQ}
& -0.0182 & 0.1526 & 0.3097 & 0.2585
& 0.0060 & 0.0323 & 0.1420 & 0.1171
& -0.0091 & 0.0549 & 0.1880 & 0.1601 \\
\textbf{SCLS}
& 0.0157 & 0.0977 & 0.2450 & 0.1998
& -0.0095 & 0.0217 & 0.1144 & 0.0924
& 0.0105 & 0.0328 & 0.1429 & 0.1176 \\
\textbf{WME}
& 0.0157 & 0.1093 & 0.2619 & 0.2161
& -0.0098 & 0.0243 & 0.1226 & 0.1055
& 0.0099 & 0.0368 & 0.1515 & 0.1284 \\
\textbf{HME}
& 0.0177 & 0.2208 & 0.3692 & 0.3066
& -0.0101 & 0.0478 & 0.1727 & 0.1439
& 0.0086 & 0.0784 & 0.2215 & 0.1879 \\
\textbf{TME}
& 0.0107 & 0.3191 & 0.4495 & 0.3747
& -0.0040 & 0.0668 & 0.2032 & 0.1637
& 0.0078 & 0.1119 & 0.2606 & 0.2117 \\
\hline
& \multicolumn{12}{c|}{\textbf{$n=200$, C.P.=18\%}} \\
\cline{2-13}
\textbf{CRQ}
& -0.0149 & 0.0804 & 0.2219 & 0.1839
& 0.0098 & 0.0172 & 0.1034 & 0.0856
& -0.0058 & 0.0275 & 0.1317 & 0.1094 \\
\textbf{SCLS}
& 0.0052 & 0.0489 & 0.1752 & 0.1472
& -0.0015 & 0.0105 & 0.0805 & 0.0670
& 0.0030 & 0.0167 & 0.1029 & 0.0846 \\
\textbf{WME}
& 0.0060 & 0.0552 & 0.1864 & 0.1548
& -0.0019 & 0.0118 & 0.0852 & 0.0703
& 0.0034 & 0.0187 & 0.1092 & 0.0900 \\
\textbf{HME}
& 0.0159 & 0.1165 & 0.2689 & 0.2232
& -0.0060 & 0.0254 & 0.1266 & 0.1042
& 0.0063 & 0.0401 & 0.1585 & 0.1330 \\
\textbf{TME}
& 0.0212 & 0.1706 & 0.3231 & 0.2698
& -0.0084 & 0.0376 & 0.1526 & 0.1258
& 0.0133 & 0.0614 & 0.1969 & 0.1678 \\
\hline
& \multicolumn{12}{c|}{\textbf{$n=400$, C.P.=18\%}} \\
\cline{2-13}
\textbf{CRQ}
& -0.0008 & 0.0402 & 0.1581 & 0.1321
& 0.0003 & 0.0088 & 0.0743 & 0.0622
& -0.0027 & 0.0142 & 0.0954 & 0.0822 \\
\textbf{SCLS}
& 0.0063 & 0.0246 & 0.1250 & 0.1042
& -0.0023 & 0.0054 & 0.0586 & 0.0506
& 0.0009 & 0.0085 & 0.0733 & 0.0637 \\
\textbf{WME}
& 0.0048 & 0.0275 & 0.1319 & 0.1090
& -0.0015 & 0.0060 & 0.0616 & 0.0525
& -0.0003 & 0.0095 & 0.0774 & 0.0651 \\
\textbf{HME}
& 0.0008 & 0.0590 & 0.1930 & 0.1623
& -0.0004 & 0.0127 & 0.0896 & 0.0740
& -0.0030 & 0.0211 & 0.1160 & 0.0962 \\
\textbf{TME}
& 0.0006 & 0.0859 & 0.2320 & 0.1951
& 0.0009 & 0.0189 & 0.1085 & 0.0901
& -0.0018 & 0.0313 & 0.1405 & 0.1187 \\
\hline
& \multicolumn{12}{c|}{\textbf{$n=500$, C.P.=18\%}} \\
\cline{2-13}
\textbf{CRQ}
& 0.0032 & 0.0335 & 0.1455 & 0.1218
& 0.0001 & 0.0073 & 0.0675 & 0.0550
& -0.0013 & 0.0118 & 0.0867 & 0.0745 \\
\textbf{SCLS}
& 0.0085 & 0.0206 & 0.1137 & 0.0967
& -0.0027 & 0.0044 & 0.0524 & 0.0437
& 0.0020 & 0.0067 & 0.0656 & 0.0565 \\
\textbf{WME}
& 0.0080 & 0.0230 & 0.1198 & 0.1010
& -0.0023 & 0.0049 & 0.0551 & 0.0452
& 0.0017 & 0.0075 & 0.0697 & 0.0600 \\
\textbf{HME}
& 0.0065 & 0.0490 & 0.1748 & 0.1442
& -0.0008 & 0.0104 & 0.0809 & 0.0686
& 0.0009 & 0.0167 & 0.1032 & 0.0867 \\
\textbf{TME}
& 0.0051 & 0.0686 & 0.2089 & 0.1725
& 0.0003 & 0.0148 & 0.0963 & 0.0802
& 0.0014 & 0.0245 & 0.1227 & 0.1005 \\
\hline
& \multicolumn{12}{c|}{\textbf{$n=600$, C.P.=18\%}} \\
\cline{2-13}
\textbf{CRQ}
& -0.0023 & 0.0257 & 0.1281 & 0.1081
& 0.0012 & 0.0056 & 0.0600 & 0.0501
& -0.0015 & 0.0092 & 0.0769 & 0.0647 \\
\textbf{SCLS}
& 0.0019 & 0.0162 & 0.1012 & 0.0852
& -0.0010 & 0.0036 & 0.0480 & 0.0404
& 0.0011 & 0.0055 & 0.0584 & 0.0498 \\
\textbf{WME}
& 0.0017 & 0.0175 & 0.1052 & 0.0882
& -0.0015 & 0.0040 & 0.0499 & 0.0419
& 0.0015 & 0.0061 & 0.0620 & 0.0518 \\
\textbf{HME}
& 0.0058 & 0.0356 & 0.1507 & 0.1272
& -0.0037 & 0.0082 & 0.0720 & 0.0618
& 0.0042 & 0.0131 & 0.0917 & 0.0771 \\
\textbf{TME}
& 0.0038 & 0.0525 & 0.1823 & 0.1546
& -0.0026 & 0.0120 & 0.0869 & 0.0726
& 0.0024 & 0.0193 & 0.1106 & 0.0939 \\
\hline
& \multicolumn{12}{c|}{\textbf{$n=800$, C.P.=18\%}} \\
\cline{2-13}
\textbf{CRQ}
& 0.0066 & 0.0192 & 0.1098 & 0.0913
& -0.0019 & 0.0042 & 0.0517 & 0.0442
& 0.0021 & 0.0069 & 0.0660 & 0.0558 \\
\textbf{SCLS}
& 0.0076 & 0.0116 & 0.0860 & 0.0723
& -0.0030 & 0.0025 & 0.0398 & 0.0332
& 0.0032 & 0.0040 & 0.0503 & 0.0427 \\
\textbf{WME}
& 0.0080 & 0.0128 & 0.0901 & 0.0747
& -0.0033 & 0.0028 & 0.0421 & 0.0348
& 0.0031 & 0.0045 & 0.0533 & 0.0440 \\
\textbf{HME}
& 0.0107 & 0.0276 & 0.1316 & 0.1101
& -0.0045 & 0.0062 & 0.0627 & 0.0536
& 0.0031 & 0.0099 & 0.0794 & 0.0667 \\
\textbf{TME}
& 0.0143 & 0.0413 & 0.1618 & 0.1381
& -0.0058 & 0.0092 & 0.0762 & 0.0640
& 0.0038 & 0.0149 & 0.0973 & 0.0828 \\
\hline
& \multicolumn{12}{c|}{\textbf{$n=1000$, C.P.=18\%}} \\
\cline{2-13}
\textbf{CRQ}
& -0.0023 & 0.0159 & 0.1016 & 0.0885
& 0.0011 & 0.0035 & 0.0473 & 0.0399
& -0.0015 & 0.0054 & 0.0591 & 0.0501 \\
\textbf{SCLS}
& 0.0019 & 0.0092 & 0.0768 & 0.0647
& -0.0009 & 0.0022 & 0.0370 & 0.0316
& 0.0007 & 0.0033 & 0.0458 & 0.0388 \\
\textbf{WME}
& 0.0023 & 0.0103 & 0.0814 & 0.0695
& -0.0012 & 0.0024 & 0.0394 & 0.0344
& 0.0011 & 0.0037 & 0.0487 & 0.0406 \\
\textbf{HME}
& 0.0040 & 0.0216 & 0.1178 & 0.1003
& -0.0022 & 0.0049 & 0.0559 & 0.0471
& 0.0034 & 0.0081 & 0.0712 & 0.0607 \\
\textbf{TME}
& 0.0073 & 0.0330 & 0.1449 & 0.1237
& -0.0027 & 0.0073 & 0.0680 & 0.0572
& 0.0048 & 0.0126 & 0.0895 & 0.0757 \\
\hline
\end{tabular}%
}
\end{table}

\begin{table}[H]
\centering
\small
\caption{Results for the model error $\varepsilon \sim \mathbb{C}(0,1)$.}
\label{table:2}
\resizebox{\textwidth}{!}{%
\begin{tabular}{|c|c|c|c|c|c|c|c|c|c|c|c|c|}
\hline
\textbf{Estimators}
& \multicolumn{12}{c|}{\textbf{$\mathbb{C}(0,1)$, $n=100$, C.P.=20\%}} \\
\cline{2-13}
& \multicolumn{4}{c|}{$\hat{\beta}_{1}$}
& \multicolumn{4}{c|}{$\hat{\beta}_{2}$}
& \multicolumn{4}{c|}{$\hat{\beta}_{3}$} \\
\cline{2-13}
& \textbf{Mean Bias} & \textbf{MSE} & \textbf{MAE} & \textbf{MedAE}
& \textbf{Mean Bias} & \textbf{MSE} & \textbf{MAE} & \textbf{MedAE}
& \textbf{Mean Bias} & \textbf{MSE} & \textbf{MAE} & \textbf{MedAE} \\
\hline
\textbf{CRQ}
& -0.0529 & 0.5703 & 0.5818 & 0.4697
& 0.0463 & 0.1207 & 0.2673 & 0.2142
& -0.0307 & 0.2158 & 0.3594 & 0.2885 \\
\textbf{SCLS}
& -0.2547 & 365.7863 & 1.6583 & 0.7408
& 0.3697 & 124.5796 & 0.8094 & 0.3589
& -0.2964 & 87.9058 & 0.9197 & 0.4577 \\
\textbf{WME}
& 0.0128 & 0.3735 & 0.4691 & 0.3778
& 0.0032 & 0.0791 & 0.2161 & 0.1706
& 0.0053 & 0.1365 & 0.2864 & 0.2289 \\
\textbf{HME}
& -0.0033 & 0.4687 & 0.5023 & 0.4022
& 0.0155 & 0.1006 & 0.2318 & 0.1792
& -0.0034 & 0.1719 & 0.3024 & 0.2375 \\
\textbf{TME}
& -0.0096 & 0.4636 & 0.4935 & 0.3860
& 0.0174 & 0.1031 & 0.2302 & 0.1781
& -0.0051 & 0.1705 & 0.2939 & 0.2269 \\
\hline
& \multicolumn{12}{c|}{\textbf{$n=200$, C.P.=20\%}} \\
\cline{2-13}
\textbf{CRQ}
& -0.0351 & 0.2763 & 0.4103 & 0.3351
& 0.0271 & 0.0618 & 0.1936 & 0.1608
& -0.0146 & 0.1064 & 0.2582 & 0.2169 \\
\textbf{SCLS}
& 0.0316 & 0.7354 & 0.6672 & 0.5410
& 0.0059 & 0.1703 & 0.3245 & 0.2670
& 0.0132 & 0.2763 & 0.4109 & 0.3342 \\
\textbf{WME}
& 0.0052 & 0.1692 & 0.3242 & 0.2664
& 0.0019 & 0.0366 & 0.1513 & 0.1273
& 0.0080 & 0.0596 & 0.1916 & 0.1596 \\
\textbf{HME}
& 0.0059 & 0.2489 & 0.3203 & 0.2555
& 0.0072 & 0.0384 & 0.1472 & 0.1203
& 0.0093 & 0.1212 & 0.1936 & 0.1556 \\
\textbf{TME}
& 0.0059 & 0.1849 & 0.3049 & 0.2475
& 0.0060 & 0.0347 & 0.1417 & 0.1130
& 0.0098 & 0.1086 & 0.1874 & 0.1473 \\
\hline
& \multicolumn{12}{c|}{\textbf{$n=400$, C.P.=20\%}} \\
\cline{2-13}
\textbf{CRQ}
& -0.0377 & 0.1557 & 0.3150 & 0.2655
& 0.0203 & 0.0327 & 0.1444 & 0.1233
& -0.0177 & 0.0526 & 0.1814 & 0.1522 \\
\textbf{SCLS}
& -0.0060 & 0.3506 & 0.4713 & 0.3896
& 0.0079 & 0.0770 & 0.2226 & 0.1936
& -0.0116 & 0.1273 & 0.2826 & 0.2348 \\
\textbf{WME}
& -0.0061 & 0.0784 & 0.2244 & 0.1935
& 0.0020 & 0.0177 & 0.1068 & 0.0914
& -0.0019 & 0.0274 & 0.1312 & 0.1064 \\
\textbf{HME}
& -0.0044 & 0.0737 & 0.2168 & 0.1812
& 0.0025 & 0.0169 & 0.1030 & 0.0874
& 0.0014 & 0.0263 & 0.1291 & 0.1062 \\
\textbf{TME}
& -0.0045 & 0.0679 & 0.2074 & 0.1747
& 0.0025 & 0.0158 & 0.0996 & 0.0836
& 0.0015 & 0.0242 & 0.1235 & 0.1026 \\
\hline
& \multicolumn{12}{c|}{\textbf{$n=500$, C.P.=20\%}} \\
\cline{2-13}
\textbf{CRQ}
& -0.0205 & 0.1239 & 0.2824 & 0.2404
& 0.0124 & 0.0279 & 0.1335 & 0.1134
& -0.0062 & 0.0451 & 0.1683 & 0.1423 \\
\textbf{SCLS}
& -0.0057 & 0.2923 & 0.4285 & 0.3699
& 0.0080 & 0.0655 & 0.2035 & 0.1723
& 0.0038 & 0.1046 & 0.2584 & 0.2184 \\
\textbf{WME}
& 0.0014 & 0.0619 & 0.1976 & 0.1693
& 0.0009 & 0.0137 & 0.0938 & 0.0811
& 0.0021 & 0.0218 & 0.1171 & 0.0938 \\
\textbf{HME}
& 0.0033 & 0.0552 & 0.1855 & 0.1569
& 0.0014 & 0.0127 & 0.0891 & 0.0746
& 0.0031 & 0.0196 & 0.1107 & 0.0927 \\
\textbf{TME}
& 0.0025 & 0.0511 & 0.1782 & 0.1507
& 0.0019 & 0.0118 & 0.0856 & 0.0717
& 0.0032 & 0.0183 & 0.1066 & 0.0887 \\
\hline
& \multicolumn{12}{c|}{\textbf{$n=600$, C.P.=20\%}} \\
\cline{2-13}
\textbf{CRQ}
& -0.0179 & 0.0977 & 0.2496 & 0.2098
& 0.0130 & 0.0219 & 0.1187 & 0.0981
& -0.0137 & 0.0364 & 0.1513 & 0.1255 \\
\textbf{SCLS}
& 0.0095 & 0.2107 & 0.3654 & 0.3091
& 0.0035 & 0.0481 & 0.1743 & 0.1476
& 0.0063 & 0.0791 & 0.2236 & 0.1869 \\
\textbf{WME}
& 0.0003 & 0.0466 & 0.1709 & 0.1393
& 0.0009 & 0.0103 & 0.0802 & 0.0667
& -0.0002 & 0.0171 & 0.1034 & 0.0876 \\
\textbf{HME}
& 0.0001 & 0.0430 & 0.1658 & 0.1361
& 0.0016 & 0.0098 & 0.0783 & 0.0645
& -0.0018 & 0.0158 & 0.0991 & 0.0819 \\
\textbf{TME}
& -0.0004 & 0.0402 & 0.1599 & 0.1356
& 0.0017 & 0.0092 & 0.0758 & 0.0639
& -0.0020 & 0.0148 & 0.0961 & 0.0798 \\
\hline
& \multicolumn{12}{c|}{\textbf{$n=800$, C.P.=20\%}} \\
\cline{2-13}
\textbf{CRQ}
& -0.0142 & 0.0793 & 0.2238 & 0.1883
& 0.0080 & 0.0173 & 0.1054 & 0.0917
& -0.0047 & 0.0278 & 0.1339 & 0.1133 \\
\textbf{SCLS}
& -0.0030 & 0.1555 & 0.3120 & 0.2627
& 0.0073 & 0.0370 & 0.1527 & 0.1313
& 0.0013 & 0.0568 & 0.1908 & 0.1620 \\
\textbf{WME}
& -0.0005 & 0.0360 & 0.1526 & 0.1300
& 0.0001 & 0.0080 & 0.0718 & 0.0612
& 0.0007 & 0.0126 & 0.0901 & 0.0782 \\
\textbf{HME}
& -0.0009 & 0.0326 & 0.1439 & 0.1241
& -0.0008 & 0.0075 & 0.0690 & 0.0573
& 0.0006 & 0.0117 & 0.0861 & 0.0727 \\
\textbf{TME}
& -0.0005 & 0.0299 & 0.1379 & 0.1153
& -0.0010 & 0.0069 & 0.0661 & 0.0542
& 0.0003 & 0.0108 & 0.0828 & 0.0694 \\
\hline
& \multicolumn{12}{c|}{\textbf{$n=1000$, C.P.=20\%}} \\
\cline{2-13}
\textbf{CRQ}
& 0.0028 & 0.0617 & 0.1985 & 0.1698
& -0.0001 & 0.0136 & 0.0941 & 0.0804
& 0.0011 & 0.0217 & 0.1171 & 0.0988 \\
\textbf{SCLS}
& 0.0182 & 0.1280 & 0.2852 & 0.2368
& -0.0044 & 0.0299 & 0.1368 & 0.1119
& 0.0086 & 0.0483 & 0.1737 & 0.1449 \\
\textbf{WME}
& 0.0079 & 0.0298 & 0.1382 & 0.1171
& -0.0043 & 0.0065 & 0.0642 & 0.0540
& 0.0043 & 0.0104 & 0.0804 & 0.0669 \\
\textbf{HME}
& 0.0066 & 0.0278 & 0.1319 & 0.1078
& -0.0033 & 0.0062 & 0.0626 & 0.0525
& 0.0038 & 0.0094 & 0.0764 & 0.0625 \\
\textbf{TME}
& 0.0055 & 0.0258 & 0.1275 & 0.1046
& -0.0030 & 0.0058 & 0.0604 & 0.0514
& 0.0032 & 0.0088 & 0.0737 & 0.0603 \\
\hline
\end{tabular}%
}
\end{table}
 
\begin{table}[H]
\centering
\small
\caption{Results for the model error $\varepsilon \sim t_{3}$.}
\label{table:3}
\resizebox{\textwidth}{!}{%
\begin{tabular}{|c|c|c|c|c|c|c|c|c|c|c|c|c|}
\hline
\textbf{Estimators}
& \multicolumn{12}{c|}{\textbf{$t_{3}$, $n=100$, C.P.=18\%}} \\
\cline{2-13}
& \multicolumn{4}{c|}{$\hat{\beta}_{1}$}
& \multicolumn{4}{c|}{$\hat{\beta}_{2}$}
& \multicolumn{4}{c|}{$\hat{\beta}_{3}$} \\
\cline{2-13}
& \textbf{Mean Bias} & \textbf{MSE} & \textbf{MAE} & \textbf{MedAE}
& \textbf{Mean Bias} & \textbf{MSE} & \textbf{MAE} & \textbf{MedAE}
& \textbf{Mean Bias} & \textbf{MSE} & \textbf{MAE} & \textbf{MedAE} \\
\hline
\textbf{CRQ}
& -0.0129 & 0.2157 & 0.3679 & 0.3137
& 0.0119 & 0.0483 & 0.1718 & 0.1390
& -0.0079 & 0.0832 & 0.2292 & 0.1964 \\
\textbf{SCLS}
& 0.0409 & 0.2227 & 0.3700 & 0.3061
& -0.0177 & 0.0486 & 0.1738 & 0.1470
& 0.0150 & 0.0757 & 0.2182 & 0.1827 \\
\textbf{WME}
& 0.0323 & 0.1515 & 0.3056 & 0.2492
& -0.0134 & 0.0328 & 0.1425 & 0.1176
& 0.0134 & 0.0520 & 0.1826 & 0.1539 \\
\textbf{HME}
& 0.0263 & 0.2114 & 0.3570 & 0.2894
& -0.0093 & 0.0473 & 0.1706 & 0.1417
& 0.0111 & 0.0767 & 0.2177 & 0.1799 \\
\textbf{TME}
& 0.0299 & 0.2583 & 0.3921 & 0.3141
& -0.0104 & 0.0586 & 0.1888 & 0.1577
& 0.0105 & 0.0975 & 0.2413 & 0.1933 \\
\hline
& \multicolumn{12}{c|}{\textbf{$n=200$, C.P.=18\%}} \\
\cline{2-13}
\textbf{CRQ}
& -0.0218 & 0.1185 & 0.2753 & 0.2301
& 0.0112 & 0.0263 & 0.1283 & 0.1083
& -0.0075 & 0.0427 & 0.1652 & 0.1412 \\
\textbf{SCLS}
& 0.0049 & 0.1144 & 0.2651 & 0.2198
& -0.0031 & 0.0249 & 0.1230 & 0.1002
& 0.0030 & 0.0392 & 0.1569 & 0.1322 \\
\textbf{WME}
& 0.0053 & 0.0771 & 0.2199 & 0.1859
& -0.0043 & 0.0172 & 0.1033 & 0.0865
& 0.0025 & 0.0260 & 0.1269 & 0.1049 \\
\textbf{HME}
& 0.0032 & 0.1028 & 0.2512 & 0.2051
& -0.0037 & 0.0236 & 0.1204 & 0.0964
& 0.0010 & 0.0364 & 0.1515 & 0.1262 \\
\textbf{TME}
& 0.0000 & 0.1181 & 0.2682 & 0.2172
& -0.0017 & 0.0273 & 0.1293 & 0.1052
& -0.0020 & 0.0433 & 0.1630 & 0.1304 \\
\hline
& \multicolumn{12}{c|}{\textbf{$n=400$, C.P.=18\%}} \\
\cline{2-13}
\textbf{CRQ}
& -0.0154 & 0.0559 & 0.1884 & 0.1560
& 0.0079 & 0.0130 & 0.0914 & 0.0787
& -0.0088 & 0.0207 & 0.1153 & 0.0979 \\
\textbf{SCLS}
& 0.0042 & 0.0538 & 0.1827 & 0.1524
& -0.0013 & 0.0121 & 0.0867 & 0.0726
& -0.0002 & 0.0192 & 0.1094 & 0.0909 \\
\textbf{WME}
& 0.0078 & 0.0335 & 0.1451 & 0.1235
& -0.0040 & 0.0079 & 0.0709 & 0.0599
& 0.0040 & 0.0125 & 0.0891 & 0.0735 \\
\textbf{HME}
& 0.0129 & 0.0466 & 0.1705 & 0.1415
& -0.0061 & 0.0110 & 0.0839 & 0.0724
& 0.0069 & 0.0175 & 0.1050 & 0.0887 \\
\textbf{TME}
& 0.0134 & 0.0551 & 0.1838 & 0.1528
& -0.0064 & 0.0129 & 0.0903 & 0.0762
& 0.0058 & 0.0211 & 0.1148 & 0.0950 \\
\hline
& \multicolumn{12}{c|}{\textbf{$n=500$, C.P.=18\%}} \\
\cline{2-13}
\textbf{CRQ}
& -0.0053 & 0.0462 & 0.1715 & 0.1449
& 0.0034 & 0.0102 & 0.0815 & 0.0704
& -0.0046 & 0.0168 & 0.1047 & 0.0905 \\
\textbf{SCLS}
& 0.0026 & 0.0455 & 0.1692 & 0.1412
& -0.0005 & 0.0100 & 0.0797 & 0.0696
& 0.0023 & 0.0158 & 0.1008 & 0.0866 \\
\textbf{WME}
& 0.0030 & 0.0289 & 0.1342 & 0.1115
& -0.0002 & 0.0065 & 0.0643 & 0.0532
& 0.0009 & 0.0102 & 0.0807 & 0.0693 \\
\textbf{HME}
& 0.0041 & 0.0397 & 0.1586 & 0.1358
& 0.0000 & 0.0089 & 0.0757 & 0.0653
& -0.0005 & 0.0139 & 0.0928 & 0.0765 \\
\textbf{TME}
& 0.0023 & 0.0454 & 0.1697 & 0.1442
& 0.0008 & 0.0101 & 0.0804 & 0.0683
& -0.0011 & 0.0163 & 0.0997 & 0.0831 \\
\hline
& \multicolumn{12}{c|}{\textbf{$n=600$, C.P.=18\%}} \\
\cline{2-13}
\textbf{CRQ}
& -0.0065 & 0.0391 & 0.1561 & 0.1281
& 0.0039 & 0.0084 & 0.0729 & 0.0630
& -0.0018 & 0.0137 & 0.0920 & 0.0743 \\
\textbf{SCLS}
& 0.0036 & 0.0341 & 0.1459 & 0.1210
& -0.0001 & 0.0074 & 0.0688 & 0.0577
& 0.0010 & 0.0119 & 0.0867 & 0.0728 \\
\textbf{WME}
& 0.0015 & 0.0234 & 0.1214 & 0.1027
& 0.0002 & 0.0051 & 0.0573 & 0.0477
& -0.0002 & 0.0082 & 0.0722 & 0.0605 \\
\textbf{HME}
& 0.0004 & 0.0312 & 0.1411 & 0.1193
& 0.0002 & 0.0069 & 0.0660 & 0.0554
& -0.0021 & 0.0118 & 0.0861 & 0.0742 \\
\textbf{TME}
& -0.0003 & 0.0357 & 0.1521 & 0.1297
& 0.0005 & 0.0080 & 0.0715 & 0.0610
& -0.0021 & 0.0135 & 0.0920 & 0.0765 \\
\hline
& \multicolumn{12}{c|}{\textbf{$n=800$, C.P.=18\%}} \\
\cline{2-13}
\textbf{CRQ}
& -0.0048 & 0.0301 & 0.1379 & 0.1161
& 0.0023 & 0.0065 & 0.0638 & 0.0538
& -0.0037 & 0.0097 & 0.0781 & 0.0671 \\
\textbf{SCLS}
& 0.0071 & 0.0272 & 0.1308 & 0.1085
& -0.0019 & 0.0057 & 0.0597 & 0.0498
& 0.0009 & 0.0091 & 0.0750 & 0.0624 \\
\textbf{WME}
& 0.0042 & 0.0184 & 0.1074 & 0.0887
& -0.0017 & 0.0039 & 0.0492 & 0.0413
& 0.0007 & 0.0059 & 0.0612 & 0.0515 \\
\textbf{HME}
& 0.0037 & 0.0254 & 0.1264 & 0.1052
& -0.0022 & 0.0055 & 0.0588 & 0.0489
& 0.0012 & 0.0083 & 0.0726 & 0.0617 \\
\textbf{TME}
& 0.0056 & 0.0288 & 0.1345 & 0.1102
& -0.0027 & 0.0062 & 0.0620 & 0.0529
& 0.0020 & 0.0096 & 0.0775 & 0.0653 \\
\hline
& \multicolumn{12}{c|}{\textbf{$n=1000$, C.P.=18\%}} \\
\cline{2-13}
\textbf{CRQ}
& -0.0064 & 0.0224 & 0.1195 & 0.1015
& 0.0041 & 0.0049 & 0.0565 & 0.0486
& -0.0025 & 0.0081 & 0.0716 & 0.0601 \\
\textbf{SCLS}
& 0.0007 & 0.0212 & 0.1165 & 0.1008
& -0.0001 & 0.0046 & 0.0545 & 0.0461
& -0.0007 & 0.0074 & 0.0686 & 0.0577 \\
\textbf{WME}
& 0.0018 & 0.0140 & 0.0952 & 0.0825
& -0.0014 & 0.0031 & 0.0445 & 0.0379
& 0.0011 & 0.0049 & 0.0557 & 0.0476 \\
\textbf{HME}
& 0.0042 & 0.0191 & 0.1099 & 0.0917
& -0.0023 & 0.0043 & 0.0517 & 0.0432
& 0.0024 & 0.0068 & 0.0660 & 0.0560 \\
\textbf{TME}
& 0.0064 & 0.0220 & 0.1187 & 0.0994
& -0.0042 & 0.0050 & 0.0562 & 0.0471
& 0.0046 & 0.0080 & 0.0711 & 0.0606 \\
\hline
\end{tabular}%
}
\end{table}
\begin{table}[H]
\centering
\small
\caption{Results for the model error $\varepsilon \sim 0.8\,\mathbb{DE}(0,1)+0.2\,\mathbb{DE}(0,5)$.}
\label{table:4}
\resizebox{\textwidth}{!}{%
\begin{tabular}{|c|c|c|c|c|c|c|c|c|c|c|c|c|}
\hline
\textbf{Estimators}
& \multicolumn{12}{c|}{\textbf{$0.8\,\mathbb{DE}(0,1)+0.2\,\mathbb{DE}(0,5)$, $n=100$, C.P.=19\%}} \\
\cline{2-13}
& \multicolumn{4}{c|}{$\hat{\beta}_{1}$}
& \multicolumn{4}{c|}{$\hat{\beta}_{2}$}
& \multicolumn{4}{c|}{$\hat{\beta}_{3}$} \\
\cline{2-13}
& \textbf{Mean Bias} & \textbf{MSE} & \textbf{MAE} & \textbf{MedAE}
& \textbf{Mean Bias} & \textbf{MSE} & \textbf{MAE} & \textbf{MedAE}
& \textbf{Mean Bias} & \textbf{MSE} & \textbf{MAE} & \textbf{MedAE} \\
\hline
\textbf{CRQ}
& -0.0345 & 0.4431 & 0.5161 & 0.4217
& 0.0266 & 0.0924 & 0.2372 & 0.1950
& -0.0242 & 0.1516 & 0.3067 & 0.2552 \\
\textbf{SCLS}
& 0.0378 & 0.7841 & 0.6872 & 0.5699
& -0.0056 & 0.1662 & 0.3202 & 0.2579
& 0.0062 & 0.2626 & 0.4039 & 0.3319 \\
\textbf{WME}
& 0.0178 & 0.2849 & 0.4151 & 0.3377
& -0.0072 & 0.0616 & 0.1938 & 0.1580
& 0.0091 & 0.0929 & 0.2398 & 0.1978 \\
\textbf{HME}
& 0.0106 & 0.2892 & 0.4189 & 0.3471
& -0.0010 & 0.0649 & 0.1989 & 0.1604
& 0.0038 & 0.0998 & 0.2483 & 0.2069 \\
\textbf{TME}
& 0.0081 & 0.2690 & 0.4035 & 0.3279
& -0.0008 & 0.0606 & 0.1912 & 0.1523
& 0.0031 & 0.0921 & 0.2386 & 0.1972 \\
\hline
& \multicolumn{12}{c|}{\textbf{$n=200$, C.P.=19\%}} \\
\cline{2-13}
\textbf{CRQ}
& -0.0201 & 0.2124 & 0.3612 & 0.2986
& 0.0145 & 0.0469 & 0.1718 & 0.1427
& -0.0149 & 0.0768 & 0.2185 & 0.1817 \\
\textbf{SCLS}
& 0.0139 & 0.3354 & 0.4551 & 0.3690
& 0.0009 & 0.0727 & 0.2127 & 0.1760
& 0.0042 & 0.1231 & 0.2761 & 0.2292 \\
\textbf{WME}
& 0.0124 & 0.1143 & 0.2667 & 0.2226
& -0.0034 & 0.0246 & 0.1237 & 0.1014
& 0.0016 & 0.0403 & 0.1571 & 0.1282 \\
\textbf{HME}
& 0.0102 & 0.1203 & 0.2710 & 0.2233
& -0.0044 & 0.0270 & 0.1293 & 0.1061
& 0.0028 & 0.0413 & 0.1602 & 0.1326 \\
\textbf{TME}
& 0.0108 & 0.1102 & 0.2593 & 0.2189
& -0.0046 & 0.0245 & 0.1225 & 0.0991
& 0.0026 & 0.0376 & 0.1526 & 0.1256 \\
\hline
& \multicolumn{12}{c|}{\textbf{$n=400$, C.P.=19\%}} \\
\cline{2-13}
\textbf{CRQ}
& -0.0135 & 0.1042 & 0.2575 & 0.2156
& 0.0076 & 0.0231 & 0.1208 & 0.1002
& -0.0068 & 0.0388 & 0.1556 & 0.1299 \\
\textbf{SCLS}
& 0.0116 & 0.1599 & 0.3162 & 0.2605
& -0.0003 & 0.0356 & 0.1488 & 0.1242
& 0.0066 & 0.0607 & 0.1962 & 0.1621 \\
\textbf{WME}
& 0.0102 & 0.0523 & 0.1813 & 0.1524
& -0.0033 & 0.0117 & 0.0868 & 0.0755
& 0.0050 & 0.0193 & 0.1108 & 0.0913 \\
\textbf{HME}
& 0.0090 & 0.0551 & 0.1846 & 0.1544
& -0.0030 & 0.0122 & 0.0886 & 0.0756
& 0.0043 & 0.0193 & 0.1113 & 0.0933 \\
\textbf{TME}
& 0.0083 & 0.0507 & 0.1773 & 0.1489
& -0.0031 & 0.0110 & 0.0843 & 0.0722
& 0.0045 & 0.0174 & 0.1057 & 0.0888 \\
\hline
& \multicolumn{12}{c|}{\textbf{$n=500$, C.P.=19\%}} \\
\cline{2-13}
\textbf{CRQ}
& -0.0229 & 0.0835 & 0.2316 & 0.1975
& 0.0126 & 0.0182 & 0.1089 & 0.0965
& -0.0094 & 0.0311 & 0.1411 & 0.1202 \\
\textbf{SCLS}
& -0.0051 & 0.1303 & 0.2830 & 0.2279
& 0.0078 & 0.0285 & 0.1324 & 0.1097
& -0.0032 & 0.0470 & 0.1713 & 0.1412 \\
\textbf{WME}
& -0.0017 & 0.0412 & 0.1608 & 0.1331
& 0.0017 & 0.0091 & 0.0759 & 0.0642
& 0.0008 & 0.0149 & 0.0974 & 0.0830 \\
\textbf{HME}
& -0.0026 & 0.0426 & 0.1643 & 0.1351
& 0.0014 & 0.0096 & 0.0774 & 0.0644
& 0.0029 & 0.0153 & 0.0988 & 0.0825 \\
\textbf{TME}
& -0.0022 & 0.0385 & 0.1560 & 0.1285
& 0.0011 & 0.0086 & 0.0731 & 0.0614
& 0.0032 & 0.0140 & 0.0944 & 0.0770 \\
\hline
& \multicolumn{12}{c|}{\textbf{$n=600$, C.P.=19\%}} \\
\cline{2-13}
\textbf{CRQ}
& -0.0043 & 0.0707 & 0.2131 & 0.1816
& 0.0061 & 0.0155 & 0.0998 & 0.0859
& -0.0095 & 0.0260 & 0.1286 & 0.1070 \\
\textbf{SCLS}
& 0.0075 & 0.1132 & 0.2670 & 0.2215
& -0.0016 & 0.0250 & 0.1253 & 0.1075
& -0.0003 & 0.0389 & 0.1566 & 0.1287 \\
\textbf{WME}
& 0.0060 & 0.0352 & 0.1510 & 0.1311
& -0.0013 & 0.0078 & 0.0705 & 0.0601
& -0.0017 & 0.0125 & 0.0889 & 0.0743 \\
\textbf{HME}
& 0.0025 & 0.0346 & 0.1482 & 0.1285
& 0.0004 & 0.0078 & 0.0701 & 0.0593
& -0.0038 & 0.0128 & 0.0899 & 0.0739 \\
\textbf{TME}
& 0.0032 & 0.0316 & 0.1414 & 0.1205
& -0.0001 & 0.0071 & 0.0669 & 0.0555
& -0.0033 & 0.0114 & 0.0845 & 0.0717 \\
\hline
& \multicolumn{12}{c|}{\textbf{$n=800$, C.P.=19\%}} \\
\cline{2-13}
\textbf{CRQ}
& -0.0002 & 0.0541 & 0.1870 & 0.1599
& 0.0015 & 0.0123 & 0.0893 & 0.0775
& 0.0005 & 0.0194 & 0.1121 & 0.0980 \\
\textbf{SCLS}
& 0.0121 & 0.0788 & 0.2227 & 0.1879
& -0.0066 & 0.0182 & 0.1071 & 0.0911
& 0.0084 & 0.0298 & 0.1355 & 0.1117 \\
\textbf{WME}
& 0.0061 & 0.0265 & 0.1286 & 0.1063
& -0.0038 & 0.0061 & 0.0622 & 0.0525
& 0.0051 & 0.0097 & 0.0784 & 0.0652 \\
\textbf{HME}
& 0.0048 & 0.0266 & 0.1287 & 0.1070
& -0.0024 & 0.0060 & 0.0616 & 0.0524
& 0.0027 & 0.0096 & 0.0779 & 0.0663 \\
\textbf{TME}
& 0.0046 & 0.0242 & 0.1228 & 0.1006
& -0.0021 & 0.0054 & 0.0588 & 0.0499
& 0.0029 & 0.0087 & 0.0743 & 0.0620 \\
\hline
& \multicolumn{12}{c|}{\textbf{$n=1000$, C.P.=19\%}} \\
\cline{2-13}
\textbf{CRQ}
& -0.0102 & 0.0468 & 0.1741 & 0.1486
& 0.0052 & 0.0103 & 0.0809 & 0.0678
& -0.0061 & 0.0179 & 0.1063 & 0.0922 \\
\textbf{SCLS}
& -0.0003 & 0.0672 & 0.2069 & 0.1750
& 0.0008 & 0.0150 & 0.0984 & 0.0846
& -0.0033 & 0.0250 & 0.1262 & 0.1093 \\
\textbf{WME}
& -0.0019 & 0.0220 & 0.1179 & 0.0994
& -0.0002 & 0.0049 & 0.0556 & 0.0474
& -0.0020 & 0.0081 & 0.0717 & 0.0614 \\
\textbf{HME}
& 0.0007 & 0.0217 & 0.1171 & 0.0972
& -0.0015 & 0.0049 & 0.0560 & 0.0471
& -0.0012 & 0.0080 & 0.0719 & 0.0621 \\
\textbf{TME}
& 0.0003 & 0.0192 & 0.1099 & 0.0918
& -0.0010 & 0.0044 & 0.0530 & 0.0454
& -0.0016 & 0.0071 & 0.0678 & 0.0588 \\
\hline
\end{tabular}%
}
\end{table}

\begin{table}[H]
\centering
\small
\caption{Results for the model error $ \varepsilon \sim 0.9\,t_{1}+0.1\,t_{2}$.}
\label{table:5}
\resizebox{\textwidth}{!}{%
\begin{tabular}{|c|c|c|c|c|c|c|c|c|c|c|c|c|}
\hline
\textbf{Estimators}
& \multicolumn{12}{c|}{\textbf{$0.9\,t_{1}+0.1\,t_{2}$, $n=100$, C.P.=20\%}} \\
\cline{2-13}
& \multicolumn{4}{c|}{$\hat{\beta}_{1}$}
& \multicolumn{4}{c|}{$\hat{\beta}_{2}$}
& \multicolumn{4}{c|}{$\hat{\beta}_{3}$} \\
\cline{2-13}
& \textbf{Mean Bias} & \textbf{MSE} & \textbf{MAE} & \textbf{MedAE}
& \textbf{Mean Bias} & \textbf{MSE} & \textbf{MAE} & \textbf{MedAE}
& \textbf{Mean Bias} & \textbf{MSE} & \textbf{MAE} & \textbf{MedAE} \\
\hline
\textbf{CRQ}
& -0.0406 & 0.5265 & 0.5586 & 0.4431
& 0.0392 & 0.1147 & 0.2619 & 0.2095
& -0.0241 & 0.1963 & 0.3423 & 0.2723 \\
\textbf{SCLS}
& 0.6490 & 752.4896 & 1.5908 & 0.7735
& -0.1733 & 93.7512 & 0.6832 & 0.3675
& 0.3277 & 229.0733 & 0.9180 & 0.4329 \\
\textbf{WME}
& 0.0079 & 0.3482 & 0.4600 & 0.3783
& 0.0012 & 0.0755 & 0.2107 & 0.1706
& -0.0064 & 0.1207 & 0.2725 & 0.2248 \\
\textbf{HME}
& 0.0119 & 0.3765 & 0.4621 & 0.3806
& 0.0011 & 0.0805 & 0.2185 & 0.1768
& -0.0077 & 0.1353 & 0.2765 & 0.2232 \\
\textbf{TME}
& 0.0086 & 0.3553 & 0.4476 & 0.3550
& 0.0036 & 0.0861 & 0.2138 & 0.1738
& -0.0125 & 0.1366 & 0.2698 & 0.2156 \\
\hline
& \multicolumn{12}{c|}{\textbf{$n=200$, C.P.=20\%}} \\
\cline{2-13}
\textbf{CRQ}
& -0.0287 & 0.2732 & 0.4117 & 0.3342
& 0.0223 & 0.0601 & 0.1932 & 0.1609
& -0.0169 & 0.0998 & 0.2475 & 0.2046 \\
\textbf{SCLS}
& -0.0013 & 0.6580 & 0.6314 & 0.5148
& 0.0196 & 0.1564 & 0.3104 & 0.2601
& -0.0069 & 0.2510 & 0.3923 & 0.3226 \\
\textbf{WME}
& -0.0003 & 0.1431 & 0.2988 & 0.2476
& 0.0022 & 0.0323 & 0.1419 & 0.1160
& -0.0033 & 0.0536 & 0.1838 & 0.1539 \\
\textbf{HME}
& 0.0024 & 0.1409 & 0.2964 & 0.2483
& 0.0026 & 0.0329 & 0.1431 & 0.1193
& -0.0044 & 0.0510 & 0.1775 & 0.1427 \\
\textbf{TME}
& 0.0022 & 0.1338 & 0.2892 & 0.2425
& 0.0025 & 0.0316 & 0.1392 & 0.1124
& -0.0030 & 0.0482 & 0.1726 & 0.1404 \\
\hline
& \multicolumn{12}{c|}{\textbf{$n=400$, C.P.=20\%}} \\
\cline{2-13}
\textbf{CRQ}
& -0.0141 & 0.1351 & 0.2902 & 0.2407
& 0.0120 & 0.0297 & 0.1360 & 0.1121
& -0.0090 & 0.0491 & 0.1761 & 0.1495 \\
\textbf{SCLS}
& 0.0194 & 0.3118 & 0.4443 & 0.3771
& -0.0006 & 0.0734 & 0.2158 & 0.1813
& 0.0183 & 0.1134 & 0.2627 & 0.2119 \\
\textbf{WME}
& 0.0100 & 0.0707 & 0.2093 & 0.1727
& -0.0029 & 0.0156 & 0.0980 & 0.0793
& 0.0092 & 0.0250 & 0.1236 & 0.1009 \\
\textbf{HME}
& 0.0095 & 0.0648 & 0.2020 & 0.1658
& -0.0034 & 0.0147 & 0.0960 & 0.0808
& 0.0070 & 0.0228 & 0.1184 & 0.0942 \\
\textbf{TME}
& 0.0082 & 0.0604 & 0.1943 & 0.1618
& -0.0029 & 0.0137 & 0.0926 & 0.0773
& 0.0062 & 0.0212 & 0.1138 & 0.0912 \\
\hline
& \multicolumn{12}{c|}{\textbf{$n=500$, C.P.=20\%}} \\
\cline{2-13}
\textbf{CRQ}
& -0.0118 & 0.1070 & 0.2607 & 0.2275
& 0.0068 & 0.0240 & 0.1237 & 0.1037
& -0.0042 & 0.0394 & 0.1594 & 0.1336 \\
\textbf{SCLS}
& 0.0066 & 0.2270 & 0.3784 & 0.3177
& -0.0002 & 0.0547 & 0.1850 & 0.1513
& 0.0053 & 0.0860 & 0.2332 & 0.1966 \\
\textbf{WME}
& 0.0062 & 0.0511 & 0.1823 & 0.1615
& -0.0030 & 0.0114 & 0.0861 & 0.0755
& 0.0024 & 0.0183 & 0.1074 & 0.0899 \\
\textbf{HME}
& 0.0079 & 0.0484 & 0.1762 & 0.1531
& -0.0035 & 0.0110 & 0.0835 & 0.0702
& 0.0022 & 0.0173 & 0.1045 & 0.0887 \\
\textbf{TME}
& 0.0075 & 0.0451 & 0.1703 & 0.1464
& -0.0032 & 0.0102 & 0.0807 & 0.0675
& 0.0013 & 0.0162 & 0.1012 & 0.0861 \\
\hline
& \multicolumn{12}{c|}{\textbf{$n=600$, C.P.=20\%}} \\
\cline{2-13}
\textbf{CRQ}
& -0.0103 & 0.0830 & 0.2308 & 0.1931
& 0.0076 & 0.0182 & 0.1079 & 0.0895
& -0.0098 & 0.0305 & 0.1380 & 0.1138 \\
\textbf{SCLS}
& 0.0224 & 0.2099 & 0.3601 & 0.3006
& -0.0057 & 0.0480 & 0.1736 & 0.1462
& 0.0102 & 0.0773 & 0.2178 & 0.1819 \\
\textbf{WME}
& 0.0085 & 0.0440 & 0.1665 & 0.1382
& -0.0032 & 0.0097 & 0.0778 & 0.0644
& -0.0001 & 0.0157 & 0.0990 & 0.0830 \\
\textbf{HME}
& 0.0040 & 0.0412 & 0.1625 & 0.1372
& -0.0011 & 0.0089 & 0.0755 & 0.0627
& -0.0016 & 0.0142 & 0.0945 & 0.0797 \\
\textbf{TME}
& 0.0048 & 0.0379 & 0.1568 & 0.1325
& -0.0017 & 0.0083 & 0.0730 & 0.0614
& -0.0017 & 0.0131 & 0.0908 & 0.0763 \\
\hline
& \multicolumn{12}{c|}{\textbf{$n=800$, C.P.=20\%}} \\
\cline{2-13}
\textbf{CRQ}
& -0.0029 & 0.0651 & 0.2012 & 0.1635
& 0.0034 & 0.0146 & 0.0955 & 0.0792
& -0.0009 & 0.0248 & 0.1256 & 0.1074 \\
\textbf{SCLS}
& 0.0158 & 0.1531 & 0.3123 & 0.2645
& -0.0019 & 0.0358 & 0.1492 & 0.1240
& 0.0041 & 0.0565 & 0.1873 & 0.1540 \\
\textbf{WME}
& 0.0036 & 0.0334 & 0.1463 & 0.1253
& -0.0018 & 0.0073 & 0.0679 & 0.0574
& 0.0033 & 0.0119 & 0.0869 & 0.0731 \\
\textbf{HME}
& 0.0026 & 0.0310 & 0.1398 & 0.1185
& -0.0007 & 0.0068 & 0.0656 & 0.0567
& 0.0023 & 0.0105 & 0.0819 & 0.0699 \\
\textbf{TME}
& 0.0013 & 0.0291 & 0.1356 & 0.1159
& 0.0000 & 0.0064 & 0.0638 & 0.0550
& 0.0012 & 0.0098 & 0.0790 & 0.0666 \\
\hline
& \multicolumn{12}{c|}{\textbf{$n=1000$, C.P.=20\%}} \\
\cline{2-13}
\textbf{CRQ}
& -0.0068 & 0.0573 & 0.1915 & 0.1622
& 0.0033 & 0.0129 & 0.0910 & 0.0779
& -0.0023 & 0.0212 & 0.1168 & 0.1003 \\
\textbf{SCLS}
& 0.0137 & 0.1150 & 0.2686 & 0.2246
& -0.0016 & 0.0269 & 0.1273 & 0.1051
& 0.0043 & 0.0439 & 0.1675 & 0.1413 \\
\textbf{WME}
& 0.0036 & 0.0270 & 0.1302 & 0.1092
& -0.0019 & 0.0061 & 0.0615 & 0.0509
& 0.0015 & 0.0100 & 0.0788 & 0.0650 \\
\textbf{HME}
& 0.0019 & 0.0248 & 0.1262 & 0.1045
& -0.0010 & 0.0057 & 0.0595 & 0.0490
& 0.0000 & 0.0088 & 0.0745 & 0.0624 \\
\textbf{TME}
& 0.0026 & 0.0231 & 0.1221 & 0.1041
& -0.0012 & 0.0052 & 0.0574 & 0.0481
& 0.0003 & 0.0081 & 0.0716 & 0.0599 \\
\hline
\end{tabular}%
}
\end{table}

In essence, the results of Tables~\ref{table:1}-\ref{table:5}   demonstrate that the performance of estimators depends strongly on the error distributions. In particular, the SCLS estimator performs favorably under normal errors; however, it becomes substantially less competitive under heavy-tailed and contaminated distributions. In contrast, WME, HME, and TME maintain comparatively stable performance across the non-normal error distributions. More specifically, WME performs particularly well for the model error $\varepsilon \sim t_{3}$. Moreover, WME performs particularly well for Cauchy errors at smaller sample sizes. On the other hand, the TME generally provides the most favorable results for moderate and large sample sizes under contaminated error distributions. The HME also performs competitively, often yielding results close to those of TME.

Thus, the preferred estimator depends on the tail behaviour of the error distribution. SCLS is most efficient under near-normal conditions, whereas CRQ, WME, HME, and TME provide greater stability under heavy-tailed or contaminated errors. Among these robust alternatives, WME is the most consistent estimator across all five error distributions. On the other hand, the TME and HME perform comparably to the WME and are preferable for moderate and large sample sizes in the presence of contaminated or heavy-tailed error distributions. Overall, the simulation results demonstrate the advantages of using general M-estimators, which offer a favorable compromise between robustness, low bias, and prediction accuracy, as reflected in lower MAE, MedAE, and MSE values.  Consequently, the choice of estimator should balance efficiency under near-normal conditions against robustness to extreme observations.

We remark that the \textbf{optim} function depends on the initial values chosen. We used the CLAD estimates as initial values, but choosing other initial estimates, as well as other tuning parameters for the Tukey \ref{Tukey Biweight Loss Function}, Huber \ref{Huber Loss Function}, and Hampel loss functions \ref{Hampel Loss Function}, might lead to better results for these estimators. In any case, the current results already demonstrate the advantage of using M-estimators. These results confirm that M-estimators are robust and offer a good compromise between robustness (low MAE), low bias, and a low median absolute error (MedAE), generally resulting in a low MSE.

\section{Conclusion} \label{Conclusion}
This paper develops a general M-estimation framework for censored regression models under endogeneity. The proposed framework provides the flexibility to incorporate different loss functions, offering alternatives to existing robust estimation methods. We establish the strong consistency and asymptotic normality of the proposed estimators and derive a consistent estimator of the asymptotic covariance matrix, providing a rigorous theoretical basis for statistical inference. In particular, the asymptotic normality result can be utilized to construct asymptotic confidence intervals or to perform testing of hypothesis for the parameters of a linear censored regression model with endogeneity. As a direction for future research, it would be interesting to extend this study to panel data settings.

\subsubsection*{Code Availability}
\textit{The R code used for the empirical study is publicly available at \url{https://github.com/swati-1602/CRM}.}
\subsubsection*{Author contributions and use of AI}
The authors independently developed the methodology and derived all mathematical results presented in this paper. AI tools were used only for coding assistance and language editing to improve the clarity and readability of the manuscript. All AI-assisted outputs were reviewed and verified by the authors, who take full responsibility for the content of the work.
\appendix
\section{Appendix }\label{appendix}
Throughout the appendix, we use the following notation. $\mathbb{N}$ denotes the set of natural numbers. For any $d \geq 1$, $\mathbb{R}^d$ denotes the $d$-dimensional Euclidean space. For any vector $\xi \in \mathbb{R}^d$, $\|\xi\|$ denotes the standard Euclidean norm in $\mathbb{R}^d$. For any matrix $A$,
$\|A\|_F=\left\{\operatorname{trace}(A^\top A)\right\}^{1/2}$
denotes the Frobenius norm of $A$. We denote $\boldsymbol{\hat{\beta}}_n(\rho, \boldsymbol{\hat{\delta}}_n)$ by $\boldsymbol{\hat{\beta}}_n$ and the corresponding population counterpart by $\boldsymbol{\beta}_0$ for any fixed $\rho \in \mathcal{L}$ (see Equation~\eqref{Eq:2.8}).  

\subsection{ Proof of Consistency}\label{app:B.1}
We now present the proof of Theorem~\ref{th1} which relies on the following results from \cite{Jennrich1969} and \cite{Amemiya1973} 
 \begin{t1}[\textbf{Theorem 2 of }\cite{Jennrich1969}]\label{T2} Let $g(\mathcal{Z},\boldsymbol{\theta})$ be a real-valued function such that it is a measurable function of $\mathcal{Z}$ for each $\boldsymbol{\theta} \in \Theta$, and a continuous function of $\boldsymbol{\theta}$ for each $\mathcal{Z} \in \mathbb{R}^d$, for some $d \geq 1$. Here, $\Theta$ is a compact subset of $\mathbb{R}^{k}$ for some $ k \geq 1$. Then, given a sequence of $i.i.d.$ random vectors $\{\mathcal{Z}_{i}\}$ such that $\mathbb{E}[\sup_{\boldsymbol{\theta}\in \Theta}|g(\mathcal{Z}_{i},\boldsymbol{\theta})|]<\infty,$ we have 
$$ \sup_{\boldsymbol{\theta} \in \Theta} \left|n^{-1}\sum\limits_{i=1}^{n} g(\mathcal{Z}_{i},\boldsymbol{\theta}) - \mathbb{E}[g(\mathcal{Z}_{i},\boldsymbol{\theta})] \right| \xrightarrow{a.s} 0 \text{ as } n \to \infty.$$

 \end{t1}

\begin{t1}[\textbf{Lemma 3 of }\cite{Amemiya1973}]\label{T3} Let $\mathcal{Z}:= \{\mathcal{Z}_1, \ldots, \mathcal{Z}_n\}$, where $\mathcal{Z}_i$'s are i.i.d random vectors in Euclidean space, and define $\boldsymbol{\hat{\theta}}_n := \arg\min_{\boldsymbol{\theta} \in \Theta} q_n(\mathcal{Z}, \boldsymbol{\theta})$  
such that $q_n(\mathcal{Z}, \boldsymbol{\theta})$ and $\Theta$ satisfy the following conditions.  
\begin{enumerate}
    \item $\Theta$ is a compact subset of $\mathbb{R}^{k}$ for some $k \geq 1$. 
    
    \item $q_{n}(\mathcal{Z},\boldsymbol{\theta})$ is a measurable function of $\mathcal{Z}$ for all $\boldsymbol{\theta} \in \Theta$ and a continuous function in $\boldsymbol{\theta}$ for all $\mathcal{Z} \in \mathbb{R}^d$ for some $d \geq 1$. 
    
    \item There exists a non-stochastic continuous function $q(\boldsymbol{\theta})$ such that  
    $$\sup_{\boldsymbol{\theta} \in \Theta} \left| n^{-1}q_n(\mathcal{Z}, \boldsymbol{\theta}) -q(\boldsymbol{\theta})\right| \xrightarrow{a.s} 0 \text{ as } n \to \infty.$$ 
\end{enumerate}
If $\boldsymbol{\theta}_0 := \inf_{\boldsymbol{\theta} \in \Theta} q(\boldsymbol{\theta})$, and $\boldsymbol{\theta}_0$ is unique, then $\hat{\boldsymbol{\theta}}_{n} \xrightarrow{a.s} \boldsymbol{\theta}_{0}$ as $n \to \infty$.
\end{t1}

In addition, we require the following two lemmas to establish Theorem~\ref{th1}. The following lemma establish strong consistency of $\boldsymbol{\hat{\delta}}_n$ (see Equation~\eqref{Eq:2.6}) to its population counterpart $\boldsymbol{\delta}_0$. 

\begin{l1}\label{l:1} 
Under the assumptions \ref{a:3.5} and \ref{a:3.8}, we have ${\boldsymbol{\hat{\delta}}}_{n}\xrightarrow{a.s.} \boldsymbol{\delta}_{0}$ as $n \to \infty.$
\end{l1}

\begin{proof}[\textbf{Proof}] Recall that $\hat{\boldsymbol{\Sigma}}_{1,\boldsymbol{\delta}} := n^{-1}\sum\limits_{i=1}^{n}\boldsymbol{z}_{i}\boldsymbol{z}^{\top}_{i}$ (see Equation~\eqref{Eq:3.16}) and  $\boldsymbol{\hat{\delta}}_n$ is the least squares estimator (LSE) of $\boldsymbol{\delta}_0$ obtained from the first-stage model. Then, we have
$$\boldsymbol{\hat{\delta}}_n- \boldsymbol{\delta}_{0} =\hat{\boldsymbol{\Sigma}}^{-1}_{1,\boldsymbol{\delta}}\; n^{-1}\sum\limits_{i=1}^{n}\boldsymbol{z}_{i} \vartheta_{i}.$$ 
Now, first, using Assumption~\ref{a:3.8}, we deduce the fact that there exists a matrix $\boldsymbol{\Sigma}_{1,\boldsymbol{\delta}_{0}}$ such that $\lim_{n \to \infty} \hat{\boldsymbol{\Sigma}}^{-1}_{1,\boldsymbol{\delta}} = \boldsymbol{\Sigma}^{-1}_{1,\boldsymbol{\delta}_{0}}$. Next, using Assumption~\ref{a:3.5} and the error assumption that $var(\vartheta_i|\boldsymbol{z}_i) < \infty$, we have $\sum_{n = 1}^\infty n^2 \mathbb{E}\| \boldsymbol{z}_n \vartheta_n\|^2 < \infty$. This condition is required to invoke the strong law of large numbers (SLLN) for the independent random vectors $\{\boldsymbol{z_i} \vartheta_i\}$ (see, e.g., Theorem 8.4.6 of \cite{athreya2006measure}). Thus, using the aforementioned conditions along with $\mathbb{E}(\vartheta_i|\boldsymbol{z}_i) = 0$, the SLLN yields $\hat{\boldsymbol{\delta}}_{n} \xrightarrow{a.s.} \boldsymbol{\delta}_{0}$. 
\end{proof}

 
\begin{l1}\label{l:2} Define the following 
 \begin{equation}\label{eq:A.1}
     \mathbb{S}_{n}(\boldsymbol{\beta},\boldsymbol{\hat{\delta}}_{n})= Q_{n}(\boldsymbol{\beta},\boldsymbol{\hat{\delta}}_{n})-Q_{n}(\boldsymbol{\beta}_{0},\boldsymbol{\delta}_{0}),
 \end{equation}
where $Q_{n}(\boldsymbol{\beta},\boldsymbol{\hat{\delta}}_{n})$ denotes the objective function defined in  Equation~\eqref{Eq:3.4}. Further, let 
 \begin{equation}\label{eq:A.2}
     \mathbb{S}(\boldsymbol{\beta},\boldsymbol{\delta}_{0})= Q(\boldsymbol{\beta},\boldsymbol{\delta}_{0})-Q(\boldsymbol{\beta}_{0},\boldsymbol{\delta}_{0}),
 \end{equation}
where $ Q(\boldsymbol{\beta},\boldsymbol{\delta})|_{\boldsymbol{\delta}=\boldsymbol{\delta}_{0}}=Q(\boldsymbol{\beta},\boldsymbol{\delta}_{0})$ denotes the population objective function defined in Equation~\eqref{Eq:3.1}. Then, under the assumptions \ref{a:3.2}-\ref{a:3.5}, for any consistent estimator $\boldsymbol{\hat{\delta}}_{n}$ of $\boldsymbol{\delta}_{0},$ we have
 $$\sup_{\boldsymbol{\beta}\in\mathcal{B}}
\left|
\mathbb{S}_{n}(\boldsymbol{\beta},\boldsymbol{\hat{\delta}}_{n})
-
\mathbb{S}(\boldsymbol{\beta},\boldsymbol{\delta}_{0})
\right|
\xrightarrow{a.s.}0 \text{ as } n \to \infty.$$
 \end{l1}
 \begin{proof}[\textbf{Proof}]
 Consider the following inequality obtained by applying the triangle inequality, 
\begin{equation}\label{eq:A.3}
    \sup_{\boldsymbol{\beta} \in \mathcal{B}} \left|\mathbb{S}_n(\boldsymbol{\beta}, \boldsymbol{\hat{\delta}}_n) - \mathbb{S}(\boldsymbol{\beta}, \boldsymbol{\delta}_0) \right|\leq \sup_{\boldsymbol{\beta} \in \mathcal{B}} \left|\mathbb{S}_n(\boldsymbol{\beta}, \boldsymbol{\hat{\delta}}_n) - \mathbb{S}_n(\boldsymbol{\beta}, \boldsymbol{\delta}_0) \right| + \sup_{\boldsymbol{\beta} \in \mathcal{B}} \left|\mathbb{S}_n(\boldsymbol{\beta}, \boldsymbol{\delta}_0) -\mathbb{S}(\boldsymbol{\beta}, \boldsymbol{\delta}_0) \right|.
\end{equation}
We need to show that both the terms on the right-hand side of Equation~\eqref{eq:A.3} converge to 0 almost surely. Now, consider the first term
\begin{align*}
\sup_{\boldsymbol{\beta} \in \mathcal{B}}
\left|\mathbb{S}_n(\boldsymbol{\beta},\boldsymbol{\hat{\delta}}_n)
-\mathbb{S}_n(\boldsymbol{\beta},\boldsymbol{\delta}_0)\right|
&\leq\sup_{\boldsymbol{\beta} \in \mathcal{B}}
\left|Q_n(\boldsymbol{\beta},\boldsymbol{\hat{\delta}}_n)
-Q_n(\boldsymbol{\beta}_{0},\boldsymbol{\delta}_0)-Q_n(\boldsymbol{\beta},\boldsymbol{\delta}_0)+Q_n(\boldsymbol{\beta}_{0},\boldsymbol{\delta}_0)\right|\\
&\leq\sup_{\boldsymbol{\beta} \in \mathcal{B}}
\left|Q_n(\boldsymbol{\beta},\boldsymbol{\hat{\delta}}_n)
-Q_n(\boldsymbol{\beta},\boldsymbol{\delta}_0)\right|\\
&\leq\sup_{\boldsymbol{\beta} \in \mathcal{B}}
\left|\frac{1}{n}\sum_{i=1}^{n}\left[
\rho (Y_i-\max\{c_{\rho},\hat{\boldsymbol{x}}_i^{\top}\boldsymbol{\beta}\}
)-\rho(Y_i-\max\{c_{\rho},\boldsymbol{x}_i^{\top}\boldsymbol{\beta}\})\right]\right|\\
&\leq \sup_{\boldsymbol{\beta} \in \mathcal{B}} \frac{1}{n} \sum_{i = 1}^n k_{\rho}\left|\max\{c_{\rho},\hat{\boldsymbol{x}}_i^{\top}\boldsymbol{\beta}\}- \max\{c_{\rho},\boldsymbol{x}_i^{\top}\boldsymbol{\beta}\}\right| \tag{E.1} \label{eq:E.1}\\
& \leq \sup_{\boldsymbol{\beta} \in \mathcal{B}} \frac{1}{n} \sum_{i =1}^n k_{\rho}\left|\hat{\boldsymbol{x}}_i^{\top}\boldsymbol{\beta}- \boldsymbol{x}_i^{\top}\boldsymbol{\beta}\right| \\
&\leq k_{\rho}\;\sup_{\boldsymbol{\beta}\in\mathcal{B}} 
\|\boldsymbol{\beta}\|\,
\|\boldsymbol{\hat{\delta}}_n-\boldsymbol{\delta}_0\|\;
\frac{1}{n}\sum_{i=1}^{n}\|\boldsymbol{z}_i\|, \tag{E.2} \label{eq:E.2}\\
&\xrightarrow{a.s.} 0 \text{ as } n \to \infty, \tag{E.3} \label{eq:E.3}
\end{align*}
where Equation~\eqref{eq:E.1} follows from Equation~\eqref{Eq:2.8} and Equation~\eqref{eq:E.2} follows from applying the Cauchy–Schwarz inequality along with $||\hat{\boldsymbol{x}}_{i}-\boldsymbol{x}_{i}|| =|\boldsymbol{z}^\top_{i}(\boldsymbol{\hat{\delta}}_{n}-\boldsymbol{\delta}_{0})|$. Lastly, Equation~\eqref{eq:E.3} follows from using the following; $\sup_{\boldsymbol{\beta}\in\mathcal{B}}
\|\boldsymbol{\beta}\|\leq C_{0}<\infty$ (by Assumption \ref{a:3.4}), $\|\boldsymbol{\hat{\delta}}_n-\boldsymbol{\delta}_0\|\xrightarrow{a.s.} 0$ (by Lemma~\ref{l:1}), and $n^{-1}\sum_{i=1}^{n}\|\boldsymbol{z}_{i}\|=O(1)$ almost surely, follows from Assumption \ref{a:3.5} along with the application of the strong law of large numbers (SLLN) on i.i.d random variables $\{||\boldsymbol{z}_i||, i = 1, \ldots,n\}$.

Next, consider the second term on the right-hand side of Equation~\eqref{eq:A.3} and apply Theorem~\ref{T2} for $g\left(\mathcal{Z}_i, \boldsymbol{\theta}\right) := \rho\left(Y_{i}-  \max \{c_{\rho},\boldsymbol{x}^\top_{i}\boldsymbol{\beta}\}\right) - \rho\left(Y_{i}-  \max \{c_{\rho},\boldsymbol{x}^\top_{i}\boldsymbol{\beta}_0\}\right)$ where the random vector $(Y_i, \boldsymbol{x}^\top_{i})^\top$ and $\boldsymbol{\beta}$ playing the same role as $Z_i$ and $\boldsymbol{\theta}$, respectively. If $g\left(\mathcal{Z}_i, \boldsymbol{\theta}\right)$ satisfies all the required conditions of Theorem~\ref{T2}, then we have 
 \begin{equation} \label{Eq:A.4}
     \sup_{\boldsymbol{\beta} \in \mathcal{B}} \left|\mathbb{S}_n(\boldsymbol{\beta}, \boldsymbol{\delta}_0) - \mathbb{S}(\boldsymbol{\beta}, \boldsymbol{\delta}_0) \right| \xrightarrow{a.s.} 0 \text{ as } n \to \infty
 \end{equation}
where $\mathbb{S}_n(\boldsymbol{\beta}, \boldsymbol{\delta}_0)= n^{-1} \sum_{i = 1}^n g(\mathcal{Z}_i, \boldsymbol{\theta})$ and $\mathbb{E}[\mathbb{S}_n(\boldsymbol{\beta}, \boldsymbol{\delta}_0)] = \mathbb{S}(\boldsymbol{\beta}, \boldsymbol{\delta}_0)$.

Now, it remains to verify all the required conditions for the function $g(\mathcal{Z}_i, \boldsymbol{\theta})$ in Theorem~\ref{T2} for the validity of Equation~\eqref{Eq:A.4}. It is straightforward to see that for each i, $g(\mathcal{Z}_i, \boldsymbol{\theta})$ is measurable in $\mathcal{Z}_i$ for each $\boldsymbol{\theta}$, and continuous in $\boldsymbol{\theta}$ for each $\mathcal{Z}_i$. This follows from the fact that the function $\rho(Y_{i}-\max\{c_{\rho},\boldsymbol{x}^\top_{i}\boldsymbol{\beta}\})$ is measurable in $(Y_i, \boldsymbol{x}^\top_{i})^\top$ for each $\boldsymbol{\beta} \in \mathcal{B}$ and is continuous in $\boldsymbol{\beta}$ for each $(Y_i, \boldsymbol{x}^\top_{i})^\top \in \mathbb{R}^{p + 3}$. Next, we proceed to verify the moment condition $\mathbb{E}[\sup_{\boldsymbol{\theta} \in \Theta} |g(\mathcal{Z}_i, \boldsymbol{\theta}| ] < \infty$. Consider the following
 \begin{align*}
    \mathbb{E}\left[\sup_{\boldsymbol{\theta} \in \Theta}\left|g(\mathcal{Z}_i, \boldsymbol{\theta})\right|\right]
    &= \mathbb{E}_{\boldsymbol{X}_i}\mathbb{E}_{Y_i|\boldsymbol{X}_i}\left[\sup_{\boldsymbol{\beta} \in \mathcal{B}} \left|\rho\left(Y_{i}-  \max \{c_{\rho},\boldsymbol{x}^\top_{i}\boldsymbol{\beta}\}\right) - \rho\left(Y_{i}-  \max \{c_{\rho},\boldsymbol{x}^\top_{i}\boldsymbol{\beta}_0\}\right)\right|\right]\\
    &\leq  k_{\rho}\; \mathbb{E}_{\boldsymbol{X}_i} \left[\sup_{\boldsymbol{\beta} \in \mathcal{B}} \left|\max \{c_{\rho},\boldsymbol{X}^\top_{i}\boldsymbol{\beta}\} - \max \{c_{\rho},\boldsymbol{X}^\top_{i}\boldsymbol{\beta}_0\}\right| \right] \tag{By Equation~\eqref{Eq:2.8}} \\
    &\leq k_{\rho}\; \mathbb{E}_{\boldsymbol{X}_i} \left[\sup_{\boldsymbol{\beta} \in \mathcal{B}} \left|\boldsymbol{X}^\top_{i}\boldsymbol{\beta} - \boldsymbol{X}^\top_{i}\boldsymbol{\beta}_0\right| \right]  \leq k_{\rho}\;\mathbb{E}_{\boldsymbol{X}_i}||\boldsymbol{X}_i||\left[\sup_{\boldsymbol{\beta} \in \mathcal{B}} ||\boldsymbol{\beta} - \boldsymbol{\beta}_0||\right]\tag{E.4} \label{eq:E.4} 
    < \infty,
 \end{align*}
 where Equation~\eqref{eq:E.4} follows from applying the Cauchy–Schwarz inequality, and subsequently using Assumption~\ref{a:3.4} and Assumption~\ref{a:3.5}. This establishes the validity of Equation~\eqref{Eq:A.4}. Hence, the proof is completed. 
\end{proof}
 \subsubsection{\textbf{Proof of Theorem}~\ref{th1}}
\begin{proof}[$\textbf{Proof}$]
The consistency of $ \boldsymbol{\hat{\beta}}_{n}$ is a direct consequence of Theorem~\ref{T3} for the function $\mathbb{S}_{n}(\boldsymbol{\beta}, \boldsymbol{\hat{\delta}_n})$ and $\mathbb{S}(\boldsymbol{\beta}, \boldsymbol{\delta}_0)$ along with Lemma~\ref{l:2} under the identification condition stated in Assumption~\ref{a:3.3}. This establishes the consistency of $\boldsymbol{\hat{\beta}}_{n}$. Hence, the proof is completed.
\end{proof}
 \subsection{Proof of Asymptotic Normality}\label{app:B.2} 
 We use the following theorem to establish the asymptotic normality of the joint random vector $( \boldsymbol{\hat{\beta}}_{n}^\top,\boldsymbol{\hat{\delta}}_{n}^\top)^\top$. 
\begin{t1}[\textbf{Theorem 3.3 of} \cite{PakesandPollard}]\label{T1} Let $( \boldsymbol{\hat{\beta}}_{n}^\top,\boldsymbol{\hat{\delta}}_{n}^\top)^\top$ be a consistent estimator of  $(\boldsymbol{\beta}_{0}^\top, \boldsymbol{\delta}_{0}^\top)^\top$ which is the unique point in $\mathcal{B}\times \Delta$ such that $\mathbb{J}(\boldsymbol{\beta}_{0},\boldsymbol{\delta}_{0})=0$ (see Equation~\eqref{Eq:3.5}). Suppose that
\begin{enumerate}
     \item $\|\mathbb{J}_{n}( \boldsymbol{\hat{\beta}}_{n},\boldsymbol{\hat{\delta}}_{n})\|= o_{P}(1),$ where $\mathbb{J}_{n}(.)$ is defined as
     \begin{align}\label{Eq:A.5}
\mathbb{J}_{n}(\boldsymbol{\beta},\boldsymbol{\delta})&= \begin{bmatrix}
          \mathbb{J}_{2,n}(\boldsymbol{\beta},\boldsymbol{\delta}) \\
          \mathbb{J}_{1,n}(\boldsymbol{\delta})
          \end{bmatrix}= n^{-1}\sum_{i = 1}^n \mathbb{V}_i(\boldsymbol{\beta},\boldsymbol{\delta}), \ \mathbb{V}_i(\boldsymbol{\beta},\boldsymbol{\delta}):=
          \begin{bmatrix}
          m_{2}(\boldsymbol{u}_{i},\boldsymbol{\beta},\delta) \\
          m_{1}(\boldsymbol{v}_{i},\boldsymbol{\delta})
          \end{bmatrix},
  \end{align}
 where
 \begin{align*}
     m_{2}(\boldsymbol{u}_{i},\boldsymbol{\beta},\boldsymbol{\delta})&= \mathds{1}(\boldsymbol{x}_{i}(\boldsymbol{\delta})^\top\boldsymbol{\beta}>c_{\rho})\psi(Y_{i}-\boldsymbol{x}_{i}(\boldsymbol{\delta})^\top\boldsymbol{\beta})\boldsymbol{x}_{i}(\boldsymbol{\delta}),\quad \boldsymbol{u}_{i}=(Y_{i},\boldsymbol{x}_{i}(\boldsymbol{\delta})^\top)^\top,\\
m_{1}(\boldsymbol{v}_{i},\boldsymbol{\delta})&= (w_{i}-\boldsymbol{z}_{i}^\top\boldsymbol{\delta})\boldsymbol{z}_{i},\quad \boldsymbol{v}_{i}=(w_{i}, \boldsymbol{z}_{i}^\top)^\top.
\end{align*}
     \item $\mathbb{J}(.)$ is differentiable at $(\boldsymbol{\beta}_{0},\boldsymbol{\delta}_{0})$  with the derivative matrix $\boldsymbol{\Sigma}$ defined in  Equation~\eqref{Eq:3.9}.
     
     \item For every sequence ${a}_{n}$ and $b_{n}$ of positive numbers that converges to zero, we have

    $$ \sup_{\substack{\|\boldsymbol{\beta} - \boldsymbol{\beta}_0\| \leq a_n \\ \|\boldsymbol{\delta} - \boldsymbol{\delta}_0\| \leq b_n}} \frac{\|\mathbb{J}_{n}(\boldsymbol{\beta},\boldsymbol{\delta})-\mathbb{J}(\boldsymbol{\beta},\boldsymbol{\delta})-\mathbb{J}_{n}(\boldsymbol{\beta}_{0},\boldsymbol{\delta}_{0})\|}{n^{-1/2}+\|\mathbb{J}_{n}(\boldsymbol{\beta},\boldsymbol{\delta})\| +\|\mathbb{J}(\boldsymbol{\beta},\boldsymbol{\delta})\|} = o_{P}(1).$$
    
    \item  $\sqrt{n}\;\mathbb{J}_{n}(\boldsymbol{\beta}_{0},\boldsymbol{\delta}_{0})\xrightarrow{D} \mathbb{N}(\boldsymbol{0}, \mathbb{D})$, where $\mathbb{D}$ is defined in Equation~\eqref{Eq:3.9}. 
    
    \item $\boldsymbol{\beta}_{0}$ is an interior point of $\mathcal{B}$ and $\boldsymbol{\delta}_{0}$ is an interior point of $\Delta$.
 \end{enumerate}
 Then 
 $$
\sqrt{n}
\begin{pmatrix}
          \begin{bmatrix}
           \boldsymbol{\hat{\beta}}_{n}\\
          \boldsymbol{\hat{\delta}}_{n}
          \end{bmatrix} -
          \begin{bmatrix}
           \boldsymbol{\beta}_{0}\\
          \boldsymbol{\delta}_{0}
         \end{bmatrix}
         \end{pmatrix}\xrightarrow{D} \mathbb{N}_{2p+3}(\boldsymbol{0}, (\boldsymbol{\Sigma}^\top\boldsymbol{\Sigma})^{-1}\boldsymbol{\Sigma}^\top \mathbb{D} \boldsymbol{\Sigma}(\boldsymbol{\Sigma}^\top\boldsymbol{\Sigma})^{-1}).
  $$
 Since $\boldsymbol{\Sigma}$ is full rank by Assumption~\ref{a:3.8}, the asymptotic covariance matrix simplifies to
  $$\sqrt{n}
\begin{pmatrix}
          \begin{bmatrix}
           \boldsymbol{\hat{\beta}}_{n}\\
          \boldsymbol{\hat{\delta}}_{n}
          \end{bmatrix} -
          \begin{bmatrix}
           \boldsymbol{\beta}_{0}\\
          \boldsymbol{\delta}_{0}
         \end{bmatrix}
         \end{pmatrix}\xrightarrow{D} \mathbb{N}_{2p+3}(\boldsymbol{0}, \boldsymbol{\Sigma}^{-1} \mathbb{D} (\boldsymbol{\Sigma}^{-1})^\top).
  $$
\end{t1}
\subsubsection{Proof of Theorem~\ref{th2}} 
\begin{proof}[$\textbf{Proof}$]
First, we proceed to verify the joint first-order condition $\mathbb{J}(\boldsymbol{\beta}_{0},\boldsymbol{\delta}_{0})= \boldsymbol{0}$ (see Equation~\eqref{Eq:3.5}). We only verify that $\mathbb{J}_2(\boldsymbol{\beta}_0, \boldsymbol{\delta}_0) = \boldsymbol{0}$ as verifying $\mathbb{J}_1(\boldsymbol{\delta}_0) = \boldsymbol{0}$ is trivial (see Equation~\eqref{Eq:3.5}). Therefore, it is enough to show that $Q(\boldsymbol{\beta}, \boldsymbol{\delta}_{0})-Q(\boldsymbol{\beta}_{0}, \boldsymbol{\delta}_{0}) = o(\|\boldsymbol{\beta} - \boldsymbol{\beta}_0\|)$. Consider the following:
\begin{align*}
    Q(\boldsymbol{\beta}, \boldsymbol{\delta}_0)-Q(\boldsymbol{\beta}_{0}, \boldsymbol{\delta}_{0})&=\mathbb{E}\left[\rho(Y-\max\{c_{\rho},\boldsymbol{X}^\top\boldsymbol{\beta}\})\right]-\mathbb{E}\left[\rho(Y-\max\{c_{\rho},\boldsymbol{X}^\top\boldsymbol{\beta}_{0}\})\right] \\
        &=\mathbb{E}\left[\mathds{1}(\boldsymbol{X}^\top\boldsymbol{\beta}\leq c_{\rho},\boldsymbol{X}^\top\boldsymbol{\beta}_{0}\leq c_{\rho})\{\rho(Y-c_{\rho})-\rho(Y-c_{\rho})\}\right] \tag{E.5} \label{eq:E.5}\\
        &+\mathbb{E}\left[\mathds{1}(\boldsymbol{X}^\top\boldsymbol{\beta}_{0}\leq c_{\rho},\boldsymbol{X}^\top\boldsymbol{\beta}> c_{\rho})\{\rho(Y-\boldsymbol{X}^\top\boldsymbol{\beta})-\rho(Y-c_{\rho})\}\right] \tag{E.6} \label{eq:E.6}\\
         &+\mathbb{E}\left[\mathds{1}(\boldsymbol{X}^\top\boldsymbol{\beta}\leq c_{\rho},\boldsymbol{X}^\top\boldsymbol{\beta}_{0}> c_{\rho})\{\rho(Y-c_{\rho})-\rho(Y-\boldsymbol{X}^\top\boldsymbol{\beta}_{0})\}\right] \tag{E.7} \label{eq:E.7}\\
          &+\mathbb{E}\left[\mathds{1}(\boldsymbol{X}^\top\boldsymbol{\beta}> c_{\rho},\boldsymbol{X}^\top\boldsymbol{\beta}_{0}> c_{\rho})\{\rho(Y-\boldsymbol{X}^\top\boldsymbol{\beta})-\rho(Y-\boldsymbol{X}^\top\boldsymbol{\beta}_{0})\}\right] \tag{E.8} \label{eq:E.8},
          \end{align*}
where, note that \eqref{eq:E.5} is zero. Now, consider \eqref{eq:E.6} and invoke the Lipschitz continuity of $\rho$ together with the Cauchy–Schwarz inequality, we have
\begin{align*}
 \eqref{eq:E.6} &\leq k_{\rho} \mathbb{E}\left[\mathds{1}\left(\left|c_{\rho}-\boldsymbol{X}^\top\boldsymbol{\beta}\right|\leq \left|\boldsymbol{X}^\top\boldsymbol{\beta}-\boldsymbol{X}^\top \boldsymbol{\beta}_{0}\right|\right)
       \left|c_{\rho}-\boldsymbol{X}^\top \boldsymbol{\beta}\right|\right]\\
       &\leq k_{\rho} \mathbb{E}\left[\mathds{1}\left(\left|c_{\rho}-\boldsymbol{X}^\top\boldsymbol{\beta}\right|\leq \left|\boldsymbol{X}^\top\boldsymbol{\beta}-\boldsymbol{X}^\top \boldsymbol{\beta}_{0} \right|\right)
       \left|\boldsymbol{X}^\top \boldsymbol{\beta}-\boldsymbol{X}^\top \boldsymbol{\beta}_{0}\right|\right]\\
       &\leq k_{\rho} \mathbb{E}\left[\mathds{1}\left(\left|c_{\rho}-\boldsymbol{X}^\top\boldsymbol{\beta}_{0}\right|\leq 2\left\|\boldsymbol{X}\|\|\boldsymbol{\beta}-\boldsymbol{\beta}_{0} \right\|\right)
       \left\|\boldsymbol{X}\|\| \boldsymbol{\beta}-\boldsymbol{\beta}_{0}\right\|\right]\\
       &\leq k_{\rho}\mathcal{K}_{0} \mathbb{E}\left[\mathds{1}\left(\left|c_{\rho}-\boldsymbol{X}^\top\boldsymbol{\beta}_{0}\right|\leq 2\mathcal{K}_{0}\| \boldsymbol{\beta}-\boldsymbol{\beta}_{0}\|\right)\right] \left\| \boldsymbol{\beta}-\boldsymbol{\beta}_{0}\right\|\\
      & = o(\left\| \boldsymbol{\beta}-\boldsymbol{\beta}_{0}\right\|) \tag{By Assumptions~\ref{a:3.5} and \ref{a:3.6} },
\end{align*}
where $\ell=\|\boldsymbol{\beta}-\boldsymbol{\beta}_{0}\|$,  follows from Assumption~\ref{a:3.6}. Similarly, it can be shown that \eqref{eq:E.7} = $o(\left\| \boldsymbol{\beta}-\boldsymbol{\beta}_{0}\right\|)$. Now, it remains to show that \eqref{eq:E.8} = $o(\left\| \boldsymbol{\beta}-\boldsymbol{\beta}_{0}\right\|)$. By the first-order Taylor expansion of $\rho$, \eqref{eq:E.8} can be written as
 \begin{equation}\label{eq:A.6}
 \begin{split}
    \mathbb{E}\left[\mathds{1}(\boldsymbol{X}^\top\boldsymbol{\beta}> c_{\rho},\boldsymbol{X}^\top\boldsymbol{\beta}_{0}> c_{\rho})\{\psi(Y-\boldsymbol{X}^\top\boldsymbol{\beta}_{0}) (\boldsymbol{X}^\top\boldsymbol{\beta} - \boldsymbol{X}^\top\boldsymbol{\beta}_0)\}\right]+ o(\left\| \boldsymbol{\beta}-\boldsymbol{\beta}_{0}\right\|),
     \end{split}
 \end{equation}
 where the first term vanishes by the moment condition $\mathbb{E}[\psi(\varepsilon)|\boldsymbol{X}]=0$ (see Assumption~\ref{a:3.2}). Hence $Q(\boldsymbol{\beta}, \boldsymbol{\delta}_{0})$ is differentiable at the true parameter $\boldsymbol{\beta}_0$ with its derivative $\mathbb{J}(\boldsymbol{\beta}_{0},\boldsymbol{\delta}_{0})=0$. 
 Now, we proceed to verify all the required conditions of Theorem~\ref{T1}.  

\noindent \textbf{Step~1:} To establish Condition (1), it suffices to show that
 \begin{equation}\label{Eq:A.7}
     \begin{split}
         \|\sqrt{n}\;\mathbb{J}_{2,n}( \boldsymbol{\hat{\beta}}_{n},\boldsymbol{\hat{\delta}}_{n})\|=o_{P}(1) \text{ and }
         \|\sqrt{n}\;\mathbb{J}_{1,n}(\boldsymbol{\hat{\delta}}_{n})\|=o_{P}(1),
     \end{split}
 \end{equation}
where the functions $\mathbb{J}_{2,n}(.)$ and $\mathbb{J}_{1,n}(.)$ are defined in Equation~\eqref{Eq:A.5}. We adopt the same approach as in \cite{POWELL1984}, \cite{Fitzenberger1994note}, and \cite{FITZENBERGER1998235} to establish Equation~\eqref{Eq:A.7} which is applicable for non-differentiable sample objective function $Q_{n}(\boldsymbol{\beta},\boldsymbol{\hat{\delta}}_{n})$ (see Equation~\eqref{Eq:3.4}). Consider the right directional derivative $Q_{n,k}^{+}(\boldsymbol{\beta},\boldsymbol{\hat{\delta}}_{n})$ of $Q_{n}(\boldsymbol{\beta},\boldsymbol{\hat{\delta}}_{n})$ in the direction $r_{k}\in \mathbb{R}^{p+2}$ for some $a>0$, we have
\begin{equation}\label{Eq:A.8}
    \begin{split}
        Q_{n,k}^{+}(\boldsymbol{\beta},\boldsymbol{\hat{\delta}}_{n})&=\underset{a\to 0}{\lim}\frac{Q_{n}(\boldsymbol{\beta}+ar_{k},\boldsymbol{\hat{\delta}}_{n})-Q_{n}(\boldsymbol{\beta},\boldsymbol{\hat{\delta}}_{n})}{a}\\
        &=n^{-1}\sum_{i=1}^{n} \left(-\hat{x}_{i,k}\right) \left\{\mathds{1}(\hat{\boldsymbol{x}}^\top_{i}\boldsymbol{\beta}>c_{\rho})\psi(Y_{i}-\hat{\boldsymbol{x}}^\top_{i}\boldsymbol{\beta})
        + \mathds{1}(\hat{\boldsymbol{x}}^\top_{i}\boldsymbol{\beta}=c_{\rho})\psi(Y_{i}-c_{\rho})
        \mathds{1}(\hat{x}_{i,k}>0)\right\}\\
        &=\mathbb{J}_{2,n,k}(\boldsymbol{\beta},\boldsymbol{\hat{\delta}}_{n})
        + n^{-1} \sum_{i=1}^{n} \mathds{1}(\hat{\boldsymbol{x}}^\top_{i}\boldsymbol{\beta}=c_{\rho})\psi(Y_{i}-c_{\rho})
        (-\hat{x}_{i,k})\mathds{1}(\hat{x}_{i,k}>0),
    \end{split}
\end{equation}
where $\mathbb{J}_{2,n,k}(\boldsymbol{\beta},\boldsymbol{\hat{\delta}}_{n})$ denotes the $k$th element of $\mathbb{J}_{2,n}(\boldsymbol{\beta},\boldsymbol{\hat{\delta}}_{n})$, $\hat{x}_{i,k}$ denotes the $k$th component of the vector $\boldsymbol{\hat{x}}_i$ and $r_{k}=(0,\dots,1,\ldots,0)$ denotes the vector in $\mathbb{R}^{p+2}$ whose $k^{th}$ component is 1 and all other components are 0. Note that Equation~\eqref{Eq:A.8} is valid provided that $(Y_{i}-\hat{\boldsymbol{x}}^\top_{i}\boldsymbol{\beta}) \notin D_\rho$ where $D_\rho$ denotes the set of points where $\rho$ is not differentiable. If $D_\rho \neq \varnothing$, then we need to separate such cases and consider appropriate left or right derivatives at the points in $D_\rho$. For example, if we consider the quantile loss function, then $D_\rho = \{0\}$. Consequently, we need to consider an additional case in both terms in Equation~\eqref{Eq:A.8} for $\mathds{1}(Y_{i}-\hat{\boldsymbol{x}}^\top_{i}\boldsymbol{\beta} = 0)$ where appropriate left or right derivatives of $\rho$ are considered (see, e.g, \cite{Fitzenberger1994note}).       

Similarly, we can obtain $Q_{n,k}^{-}(\boldsymbol{\beta},\boldsymbol{\hat{\delta}}_{n})$ with the only difference in Equation~\eqref{Eq:A.8} being that $\mathds{1}(\hat{x}_{i,k}>0)$ is replaced by $\mathds{1}(\hat{x}_{i,k} < 0)$. Now, in view of the fact $ \boldsymbol{\hat{\beta}}_{n}$ is a minimizer of $Q_{n}(\boldsymbol{\beta},\boldsymbol{\hat{\delta}}_{n})$, we have $Q_{n,k}^{+}(  \boldsymbol{\hat{\beta}}_{n},\boldsymbol{\hat{\delta}}_{n})\geq 0$ and $Q_{n,k}^{-}( \boldsymbol{\hat{\beta}}_{n},\boldsymbol{\hat{\delta}}_{n})\leq 0$. This fact can be used along with Equation~\eqref{Eq:A.8} and the boundedness of \(\psi\) under Assumption \ref{a:3.7} to deduce the following:
\begin{equation}\label{Eq:A.9}
\begin{split}
    |\sqrt{n}\mathbb{J}_{2,n,k}( \boldsymbol{\hat{\beta}}_{n},\boldsymbol{\hat{\delta}}_{n})|\leq\frac{\phi_1}{\sqrt{n}}\sum_{i=1}^{n}\mathds{1}(\hat{\boldsymbol{x}}_i^\top  \boldsymbol{\hat{\beta}}_{n}=c_{\rho})|\hat{x}_{i,k}|
    \leq\frac{\phi_1\underset{i\leq n}{\max}\|\hat{\boldsymbol{x}}_{i}\|}{\sqrt{n}}\sum_{i=1}^{n}\mathds{1}(\hat{\boldsymbol{x}}_i^\top  \boldsymbol{\hat{\beta}}_{n}=c_{\rho}),
    \end{split}
\end{equation}
where $\phi_1$ is defined in Assumption \ref{a:3.7}.
By the boundedness of $\psi(\cdot)$ under Assumption \ref{a:3.7}, the consistency of $\hat{\boldsymbol{\beta}}_n$ and $\hat{\boldsymbol{\delta}}_n$, and Assumption \ref{a:3.6}, the summation term on the right-hand side of \eqref{Eq:A.9} is finite almost surely. This follows from the same argument as in \cite{POWELL1984}, p. 320, and equation (A.19) of \cite{Lee1992}. Next, observe that
\begin{align*}
  \frac{\underset{i\leq n}{\max}\|\hat{\boldsymbol{x}}_{i}\|}{\sqrt{n}}&\leq \frac{\underset{i\leq n}{\max}\|\hat{\boldsymbol{x}}_{i}-\boldsymbol{x}_{i}\|}{\sqrt{n}}+\frac{\underset{i\leq n}{\max}\|\boldsymbol{x}_{i}\|}{\sqrt{n}}\\
  &\leq \frac{\underset{i\leq n}{\max}\|\boldsymbol{z}_{i}\|}{\sqrt{n}}.\|\boldsymbol{\hat{\delta}}_{n}-\boldsymbol{\delta}_{0}\|+\frac{\underset{i\leq n}{\max}\|\boldsymbol{x}_{i}\|}{\sqrt{n}}\\
  &\leq \frac{\mathcal{K}_{1}}{\sqrt{n}}.\|\boldsymbol{\hat{\delta}}_{n}-\boldsymbol{\delta}_{0}\|+\frac{\mathcal{K}_{0}}{\sqrt{n}}
\end{align*}
by Assumption \ref{a:3.5}.
Using the consistency of \(\boldsymbol{\hat{\delta}}_{n}\), it follows that
$\frac{\max_{i\leq n}\|\hat{\boldsymbol{x}}_{i}\|}{\sqrt{n}}
=o(1)$ almost surely. Consequently,
$\sqrt{n}\|\mathbb{J}_{2,n}( \boldsymbol{\hat{\beta}}_{n},\boldsymbol{\hat{\delta}}_{n})\|
=o_{P}(1).$ Similarly, since $\hat{\boldsymbol{\delta}}_{n}$ is the least squares estimator, it satisfies
the normal equations exactly. Hence the second equation in \eqref{Eq:A.7} is identically
zero, and in particular it is $o_{P}(1)$ as $n\to\infty$. Therefore, condition (1) is proved.

\noindent \textbf{Step~2:} Next, we establish the condition (2). To prove differentiability of $\mathbb{J}(\boldsymbol{\beta},\boldsymbol{\delta})$ at $(\boldsymbol{\beta}_0^\top, \boldsymbol{\delta}_0^\top)^\top$, it is enough to show that $\mathbb{J}_{2}(\boldsymbol{\beta},\boldsymbol{\delta})-\mathbb{J}_{2}(\boldsymbol{\beta}_{0},\boldsymbol{\delta}_{0}) = o(\|\boldsymbol{\beta}-\boldsymbol{\beta}_{0}\|+\|\boldsymbol{\delta}-\boldsymbol{\delta}_{0}\|) $ and
$\mathbb{J}_1(\boldsymbol{\delta}) - \mathbb{J}_1(\boldsymbol{\delta}_0) = o(\|\boldsymbol{\delta} - \boldsymbol{\delta}_0||)$.
Consider the following:
\begin{equation}\label{eq:A.10}
\begin{split}
    \mathbb{J}_{2}(\boldsymbol{\beta},\boldsymbol{\delta})-\mathbb{J}_{2}(\boldsymbol{\beta}_{0},\boldsymbol{\delta}_{0})&=\mathbb{E}\left[\big\{\mathds{1}\left(\boldsymbol{ X}^\top\big(\boldsymbol{\delta}\right)\boldsymbol{\beta}>c_{\rho}\big)-\mathds{1}\big(\boldsymbol{X}^\top\boldsymbol{\beta}_{0}>c_{\rho}\big)\big\}\psi\big(Y-\boldsymbol{X}^\top\boldsymbol{\beta}_{0}\big)\boldsymbol{X}\right]\\
    &+\mathbb{E}\left[\mathds{1}\left(\boldsymbol{ X}^\top\big(\boldsymbol{\delta}\right)\boldsymbol{\beta}>c_{\rho}\big)\big\{\psi\big(Y-\boldsymbol{X}^\top(\boldsymbol{\delta})\boldsymbol{\beta}\big)\boldsymbol{X}(\boldsymbol{\delta})-\psi\big(Y-\boldsymbol{X}^\top\boldsymbol{\beta}_{0}\big)\boldsymbol{X}\big\}\right].
    \end{split}
\end{equation}
Since $\psi(Y-\boldsymbol X^\top\boldsymbol\beta_0)=\psi(\varepsilon)$, the first term in Equation~\eqref{eq:A.10} vanishes by the first-order moment condition
$\mathbb{E}\left[\psi(\varepsilon)\mid\boldsymbol{X}\right]=0.$
Therefore, Equation~\eqref{eq:A.10} reduces to
\begin{equation}\label{eq:A.11}
\begin{split}
    \mathbb{J}_{2}(\boldsymbol{\beta},\boldsymbol{\delta})-\mathbb{J}_{2}(\boldsymbol{\beta}_{0},\boldsymbol{\delta}_{0})&=\mathbb{E}\left[\mathds{1}\left(\boldsymbol{ X}^\top\big(\boldsymbol{\delta}\right)\boldsymbol{\beta}>c_{\rho}\big)\big\{\psi\big(Y-\boldsymbol{X}^\top(\boldsymbol{\delta})\boldsymbol{\beta}\big)\boldsymbol{X}(\boldsymbol{\delta})\big\}\right].
    \end{split}
\end{equation}
Now, using the first-order Taylor expansion around  $(\boldsymbol{\beta}_0^\top,\boldsymbol{\delta}_0^\top)$ along with $\mathbb{E}(\psi(\varepsilon)|\boldsymbol{X}) = 0$, Equation~\eqref{eq:A.11} yields
\begin{equation} \label{eq:A.12}
    \begin{split}
        \mathbb{J}_{2}(\boldsymbol{\beta},\boldsymbol{\delta})-\mathbb{J}_{2}(\boldsymbol{\beta}_{0},\boldsymbol{\delta}_{0})& = - \ \mathbb{E}
\,\left[\mathds{1}\left(\boldsymbol{ X}^\top\boldsymbol{\beta}_{0}>c_{\rho}\right)\left\{
\,\gamma_{2,0}C_{\psi}\boldsymbol{X}\boldsymbol{Z}^\top \left(\boldsymbol{\delta}-\boldsymbol{\delta}_0\right)
+C_{\psi}\boldsymbol{X}\boldsymbol{X}^\top(\boldsymbol{\beta}-\boldsymbol{\beta}_0)\right\}
\right]\\
&\qquad\qquad\qquad\qquad\qquad\qquad\qquad\qquad+ o\big(\|\boldsymbol{\delta}-\boldsymbol{\delta}_0\|+\|\boldsymbol{\beta}-\boldsymbol{\beta}_0\|\big)\\
& = o\big(\|\boldsymbol{\delta}-\boldsymbol{\delta}_0\|+\|\boldsymbol{\beta}-\boldsymbol{\beta}_0\|\big)
    \end{split}
\end{equation}
where $C_{\psi}$ is defined as in Assumption~\ref{a:3.8}.  

Next, consider $\mathbb{J}_{1} (\boldsymbol{\delta})=\mathbb{E}\big[(\boldsymbol{W}-\boldsymbol{Z}^\top\boldsymbol{\delta})\boldsymbol{Z}\big]$ and apply the first-order Taylor expansion, we have
\begin{equation}\label{eq:A.13}
   \mathbb{J}_{1}(\boldsymbol{\delta})-\mathbb{J}_{1}(\boldsymbol{\delta}_{0})
   =-\,\mathbb{E}\big[\boldsymbol{Z}\boldsymbol{Z}^\top\big]\big(\boldsymbol{\delta}-\boldsymbol{\delta}_{0}\big) + o(\|\boldsymbol{\delta}-\boldsymbol{\delta}_{0}\|).
\end{equation}
Now, combining \eqref{eq:A.12} and \eqref{eq:A.13}, we obtain
\begin{equation}
\mathbb{J}(\boldsymbol{\beta},\boldsymbol{\delta})-\mathbb{J}(\boldsymbol{\beta}_{0},\boldsymbol{\delta}_{0})
=\begin{bmatrix}
\boldsymbol{\Sigma}_{2,\boldsymbol{\beta}_{0}} & \boldsymbol{\Sigma}_{2,\boldsymbol{\delta}_{0}}\\
\boldsymbol{0} & \boldsymbol{\Sigma}_{1,\boldsymbol{\delta}_{0}}
\end{bmatrix}
\begin{pmatrix}
\boldsymbol{\beta}_{0}-\boldsymbol{\beta}\\
\boldsymbol{\delta}_{0}-\boldsymbol{\delta}
\end{pmatrix}
+\begin{pmatrix}
o\big(\|\boldsymbol{\beta}-\boldsymbol{\beta}_{0}\|+\|\boldsymbol{\delta}-\boldsymbol{\delta}_{0}\|\big)\\
o(\|\boldsymbol{\delta}-\boldsymbol{\delta}_{0}\|)
\end{pmatrix}.
\end{equation}
This implies that $\mathbb{J}$ is differentiable at $(\boldsymbol{\beta}_{0}^\top,\boldsymbol{\delta}_{0}^\top)$, with derivative matrix
\begin{equation}\label{eq:A.15}
\boldsymbol{\Sigma}
=\frac{\partial \mathbb{J}(\boldsymbol{\beta}_{0},\boldsymbol{\delta}_{0})}
{\partial(\boldsymbol{\beta}_{0}^\top,\boldsymbol{\delta}_{0}^\top)}
=\begin{bmatrix}
\boldsymbol{\Sigma}_{2,\boldsymbol{\beta}_{0}} & \boldsymbol{\Sigma}_{2,\boldsymbol{\delta}_{0}}\\
\boldsymbol{0} & \boldsymbol{\Sigma}_{1,\boldsymbol{\delta}_{0}}
\end{bmatrix}.
\end{equation}
Hence Condition (2) is satisfied.

\noindent \textbf{Step~3:} We next verify condition (3). Rather than proceeding by direct calculation, we express the relevant remainder in terms of the standardized empirical process. Let the random variables $\boldsymbol{u}_1,\ldots,\boldsymbol{u}_n$ be i.i.d. copies of $\boldsymbol{u}$, and for some measurable function $f$ define the following:
\begin{equation}\label{eq:A.16}
    \mathbb{G}_{n}(f(\boldsymbol{u}))
    :=n^{-1/2}\sum_{i=1}^{n}
    \left\{f(\boldsymbol{u}_i)-\mathbb{E}[f(\boldsymbol{u}_i)]\right\}.
\end{equation}
First, observe that
\begin{equation}\label{eq:A.17}
\frac{
\left\|
\mathbb{J}_{n}(\boldsymbol{\beta},\boldsymbol{\delta})
-\mathbb{J}(\boldsymbol{\beta},\boldsymbol{\delta})
-\mathbb{J}_{n}(\boldsymbol{\beta}_{0},\boldsymbol{\delta}_{0})
\right\|
}{
n^{-1/2}
+\|\mathbb{J}_{n}(\boldsymbol{\beta},\boldsymbol{\delta})\|
+\|\mathbb{J}(\boldsymbol{\beta},\boldsymbol{\delta})\|
}
\leq
n^{1/2}
\left\|
\mathbb{J}_{n}(\boldsymbol{\beta},\boldsymbol{\delta})
-\mathbb{J}(\boldsymbol{\beta},\boldsymbol{\delta})
-\mathbb{J}_{n}(\boldsymbol{\beta}_{0},\boldsymbol{\delta}_{0})
\right\|.
\end{equation}
Consequently, it is sufficient to show that
\begin{equation}\label{eq:A.18}
 \sup_{\substack{\|\boldsymbol{\beta}-\boldsymbol{\beta}_0\|\leq a_n\\
 \|\boldsymbol{\delta}-\boldsymbol{\delta}_0\|\leq b_n}}
 \left\|
 \mathbb{J}_{n}(\boldsymbol{\beta},\boldsymbol{\delta})
 -\mathbb{J}(\boldsymbol{\beta},\boldsymbol{\delta})
 -\mathbb{J}_{n}(\boldsymbol{\beta}_{0},\boldsymbol{\delta}_{0})
 \right\|
 =o_P(n^{-1/2}).
\end{equation}
We begin with the first-stage component; that is, consider the following:
\begin{equation} \label{eq:A.19}
    \begin{split}
        \sup_{\|\boldsymbol{\delta}-\boldsymbol{\delta}_{0}\|\leq b_n}
n^{1/2}
\left\|
\mathbb{J}_{1,n}(\boldsymbol{\delta})
-\mathbb{J}_{1}(\boldsymbol{\delta})
-\mathbb{J}_{1,n}(\boldsymbol{\delta}_{0})
\right\| 
& =
\sup_{\|\boldsymbol{\delta}-\boldsymbol{\delta}_{0}\|\leq b_n}
\left\|
\mathbb{G}_{n}\bigl(m_{1}(\boldsymbol{v},\boldsymbol{\delta})\bigr)
-\mathbb{G}_{n}\bigl(m_{1}(\boldsymbol{v},\boldsymbol{\delta}_{0})\bigr)
\right\| \\
&\leq \sup_{\|\boldsymbol{\delta}-\boldsymbol{\delta}_{0}\|\leq b_n} n^{-1/2} \sum_{i=1}^{n} 
\left\|\boldsymbol{Z}_{i}^{\top}(\boldsymbol{\delta}-\boldsymbol{\delta}_{0})\boldsymbol{Z}_{i}
- \mathbb{E}\left[
\boldsymbol{Z}_{i}^{\top}(\boldsymbol{\delta}-\boldsymbol{\delta}_{0})\boldsymbol{Z}_{i}
\right]\right\|,\\
    \end{split}
\end{equation}
where, recall that $m_1(.)$ is defined in Equation~\eqref{Eq:A.5}. Applying Chebyshev's inequality for $t > 0$,
\begin{align*}
\mathbb{P}\left[ \sum_{i=1}^{n} 
\left\|\boldsymbol{Z}_{i}^{\top}(\boldsymbol{\delta}-\boldsymbol{\delta}_{0})\boldsymbol{Z}_{i}
- \mathbb{E}\left[
\boldsymbol{Z}_{i}^{\top}(\boldsymbol{\delta}-\boldsymbol{\delta}_{0})\boldsymbol{Z}_{i}
\right]\right\|>t\sqrt{n}\right]&\leq\frac{\mathbb{E}\left[\left\|\boldsymbol{Z}_{i}^{\top}(\boldsymbol{\delta}-\boldsymbol{\delta}_{0})\boldsymbol{Z}_{i}
- \mathbb{E}\left[
\boldsymbol{Z}_{i}^{\top}(\boldsymbol{\delta}-\boldsymbol{\delta}_{0})\boldsymbol{Z}_{i}
\right]\right\|^{2}\right]}{t^{2}}\\
&\leq\frac{4\;\|\boldsymbol{\delta}-\boldsymbol{\delta}_{0}\|^{2}\; \mathbb{E}\|\boldsymbol{Z}_{i}\|^{4}}{t^{2}},
\end{align*}
where the first inequality follows from Jensen's inequality and the second inequality follows from Cauchy-Schwarz inequality. Under Assumption~\ref{a:3.5}, we have $\mathbb{E}\|\boldsymbol{Z}_{i}\|^4 < \infty$. 
Thus, we obtain
\begin{equation}
   \sup_{\|\boldsymbol{\delta}-\boldsymbol{\delta}_{0}\|\leq b_n}
\left\|
\mathbb{G}_{n}\bigl(m_{1}(\boldsymbol{v},\boldsymbol{\delta})\bigr)
-\mathbb{G}_{n}\bigl(m_{1}(\boldsymbol{v},\boldsymbol{\delta}_{0})\bigr)
\right\|\leq b_{n}^{2} O_{P}(1)=o_P(1), \text{ as } b_{n}\to 0. 
\end{equation}
Thus, condition (3) is satisfied for the first-stage component
$\mathbb{J}_{1}(\boldsymbol{\delta})$. 

We now consider the second-stage component $\mathbb{J}_2(.)$ to establish Equation~\eqref{eq:A.18}. In particular, consider the following:
\begin{equation}\label{eq:A.21}
\begin{aligned}
\sup_{\substack{\|\boldsymbol{\beta}-\boldsymbol{\beta}_0\|\leq a_n\\
\|\boldsymbol{\delta}-\boldsymbol{\delta}_0\|\leq b_n}}
n^{1/2}
\left\|
\mathbb{J}_{2,n}(\boldsymbol{\beta},\boldsymbol{\delta})
-\mathbb{J}_{2}(\boldsymbol{\beta},\boldsymbol{\delta})
-\mathbb{J}_{2,n}(\boldsymbol{\beta}_{0},\boldsymbol{\delta}_{0})
\right\| 
&\leq
\sup_{\|\boldsymbol{\beta}-\boldsymbol{\beta}_0\|\leq a_n}
n^{1/2}
\left\| S_{2,n}(\boldsymbol{\beta})
-\mathbb{J}_{2}(\boldsymbol{\beta},\boldsymbol{\delta}_{0})
\right\| + \\
& 
\sup_{
\|\boldsymbol{\delta}-\boldsymbol{\delta}_0\|\leq b_n}
n^{1/2}
\left\|
S_{2, n}(\boldsymbol{\beta},\boldsymbol{\delta})
-\mathbb{J}_{2}(\boldsymbol{\beta},\boldsymbol{\delta})
+\mathbb{J}_{2}(\boldsymbol{\beta},\boldsymbol{\delta}_{0})
\right\|,
\end{aligned}
\end{equation}
where $S_{2,n}(\boldsymbol{\beta}) := \mathbb{J}_{2,n}(\boldsymbol{\beta},\boldsymbol{\delta}_{0}) -\mathbb{J}_{2,n}(\boldsymbol{\beta}_{0},\boldsymbol{\delta}_{0})$ and $S_{2, n}(\boldsymbol{\beta},\boldsymbol{\delta}) :=\mathbb{J}_{2,n}(\boldsymbol{\beta},\boldsymbol{\delta})
-\mathbb{J}_{2,n}(\boldsymbol{\beta},\boldsymbol{\delta}_{0})$. 

We now proceed to show that each term on the right-hand side of \eqref{eq:A.21} is $o_P(1)$. Consider the $k$th component of the first term which is bounded from above by      
\begin{equation}\label{eq:A.22}
\begin{aligned}
&\sup_{\|\boldsymbol{\beta}-\boldsymbol{\beta}_0\|\leq a_n}
\left|
\mathbb{G}_{n}\bigl(
m_{2,k}(\boldsymbol{u},\boldsymbol{\beta},\boldsymbol{\delta}_{0})
\bigr)
-\mathbb{G}_{n}\bigl(
m_{2,k}(\boldsymbol{u},\boldsymbol{\beta}_{0},\boldsymbol{\delta}_{0})
\bigr)
\right|,
\end{aligned}
\end{equation}
where $m_{2,k}(\boldsymbol{u},\boldsymbol{\beta},\boldsymbol{\delta})$ denote the $k$th coordinate of
$m_2(\boldsymbol{u},\boldsymbol{\beta},\boldsymbol{\delta})$ (see Equation~\eqref{Eq:A.5}).  

Now, we proceed to show that \eqref{eq:A.22} is $o_{P}(1)$ as $a_n \to 0$ using Lemma 2.17 of \cite{PakesandPollard}. In particular, we define the class
\begin{equation}\label{eq:A.23}\mathcal{F}_{1} = \left\{  m_{2,k}(\boldsymbol{u},\boldsymbol{\beta},\boldsymbol{\delta}_{0}):=\mathds{1}(\boldsymbol{X}^\top\boldsymbol{\beta}>c_{\rho})\psi(Y-\boldsymbol{X}^\top\boldsymbol{\beta})(X_{k}): \boldsymbol{\beta} \in \mathcal{B}\right\},
\end{equation}
where $X_k$ denote the $k$th coordinate of the vector $\boldsymbol{X}$. Then, \eqref{eq:A.22} is $o_{P}(1)$ by Lemma 2.17 of \cite{PakesandPollard} if $\mathcal{F}_{1}$ satisfies the following two conditions:
\begin{enumerate}[label=(\Roman*)]
    \item $\mathcal{F}_1$ is an Euclidean class with an envelope $F_1$ satisfying $\mathbb{E}[F_1^{2}]<\infty$.
 \item $\lim_{\boldsymbol{\beta}\to\boldsymbol{\beta}_0}\mathbb{E}\!\left[
\left(m_{2,k}(\boldsymbol{u},\boldsymbol{\beta},\boldsymbol{\delta}_0)-m_{2,k}(\boldsymbol{u},\boldsymbol{\beta}_0,\boldsymbol{\delta}_0)\right)^2
\right]=0.$
\end{enumerate}

To verify condition (I), we show that each factor in $m_{2,k}$ belongs to an Euclidean class. Consider the following three classes of functions:
$$
\text{(a)}\; \left\{\mathds{1}(\boldsymbol{X}^\top\boldsymbol{\beta}>c_{\rho}):\boldsymbol{\beta}\in \mathcal{B}\right\}, \;
\text{(b)}\; \left\{\psi(Y-\boldsymbol{X}^\top\boldsymbol{\beta}):\boldsymbol{\beta}\in \mathcal{B}\right\}, \;
\text{(c)}\; \left\{X_{k} \right\}.$$
Note that the classes (a) and (b) are each Euclidean with a constant envelope, since both the indicator function $\mathds{1}(\cdot)$ and $\psi(\cdot)$ are of bounded variation. The Euclidean property of both the classes therefore follows from Lemma 22(ii) of \cite{NolanandDavidPollard}. Class (c) is trivially Euclidean with envelope $|X_{k}|$. Since the product of Euclidean classes remains Euclidean with envelope equal to the product of the individual envelopes (Lemma 2.14(ii) of \cite{PakesandPollard}), it follows that $\mathcal{F}_1$ is Euclidean with the envelope $F_1\equiv \phi_{1}|X_{k}|.$
 Moreover, $F_1$ is square-integrable by Assumption \ref{a:3.5}, since
$\mathbb{E}\left[\|\boldsymbol{X}\|^{2}\right]
 <\infty.$ This establishes condition (I).
To establish condition (II), we need to show that
\begin{equation}\label{eq:A.24}
 \mathbb{E}\big[\big\|\mathds{1}(\boldsymbol{X}^\top\boldsymbol{\beta}>c_{\rho})\psi(Y-\boldsymbol{X}^\top\boldsymbol{\beta})\boldsymbol{X}-\mathds{1}(\boldsymbol{X}^\top\boldsymbol{\beta}_{0}>c_{\rho})\psi(Y-\boldsymbol{X}^\top\boldsymbol{\beta}_{0})\boldsymbol{X}\big\|^2\big]    
\longrightarrow 0
\end{equation}
as $\|\boldsymbol{\beta}-\boldsymbol{\beta}_0\|\to0$. We add and subtract $\mathds{1}(\boldsymbol{X}^\top\boldsymbol{\beta}_{0}>c_{\rho})\psi(Y-\boldsymbol{X}^\top\boldsymbol{\beta})\boldsymbol{X}$ to bound \eqref{eq:A.24} from above, by the following two components. First,
\begin{equation}
\mathbb{E}\left[
\left|
\mathds{1}(\boldsymbol{X}^{\top}\boldsymbol{\beta}>c_{\rho})
-\mathds{1}(\boldsymbol{X}^{\top}\boldsymbol{\beta}_{0}>c_{\rho})
\right|
\right] \leq
\mathbb{E}\left[
\mathds{1}\left\{
|c_{\rho}-\boldsymbol{X}^\top\boldsymbol{\beta}_{0}|
\leq
\mathcal{K}_{0}
\|\boldsymbol{\beta}-\boldsymbol{\beta}_{0}\|
\right\}
\right],
\end{equation}
which converges to zero by Assumption~\ref{a:3.6} as $\|\boldsymbol{\beta}-\boldsymbol{\beta}_{0}\|\to 0$. Second, by the continuity and boundedness of $\psi$ under Assumption~\ref{a:3.7}, together with the dominated convergence theorem, we have
$$
\mathbb{E}\left[
\left\|
\left\{
\psi(Y-\boldsymbol{X}^{\top}\boldsymbol{\beta})
-\psi(Y-\boldsymbol{X}^{\top}\boldsymbol{\beta}_{0})
\right\}
\boldsymbol{X}
\right\|^{2}
\right]
\longrightarrow 0,$$
as $\|\boldsymbol{\beta}-\boldsymbol{\beta}_{0}\|\to 0$.
Thus, condition (II) holds for $\mathcal{F}_1$, and therefore \eqref{eq:A.22} is $o_P(1)$ as $a_n \to 0$. Hence, the first term of \eqref{eq:A.21} is $o_P(1)$ as $a_n \to 0$.  

Next, we consider the second term of \eqref{eq:A.21} and show that it is $o_{P}(1)$ as $b_n \to 0$. We again invoke Lemma 2.17 of \cite{PakesandPollard} and verify conditions (I) and (II) for the following class
\begin{equation}\label{eq:A.26}
\mathcal{H} 
=  
\bigl\{ 
f_{\boldsymbol{\beta},\boldsymbol{\delta}} 
- 
f_{\boldsymbol{\beta},\boldsymbol{\delta}_{0}} 
: 
\boldsymbol{\beta} \in \mathcal{B},\ 
\boldsymbol{\delta} \in \Delta 
\bigr\},
\end{equation}
where $f_{\boldsymbol{\beta},\boldsymbol{\delta}} 
:= m_{2,k}(\boldsymbol{u},\boldsymbol{\beta},\boldsymbol{\delta})$. To verify condition (I) for $\mathcal{H}$, we define the following class   
\begin{equation}\label{eq:A.27}
\mathcal{F}_{2} 
=
\bigl\{ 
f_{\boldsymbol{\beta},\boldsymbol{\delta}}
: 
\boldsymbol{\beta} \in \mathcal{B},\ 
\boldsymbol{\delta} \in \Delta 
\bigr\}.
\end{equation}
Now, observe that if $\mathcal{F}_2$ is an Euclidean class with a square-integrable envelope $E_2$, then the subclass
$\left\{
f_{\boldsymbol{\beta},\boldsymbol{\delta}_0}:
\boldsymbol{\beta}\in\mathcal{B}
\right\}$ is also Euclidean with an envelope $E_1$. Consequently, $\mathcal{H}$ is Euclidean since polynomial covering-number bounds are preserved under finite sums and differences of Euclidean classes. Moreover, $\mathcal{H}$ is Euclidean with the envelope $E_1+E_2$ since $\left|
f_{\boldsymbol{\beta},\boldsymbol{\delta}}
-
f_{\boldsymbol{\beta},\boldsymbol{\delta}_0}
\right|
\leq
\left|f_{\boldsymbol{\beta},\boldsymbol{\delta}}\right|
+
\left|f_{\boldsymbol{\beta},\boldsymbol{\delta}_0}\right|
\leq E_1+E_2.$ 
Thus, we only need to verify that $\mathcal{F}_2$ is Euclidean with a square-integrable envelope to verify condition (I) for $\mathcal{H}$. Therefore, using a similar argument as for $\mathcal{F}_1$, we conclude that $\mathcal{F}_2$ is an Euclidean class. This implies that $\mathcal{H}$ is an Euclidean class with envelope $\phi_1\sup_{\boldsymbol{\delta}\in\Delta}
|X_{k}(\boldsymbol{\delta})|$, where $X_{k}(\boldsymbol{\delta})$ is the $k$th element of the vector $\boldsymbol{X}(\boldsymbol{\delta})$ (see Equation~\eqref{Eq:3.5}). Further, to show that the envelope of $\mathcal{H}$ is square-integrable, consider the following:
$$
\mathbb{E}
\|\boldsymbol{X}(\boldsymbol{\delta})\|^2
\leq
\mathbb{E}\|\boldsymbol{X}\|^2
+
\mathbb{E}|\boldsymbol{Z}^\top (\boldsymbol{\delta} - \boldsymbol{\delta}_0)|^2 \leq
\mathbb{E}\|\boldsymbol{X}\|^2
+
\|\boldsymbol{\delta} - \boldsymbol{\delta}_0\|^2\mathbb{E}\|\boldsymbol{Z}\|^2
<\infty.$$
Thus, using the fact that $\|\boldsymbol{X}(\boldsymbol{\delta}) - \boldsymbol{X}\| = |\boldsymbol{Z}^\top (\boldsymbol{\delta} - \boldsymbol{\delta}_0)|$ along with Assumption~\ref{a:3.5}, we have shown that the envelope of $\mathcal{H}$ is square-integrable for $\|\boldsymbol{\delta} - \boldsymbol{\delta}_0\| \leq b_n$, where $b_n \to 0$. This establishes condition (I) for the class $\mathcal{H}$.  

Now, we proceed to establish condition (II) for $\mathcal{H}$ in a similar fashion as for the class $\mathcal{F}_1$ (see Equation~\eqref{eq:A.24}). We add and subtract $\mathds{1}(\boldsymbol{X}^\top\boldsymbol{\beta}>c_{\rho})\psi(Y-\boldsymbol{X}^\top\boldsymbol{\beta})\boldsymbol{X}$ to bound $\mathbb{E}(f_{\boldsymbol\beta,\boldsymbol\delta}-f_{\boldsymbol\beta,\boldsymbol\delta_0})^2$ from above, by the following two components. First,
\begin{equation}\label{eq:A.28}
\begin{split}
\mathbb{E}\left[ \left| \mathds{1}\bigl(\boldsymbol{X}(\boldsymbol{\delta})^\top\boldsymbol{\beta} > c_\rho\bigr) - \mathds{1}\bigl(\boldsymbol{X}^\top\boldsymbol{\beta} > c_\rho\bigr) \right| \right]
\;&\le\;
\mathbb{P}\Bigl[ \bigl|c_\rho - \boldsymbol{X}^\top\boldsymbol{\beta}\bigr| \le C_0\, \mathcal{K}_{1}\, \|\boldsymbol{\delta}-\boldsymbol{\delta}_0\| \Bigr],\\
&\le\;
\mathbb{P}\Bigl[ \bigl|c_\rho - \boldsymbol{X}^\top\boldsymbol{\beta}_{0}\bigr| \le C_0\, \mathcal{K}_{1}\, b_{n} +\mathcal{K}_{0}\,a_{n}\Bigr] \to 0,
\end{split}
\end{equation}
 as $a_{n} \to 0$ and $b_{n} \to 0$, follows from Assumption \ref{a:3.6}. Here $C_0 := \sup_{\boldsymbol{\beta}\in\mathcal{B}}\|\boldsymbol{\beta}\| < \infty$ by compactness of $\mathcal{B}$ and the inequality in \eqref{eq:A.28} follows from using Assumption \ref{a:3.5} along with the following inequality 
$$
\left| \mathds{1}\bigl(\boldsymbol{X}(\boldsymbol{\delta})^\top\boldsymbol{\beta} > c_\rho\bigr) - \mathds{1}\bigl(\boldsymbol{X}^\top\boldsymbol{\beta} > c_\rho\bigr) \right|
\;\le\;
\mathds{1}\Bigl( \bigl|c_\rho - \boldsymbol{X}^\top\boldsymbol{\beta}\bigr| \le \|\boldsymbol{Z}\|\,\|\boldsymbol{\delta}-\boldsymbol{\delta}_0\|\,\|\boldsymbol{\beta}\| \Bigr).$$
The second component is,
$$
\mathbb{E}\Big[ \big( \big(\psi(Y - \boldsymbol{X}(\boldsymbol{\delta})^\top\boldsymbol{\beta}) - \psi(Y - \boldsymbol{X}^\top\boldsymbol{\beta})\big)\, \boldsymbol{X} \big)^{2} \Big] \;\longrightarrow\; 0
\quad \text{as } \|\boldsymbol{\delta} - \boldsymbol{\delta}_0\| \to 0,$$
where the convergence follows from Assumption \ref{a:3.7} along with an application of the dominated convergence theorem. This establishes condition (II) for the class $\mathcal{H}$. This implies that the $k$th component of the second term in \eqref{eq:A.21} is $o_P(1)$ as $b_n \to 0$, that is,
\begin{equation}
    \sup_{ \|\boldsymbol{\delta} - \boldsymbol{\delta}_{0}\| \leq b_n} \big|\mathbb{G}_{n}\big(m_{2,k}(\boldsymbol{u},\boldsymbol{\beta},\boldsymbol{\delta})\big)-\mathbb{G}_{n}\big(m_{2,k}(\boldsymbol{u},\boldsymbol{\beta},\boldsymbol{\delta}_{0})\big)\big|\xrightarrow{P} 0 \text{ as } b_n \to 0.
\end{equation}
Hence, the second term in \eqref{eq:A.21} is $o_P(1)$ as $b_n \to 0$. This completes \textbf{Step 3}, and hence the condition (3) of the theorem's statement holds.  

\noindent \textbf{Step~4:} In this step, we verify condition (4) of the theorem's statement. We apply the central limit theorem (CLT) to $\mathbb{J}_n(\boldsymbol{\beta}_0, \boldsymbol{\delta}_0):= n^{-1} \sum_{i = 1}^n \mathbb{V}_i(\boldsymbol{\beta}_0, \boldsymbol{\delta}_0)$ (see Equation~\eqref{Eq:A.5}) with $\mathbb{E}(\mathbb{J}_n(\boldsymbol{\beta}_0, \boldsymbol{\delta}_0)) = 0$, we have:
 \begin{equation}\label{eq:A.30}
 \sqrt{n}\mathbb{J}_{n}(\boldsymbol{\beta}_{0},\boldsymbol{\delta}_{0})\xrightarrow{D} \mathbb{N}_{2p+3}\Bigg(\boldsymbol{0},\begin{bmatrix}
     \mathbb{E}[m_{2}m_{2}^\top] &\mathbb{E}[m_{2}m_{1}^\top]\\
     \mathbb{E}[m_{1}m_{2}^\top] &\mathbb{E}[m_{1}m_{1}^\top]
 \end{bmatrix}\Bigg),
 \end{equation} 
as $n\to \infty.$ Here, $m_{1}$ and $m_{2}$ denote the $m_{1}(\boldsymbol{v}_{i},\boldsymbol{\delta})$ and $m_{2}(\boldsymbol{u}_{i},\boldsymbol{\beta},\boldsymbol{\delta})$ (see Equation~\eqref{Eq:A.5}) respectively. We now examine the covariance matrix elements in \eqref{eq:A.30}. The diagonal elements of the covariance matrix in \eqref{eq:A.30} are given by
 \begin{equation}\label{eq:A.31}
    \mathbb{E}[m_{2}m_{2}^\top]= \mathbb{E}[\mathds{1}(\boldsymbol{X}^\top\boldsymbol{\beta}_{0}>c_{\rho})\psi^{2}(Y-\boldsymbol{X}^\top\boldsymbol{\beta}_{0})\boldsymbol{X}\boldsymbol{X}^\top]= \mathbb{D}_{2},
 \end{equation}
 which is finite under the conditions specified in Assumption \ref{a:3.8}. Similarly, we can show that the component $\mathbb{E}[m_{1}m_{1}^\top]$ has the form
 \begin{equation}\label{eq:A.32}
     \mathbb{E}[m_{1}m_{1}^\top]= \mathbb{E}[(W-\boldsymbol{Z}^\top\boldsymbol{\delta}_{0})\boldsymbol{Z}_{i}(W-\boldsymbol{Z}^\top\boldsymbol{\delta}_{0})^\top\boldsymbol{Z}^\top]\\
     =\mathbb{E}[\vartheta^{2}\boldsymbol{Z}\boldsymbol{Z}^\top]=\mathbb{D}_{1}.
 \end{equation}
  The matrix $\mathbb{D}_{1}$ is finite due to the bounded conditional moment assumption on the reduced-form error and the second-moment condition on the covariates, as specified in the last part of Assumption \ref{a:3.8}. We now consider the off-diagonal elements in \eqref{eq:A.30}. For the off-diagonal elements,
  \begin{equation}\label{eq:A.33}
  \begin{split}
   \mathbb{E}\!\left[m_{2}m_{1}^\top\right]   &=\mathbb{E}\!\left[\mathds{1}(\boldsymbol{X}^{\top}\boldsymbol{\beta}_0 > c_{\rho})\,
    \psi(Y - \boldsymbol{X}^{\top}\boldsymbol{\beta}_0)\,
   \boldsymbol{X}\,\vartheta\,\boldsymbol{Z}^{\top}\right]\\
&= \mathbb{E}\!\left[\vartheta_i\,\boldsymbol{Z}^{\top} 
    \mathbb{E}\!\left[\mathbf{1}(\boldsymbol{X}^{\top}\boldsymbol{\beta}_0 > c_{\rho})\,
    \psi(Y - \boldsymbol{X}^{\top}\boldsymbol{\beta}_0)\,
   \boldsymbol{X} \;\big|\; \boldsymbol{X},\right]\right] 
= \mathbb{E}\!\left[\vartheta\,\boldsymbol{Z}_i^{\top}  \boldsymbol{0}\right] \\
&= \boldsymbol{0}
      \end{split}
  \end{equation}
where the last equality follows from the first-order population moment condition (see Equation~\eqref{Eq:3.3}). Similarly, we have $\mathbb{E}[m_{1}m_{2}^\top]= \boldsymbol{0}$. Thus, combining all the components, we have:
\begin{equation}\label{eq:A.34}
 \sqrt{n}\;\mathbb{J}_{n}(\boldsymbol{\beta}_{0},\boldsymbol{\delta}_{0})\xrightarrow{D}  \mathbb{N}_{2p+3}(\boldsymbol{0}, \mathbb{D}),
 \end{equation}
 where $\mathbb{D}= \begin{bmatrix}
     \mathbb{D}_{2} &\boldsymbol{0}\\
    \boldsymbol{0} &\mathbb{D}_{1}
 \end{bmatrix}.$

\noindent\textbf{Step~5:} Condition (5) is satisfied by Assumption~\ref{a:3.4} and by using the fact that $\boldsymbol{\delta}_0$ is a minimizer of the squared error loss function, which is convex.

Now, under these conditions, we can derive the joint asymptotic distribution of $(\boldsymbol{\hat{\beta}}_n^\top, \boldsymbol{\hat{\delta}_n}^\top)^\top$ as follows. Consider the following:
\begin{equation}\label{eq:A.35}
    \begin{split}
        \mathbb{J}_{2,n}( \boldsymbol{\hat{\beta}}_{n}, \boldsymbol{\hat{\delta}}_{n})& =\mathbb{J}_{2,,n}(\boldsymbol{\beta}_{0},\boldsymbol{\delta}_{0})+\mathbb{J}_{2}( \boldsymbol{\hat{\beta}}_{n},\boldsymbol{\hat{\delta}}_{n})+\mathbb{J}_{2,n}( \boldsymbol{\hat{\beta}}_{n},\boldsymbol{\hat{\delta}}_{n})-\mathbb{J}_{2}( \boldsymbol{\hat{\beta}}_{n},\boldsymbol{\hat{\delta}}_{n})-\mathbb{J}_{2,n}(\boldsymbol{\beta}_{0},\boldsymbol{\delta}_{0})\\
        & = \mathbb{J}_{2,,n}(\boldsymbol{\beta}_{0},\boldsymbol{\delta}_{0})+\mathbb{J}_{2}( \boldsymbol{\hat{\beta}}_{n},\boldsymbol{\hat{\delta}}_{n}) + o_P(n^{-1/2}),
    \end{split}
\end{equation}
where the last line follows from \eqref{eq:A.21} and $\mathbb{J}_{2}( \boldsymbol{\hat{\beta}}_{n},\boldsymbol{\hat{\delta}}_{n})$ denotes $\mathbb{J}_{2}(\boldsymbol{\beta},\boldsymbol{\delta})\big|_{\boldsymbol{\beta}= \boldsymbol{\hat{\beta}}_{n},\,\boldsymbol{\delta}=\boldsymbol{\hat{\delta}}_{n}}$. Next, a Taylor expansion of $\mathbb{J}_{2}( \boldsymbol{\hat{\beta}}_{n},\boldsymbol{\hat{\delta}}_{n})$ around $(\boldsymbol{\beta}_0^\top, \boldsymbol{\delta}_0^\top)^\top$ in \eqref{eq:A.35} yields:
\begin{equation}\label{eq:A.36}
    \mathbb{J}_{2}( \boldsymbol{\hat{\beta}}_{n},\boldsymbol{\hat{\delta}}_{n})= \mathbb{J}_{2}( \boldsymbol{\beta}_{0},\boldsymbol{\delta}_{0}) +  \frac{\partial \mathbb{J}_{2}(\boldsymbol{\beta}_{0},\boldsymbol{\delta}_{0})}{\partial \boldsymbol{\beta}^\top}(\boldsymbol{\hat{\beta}}_{n}-\boldsymbol{\beta}_{0})+ \frac{\partial \mathbb{J}_{2}(\boldsymbol{\beta}_{0},\boldsymbol{\delta}_{0})}{\partial \boldsymbol{\delta}^\top}(\boldsymbol{\hat{\delta}}_{n}-\boldsymbol{\delta}_{0})+ r_{n} + o_{P}(n^{-1/2}),
\end{equation}
where $r_{n}=o_{P}(\|\boldsymbol{\hat{\beta}}_{n}-\boldsymbol{\beta}_{0}\|+\|\boldsymbol{\hat{\delta}}_{n}-\boldsymbol{\delta}_{0}\|)$. Then, using condition (1) of the theorem's statement in \eqref{eq:A.36}, we have
\begin{equation} \label{eq:A.37}
    o_{P}(1) = \sqrt{n}\;\mathbb{J}_{2,n}(\boldsymbol{\beta}_{0},\boldsymbol{\delta}_{0}) + (\boldsymbol{\Sigma}_{2,\boldsymbol{\beta}_{0}}+o_{P}(1))\sqrt{n}(\boldsymbol{\hat{\beta}}_{n}-\boldsymbol{\beta}_{0})
+(\boldsymbol{\Sigma}_{2,\boldsymbol{\delta}_{0}}+o_{P}(1))\sqrt{n}(\boldsymbol{\hat{\delta}}_{n}- \boldsymbol{\delta}_{0})+ o_{P}(1),
\end{equation}
where $\boldsymbol{\Sigma}_{2,\boldsymbol{\delta}_{0}}=\frac{\partial \mathbb{J}_{2}(\boldsymbol{\beta}_{0},\boldsymbol{\delta}_{0})}{\partial \boldsymbol{\delta}},$ and $\boldsymbol{\Sigma}_{2,\boldsymbol{\beta}_{0}}=\frac{\partial \mathbb{J}_{2}(\boldsymbol{\beta}_{0},\boldsymbol{\delta}_{0})}{\partial \boldsymbol{\beta}}$. 
Further, proceeding in a similar manner, we can derive the following expression for the first-stage estimator: 
\begin{equation}\label{eq:A.38}
   0=\sqrt{n}\;\mathbb{J}_{1,n}(\boldsymbol{\delta}_{0})+(\boldsymbol{\Sigma}_{1,\boldsymbol{\delta}_{0}}+o_{P}(1))\sqrt{n}(\boldsymbol{\hat{\delta}}_{n}- \boldsymbol{\delta}_{0}). 
\end{equation}
 Now, combining both expressions from Equation~\eqref{eq:A.37} and \eqref{eq:A.38}, we have
 \begin{equation}\label{eq:A.39}
     \left(\boldsymbol{\Sigma}+o_{P}(1)\right)\sqrt{n}
\begin{pmatrix}
          \boldsymbol{\hat{\beta}}_{n}-\boldsymbol{\beta}_{0}\\
          \boldsymbol{\hat{\delta}}_{n}-\boldsymbol{\delta}_{0}
\end{pmatrix}=-\sqrt{n}\;\mathbb{J}_{n}(\boldsymbol{\beta}_{0},\boldsymbol{\delta}_{0}) +o_{P}(1),
 \end{equation}
where $\boldsymbol{\Sigma}$ is as defined in Equation~\eqref{Eq:3.9}. Further, we simplify \eqref{eq:A.39} as
\begin{equation}\label{eq:A.42}
\begin{split}
     \sqrt{n}
\begin{pmatrix}
          \boldsymbol{\hat{\beta}}_{n}-\boldsymbol{\beta}_{0}\\
          \boldsymbol{\hat{\delta}}_{n}-\boldsymbol{\delta}_{0}
\end{pmatrix}&=-\boldsymbol{\Sigma}^{-1}\sqrt{n}\;\mathbb{J}_{n}(\boldsymbol{\beta}_{0},\boldsymbol{\delta}_{0}) +o_{P}(1),
\end{split}
 \end{equation}
where,
\begin{equation}\label{eq:A.43}
    \boldsymbol{\Sigma}^{-1} =
          \begin{bmatrix}
          \boldsymbol{\Sigma}_{2,\boldsymbol{\beta}_{0}} & \boldsymbol{\Sigma}_{2,\boldsymbol{\delta}_{0}}\\
         \boldsymbol{0} &\boldsymbol{\Sigma}_{1,\boldsymbol{\delta}_{0}}
          \end{bmatrix}^{-1} =
          \begin{bmatrix}
            \boldsymbol{\Sigma}_{2,\boldsymbol{\beta}_{0}}^{-1}&-\boldsymbol{\Sigma}_{2,\boldsymbol{\beta}_{0}}^{-1}\boldsymbol{\Sigma}_{2,\boldsymbol{\delta}_{0}}\boldsymbol{\Sigma}_{1,\boldsymbol{\delta}_{0}}^{-1}\\
          \boldsymbol{0}&\boldsymbol{\Sigma}_{1,\boldsymbol{\delta}_{0}}^{-1}
         \end{bmatrix}.    
 \end{equation}
We can derive \eqref{eq:A.42} from \eqref{eq:A.39} as follows. First, note that for any symmetric positive definite matrix $A$ and any nonzero vector $\xi$, we have $\xi^\top A \xi \geq \mu_{\min}(A)\|\xi\|,$ where $\mu_{\min}(A)$ denotes the smallest eigenvalue of $A$. Let $\mu_0$ denote the smallest eigenvalue of $\boldsymbol{\Sigma}^\top\boldsymbol{\Sigma}$ which is symmetric and positive definite by Assumption~\ref{a:3.8}. In view of the fact that for any nonzero vector $\xi$, $\|\boldsymbol{\Sigma} \xi\| = \xi^\top \boldsymbol{\Sigma}^\top\boldsymbol{\Sigma} \xi \geq \sqrt{\mu_0} \|\xi\|$, \eqref{eq:A.39} yields:
\begin{equation}\label{eq:A.40}
\left(\sqrt{\mu_0} +o_{P}(1)\right)
\left\|
\sqrt{n}
\begin{pmatrix}
\boldsymbol{\hat{\beta}}_{n}-\boldsymbol{\beta}_{0}\\
\boldsymbol{\hat{\delta}}_{n}-\boldsymbol{\delta}_{0}
\end{pmatrix}
\right\| \leq 
\left\|\sqrt{n}\,\mathbb{J}_{n}(\boldsymbol{\beta}_{0},\boldsymbol{\delta}_{0})+o_{P}(1)\right\| = O_P(1),
\end{equation}
where the last line follows from condition (3) (see Equation~\eqref{eq:A.34}) of the theorem's statement. Thus, we have
\begin{equation} \label{Eq:A.43}
    \left\|
\sqrt{n}
\begin{pmatrix}
\boldsymbol{\hat{\beta}}_{n}-\boldsymbol{\beta}_{0}\\
\boldsymbol{\hat{\delta}}_{n}-\boldsymbol{\delta}_{0}
\end{pmatrix}
\right\|
=O_{P}(1),
\end{equation}
which implies \eqref{eq:A.42} from \eqref{eq:A.39}. Hence, by \eqref{eq:A.34} along with Slutsky’s theorem in \eqref{eq:A.42}, we have
\begin{equation}
\sqrt{n}
    \begin{pmatrix}
\boldsymbol{\hat{\beta}}_{n}-\boldsymbol{\beta}_{0}\\
\boldsymbol{\hat{\delta}}_{n}-\boldsymbol{\delta}_{0}
\end{pmatrix} \xrightarrow{D} \mathbb{N}_{2p + 3}(\boldsymbol{0}, \boldsymbol{\Sigma}^{-1} \mathbb{D} (\boldsymbol{\Sigma}^{-1})^\top).
\end{equation}
This completes the proof of the theorem.
\end{proof}

\subsection{\textbf{Proof of Theorem~\ref{th3}}}\label{app:B.3}
\begin{proof}[\textbf{Proof}]
\noindent\textbf{Consistency of $\hat{\boldsymbol{\Sigma}}:$}
We establish consistency of $\hat{\boldsymbol{\Sigma}}$ (see \eqref{Eq:3.15}) by showing consistency of its blocks $\hat{\boldsymbol{\Sigma}}_{2,\boldsymbol{\beta}}$, $\hat{\boldsymbol{\Sigma}}_{2,\boldsymbol{\delta}}$, and $\hat{\boldsymbol{\Sigma}}_{1,\boldsymbol{\delta}}$. We begin by establishing that $\hat{\boldsymbol{\Sigma}}_{2,\boldsymbol{\beta}}$ is a consistent estimator of $\boldsymbol{\Sigma}_{2,\boldsymbol{\beta}_{0}}$. Let $\hat{\boldsymbol{\Sigma}}_{2,\boldsymbol{\beta}}^{(r,s)}$ and $\boldsymbol{\Sigma}_{2,\boldsymbol{\beta}_{0}}^{(r,s)}$ denote the $(r,s)$th element of $\hat{\boldsymbol{\Sigma}}_{2,\boldsymbol{\beta}}$ and $\boldsymbol{\Sigma}_{2,\boldsymbol{\beta}_{0}}$, respectively. We proceed to show that $\hat{\boldsymbol{\Sigma}}_{2,\boldsymbol{\beta}}^{(r,s)}$ is a consistent estimator of $\boldsymbol{\Sigma}_{2,\boldsymbol{\beta}_{0}}^{(r,s)}$.  Consider the following:
\begin{equation} \label{Eq:A.45}
    \begin{split}
\left|\hat{\boldsymbol{\Sigma}}_{2,\boldsymbol{\beta}}^{(r,s)} - \boldsymbol{\Sigma}_{2,\boldsymbol{\beta}_{0}}^{(r,s)}\right| &= \left|n^{-1} \sum_{i = 1}^n \mathds{1} \left(\boldsymbol{\hat{x}_i}^\top\boldsymbol{\hat{\beta}}_n > c_{\rho}\right) \hat{C}_{\psi} \hat{x}_{i, r} \hat{x}_{i, s} - \mathbb{E}\left[ \mathds{1} \left( \boldsymbol{X}^\top \boldsymbol{\beta}_0 > c_{\rho}\right) C_{\psi} X_rX_s\right]\right| \\ 
    & \leq |S_{1, n}| + |S_{2, n}|,\end{split}
\end{equation}
where \begin{equation} \label{Eq:A.46}
    S_{1, n} := n^{-1} \sum_{i = 1}^n \hat{C}_{\psi} \left\{ \mathds{1} \left(\boldsymbol{\hat{x}_i}^\top\boldsymbol{\hat{\beta}}_n > c_{\rho}  \right)  \hat{x}_{i, r} \hat{x}_{i, s} - \mathds{1} \left(\boldsymbol{x_i}^\top\boldsymbol{\hat{\beta}}_n > c_{\rho} \right)  x_{i, r} x_{i, s} \right\},
\end{equation}
and \begin{equation} \label{Eq:A.47}
    S_{2, n} := n^{-1} \sum_{i = 1}^n \mathds{1} \left(\boldsymbol{x_i}^\top\boldsymbol{\hat{\beta}}_n > c_{\rho} \right) \hat{C}_{\psi}  x_{i, r} x_{i, s} - \mathbb{E}\left[ \mathds{1} \left( \boldsymbol{X}^\top \boldsymbol{\beta}_0 > c_{\rho}\right) C_{\psi} X_rX_s\right].
\end{equation}
Here, $\hat{x}_{i, r}$ (resp. $\hat{x}_{i, s}$) denotes the $r$th (resp. $s$th) element of the vector $\boldsymbol{\hat{x}}_i$. Similarly, $x_{i, r}$ (resp. $x_{i, s}$) denotes the $r$th (resp. $s$th) element of the vector $\boldsymbol{x}_i$. Furthermore, $X_r$ (resp. $X_s$) denotes the $r$th (resp. $s$th) element of the random vector $\boldsymbol{X}$. Now, we proceed to show that $|S_{1, n}| \xrightarrow{P} 0$ and $|S_{2, n}| \xrightarrow{P} 0$ as $n \to \infty$.

 Let $D_{i, n} :=\mathds{1}(\hat{\boldsymbol{x}}_{i}^{\top}\boldsymbol{\hat{\beta}}_{n}> c_{\rho}) - \mathds{1}(\boldsymbol{x}_{i}^{\top}\boldsymbol{\hat{\beta}}_{n}>c_{\rho}), $ 
$C_{i, n} :=\mathds{1}(\hat{\boldsymbol{x}}_{i}^{\top}\boldsymbol{\hat{\beta}}_{n}> c_{\rho}) - \mathds{1}(\boldsymbol{x}_{i}^{\top}\boldsymbol{\beta}_{0}>c_{\rho}) $, and $D^x_{i, n}:= \hat{x}_{i, r} \hat{x}_{i,s} - x_{i,r} x_{i, s} $. Further note that, $|D_{i, n}| \leq \mathds{1} (A_{i,n})$ and $|C_{i, n}| \leq \mathds{1} (A_{i,n})$, where 
\begin{equation} \label{Eq:A.48}
    A_{i,n}:= \left\{\mathds{1}\Big(|\boldsymbol{x}_i^{\top}\boldsymbol{\beta}_{0}-c_{\rho}|\leq \mathcal{K}_{1}\,\|\hat{\boldsymbol{\delta}}_{n}-\boldsymbol{\delta}_{0}\|\|\hat{\boldsymbol{\beta}}_{n}-\boldsymbol{\beta}_{0}\|+\mathcal{K}_{1}\,\|\hat{\boldsymbol{\delta}}_{n}-\boldsymbol{\delta}_{0}\| C_{0}+\mathcal{K}_{0}\,\|\hat{\boldsymbol{\beta}}_{n}-\boldsymbol{\beta}_{0}\|\Big)\right\}.
\end{equation} 
In addition, $|D^x_{i,n}| \leq g_{n}(\boldsymbol{x}_i, \boldsymbol{z}_i) $, where $g_{n}(\boldsymbol{x}_i, \boldsymbol{z}_i):= 2\mathcal{K}_0\mathcal{K}_1
\|\boldsymbol{\hat{\delta}}_{n}-\boldsymbol{\delta}_{0}\|
+\mathcal{K}_1^2\|\boldsymbol{\hat{\delta}}_{n}-\boldsymbol{\delta}_{0}\|^{2}$. Here, $C_0:= \sup_{\beta \in \mathcal{B}} \|\boldsymbol{\beta\|}$, and the constants $\mathcal{K}_0$ and $\mathcal{K}_1$ are defined in Assumption~\ref{a:3.5}. We refer to Results~\ref{Re:1}-\ref{Re:3} for the derivation of the aforementioned bounds.

Next, addition and subtraction of $\mathds{1}(\boldsymbol{\hat{x}}_{i}^{\top}\boldsymbol{\hat{\beta}}_{n}>c_{\rho}) x_{i, r} x_{i,s} \hat{C}_{\psi}$ and $ \mathds{1}(\boldsymbol{x}_{i}^{\top}\boldsymbol{\beta}_{0}>c_{\rho}) D^x_{i, n} \hat{C}_{\psi}$ in \eqref{Eq:A.46} yields: 
\begin{align*}
  |S_{1, n}| &= \left|n^{-1} \sum_{i = 1}^n \hat{C}_{\psi} \left[ D_{i, n}  x_{i, r} x_{i, s} + C_{i, n} D^x_{i, n}+\mathds{1} \left(\boldsymbol{x_i}^\top\boldsymbol{\beta}_0 > c_{\rho} \right) D^x_{i, n}\right]\right|\\
  & \leq n^{-1} \sum_{i = 1}^n |\hat{C}_{\psi}| \left[ |D_{i, n}| |x_{i, r} x_{i, s}| + |C_{i, n}| |D^x_{i, n}| + | D^x_{i, n}| \right] \\
  & \leq n^{-1} | \hat{C}_{\psi}| \sum_{i = 1}^n \left[\mathds{1}(A_{i,n}) \left\{\|\boldsymbol{x}_{i}\|^2 + g_n(\boldsymbol{x}_i, \boldsymbol{z}_i)\right\} + g_n(\boldsymbol{x}_i, \boldsymbol{z}_i) \right]\\
  & \leq | \hat{C}_{\psi}| \left\{ n^{-1} \mathcal{K}_0^2 \sum_{i = 1}^n \mathds{1}(A_{i,n}) + 2 g_n(\boldsymbol{x}_i, \boldsymbol{z}_i)  \right\} \tag{E.4} \label{E.4} \\
  & \xrightarrow{P} 0, \text{ as } n \to \infty,  
\end{align*} 
provided $\hat{C}_{\psi} \xrightarrow{P} C_{\psi}$ as $n \to \infty$. This follows from the following two facts. First, $g_n(\boldsymbol{x}_i, \boldsymbol{z}_i) = o_{P}(1)$ as $n \to \infty$ by consistency of $\boldsymbol{\hat{\delta}}_n$ for $\boldsymbol{\delta}_0$ (see Lemma~\ref{l:1}). Second, in the following, we show that $n^{-1}\sum_{i = 1}^n \mathds{1}(A_{n,i}) = o_{P}(1)$. To this end, define the events $E_{1, n}:= \{\|\boldsymbol{\hat{\beta}}_{n}-\boldsymbol{\beta}_{0}\| > \ell_1\}$ and $E_{2, n}:= \{\|\boldsymbol{\hat{\delta}}_{n}-\boldsymbol{\delta}_{0}\|> \ell_2\}$ for arbitrary small positive numbers $\ell_1$ and $\ell_2$, and for any $\zeta_{0}>0$ consider the following: 
\begin{equation}\label{Eq:A.49}
\begin{split}
  &\mathbb{P} \left[n^{-1}\sum_{i=1}^{n}
\mathds{1}(A_{i,n})>\zeta_{0}\right]\\
&\leq \mathbb{P} \left[\left\{n^{-1}\sum_{i=1}^{n}
\mathds{1}\left(
|c_{\rho}-\boldsymbol{x}_{i}^{\top}\boldsymbol{\beta}_{0}|
\leq (\mathcal{K}_{0} \ell_1
+ \mathcal{K}_{1}\ell_1\ell_2 + \mathcal{K}_{1}\ell_2 C_{0})
\right)\right\}>\zeta_{0}\right]+\mathbb{P}\left[E_{1,n}\right]
+\mathbb{P}\left[E_{2,n}\right]\\
&\leq\zeta_{0}^{-1} \mathbb{E}\left[\mathds{1}\left(
|c_{\rho}-\boldsymbol{x}_{i}^{\top}\boldsymbol{\beta}_{0}|
\leq (\mathcal{K}_{0} \ell_1
+ \mathcal{K}_{1}\ell_1\ell_2 + \mathcal{K}_{1}\ell_2 C_{0})
\right)\right]+\mathbb{P}\left[E_{1,n}\right]+\mathbb{P}\left[E_{2,n}\right] \\
&\leq\zeta_{0}^{-1}(k_{1}\ell) + o(1) = o(1) \text{ as } \ell \to 0,
\end{split}
 \end{equation}
where the last line follows from consistency of $\boldsymbol{\hat{\beta}}_n$ and $\boldsymbol{\hat{\delta}}_n$, and applying Markov’s inequality along with Assumption~\ref{a:3.6} for $\ell = \min(\ell_1, \ell_2)$ and $\ell_0 = \max(\ell_1, \ell_2)$. This establishes that $|S_{1,n}| = o_P(1)$. 

Now, we proceed to show that $|S_{2,n}| = o_P(1)$. We add and subtract $n^{-1} \sum\limits_{i=1}^{n}\mathds{1}(\boldsymbol{x}_{i}^\top \boldsymbol{\hat{\beta}}_{n}>c_{\rho}) C_{\psi} x_{i,r}x_{i,s}$ and $\mathbb{E}\left[
\mathds{1}(\boldsymbol{X}^\top \boldsymbol{\hat{\beta}}_{n}>c_{\rho}) C_{\psi} X_{r}X_{s}
\right]$ to $S_{2,n}$ (see \eqref{Eq:A.47}), and apply the triangle inequality to obtain the following:
\begin{equation} \label{Eq:A.50}
\begin{split}
    |S_{2,n}| &\leq \left| \hat{C}_{\psi} - C_{\psi} \right| n^{-1} \sum_{i = 1}^n \mathds{1}(\boldsymbol{x}_{i}^\top \boldsymbol{\hat{\beta}}_{n}>c_{\rho})  \left|x_{i,r}x_{i,s}\right| + \left|C_{\psi} \right|\mathbb{E}\left[ \left|X_{r}X_{s}\right|
\left|\mathds{1}(\boldsymbol{X}^\top \boldsymbol{\hat{\beta}}_{n}>c_{\rho}) - \mathds{1}(\boldsymbol{X}^\top\boldsymbol{\beta}_{0}>c_{\rho})\right|
\right] \\
    &+ \left|n^{-1} \sum_{i = 1}^n\mathds{1}(\boldsymbol{x}_{i}^\top \boldsymbol{\hat{\beta}}_{n}>c_{\rho}) C_{\psi} x_{i,r}x_{i,s} -  \mathbb{E}\left[
\mathds{1}(\boldsymbol{X}^\top \boldsymbol{\hat{\beta}}_{n}>c_{\rho}) C_{\psi} X_{r}X_{s}\right]
 \right| \\
&= T_{1,n} + T_{2,n} + T_{3,n} \text{ (say) }.
\end{split}
\end{equation}
Now, we show that $T_{i, n} = o_P(1)$ for $i = 1, 2, 3$. Consider the following:
\begin{align*}
    T_{1,n}&:= \left| \hat{C}_{\psi} - C_{\psi} \right| n^{-1} \sum_{i = 1}^n \mathds{1}(\boldsymbol{x}_{i}^\top \boldsymbol{\hat{\beta}}_{n}>c_{\rho})  \left|x_{i,r}x_{i,s}\right| \\
        &\leq \left| \hat{C}_{\psi} - C_{\psi} \right| n^{-1} \sum_{i = 1}^n \left\{ \left|\mathds{1}(\boldsymbol{x}_{i}^\top \boldsymbol{\hat{\beta}}_{n}>c_{\rho}) - \mathds{1}(\boldsymbol{x}_{i}^\top \boldsymbol{\beta}_{0}>c_{\rho}) \right| +   \left|\mathds{1}(\boldsymbol{x}_{i}^\top \boldsymbol{\beta}_{0}>c_{\rho})\right| \right\} \left\|\boldsymbol{x}_i\right\|^2 \\
        & \leq \left| \hat{C}_{\psi} - C_{\psi} \right| \mathcal{K}_0^2 \left[n^{-1} \sum_{i = 1}^n \mathds{1} \left(\left|\boldsymbol{x}_{i}^\top \boldsymbol{\beta}_{0}>c_{\rho}\right| \leq \left\|\boldsymbol{x}_i\right\| \|\boldsymbol{\hat{\beta}}_n - \boldsymbol{\beta}_0\| \right) + 1\right] \tag{E.50} \label{E.50} \\
        & \xrightarrow{P} 0, \text{ as } n \to \infty, 
\end{align*}
where the last line follows from the consistency of $\hat{C}_{\psi}$, $\boldsymbol{\hat{\beta}}_n$, along with an application (similar to (\eqref{Eq:A.49})) of Markov’s inequality using Assumption~\ref{a:3.6}. Further, \eqref{E.50} follows from the consistency of $\hat{C}_{\psi}$. 

Next, consider $T_{2, n}$, we have:
\begin{align*}
   T_{2, n} &:=  \left|C_{\psi} \right|\mathbb{E}\left[ \left|X_{r}X_{s}\right|
\left|\mathds{1}(\boldsymbol{X}^\top \boldsymbol{\hat{\beta}}_{n}>c_{\rho}) - \mathds{1}(\boldsymbol{X}^\top\boldsymbol{\beta}_{0}>c_{\rho}\right|\right] \\
& \leq \phi_{2}\mathcal{K}_{0}^{2}\mathbb{E}\left[ 
 \mathds{1}\left(\left|c_{\rho}-\boldsymbol{X}^\top\boldsymbol{\beta}_{0}\right|\leq \|\boldsymbol{X}\|\|\boldsymbol{\hat{\beta}}_{n}-\boldsymbol{\beta}_{0}\|\right)\right] \\
 & \leq \phi_{2}\mathcal{K}_{0}\mathbb{E}\left[ 
 \mathds{1}\left(\left|c_{\rho}-\boldsymbol{X}^\top\boldsymbol{\beta}_{0}\right|\leq \mathcal{K}_{0}\ell\right)\right]+\mathbb{P}\left[\|\boldsymbol{\hat{\beta}}_n - \boldsymbol{\beta}_0\| > \ell\right] \\
& \leq \phi_{1}^{2}\mathcal{K}_{0}^{2}k_{1}\ell+o(1), \text{ as } n \to \infty, \tag{By Assumption~\ref{a:3.6}} \\
& \xrightarrow{} 0 \text{ as } \ell \to 0,
\end{align*}
where $\phi_2$ is defined in Assumption~\ref{a:3.8}.

Lastly, we show that $T_{3,n} = o_P(1)$ as follows. Consider the following:
\begin{align*}
   T_{3,n}&:= \left| n^{-1} \sum_{i = 1}^n\mathds{1}(\boldsymbol{x}_{i}^\top \boldsymbol{\hat{\beta}}_{n}>c_{\rho}) C_{\psi} x_{i,r}x_{i,s} -  \mathbb{E}\left[
\mathds{1}(\boldsymbol{X}^\top \boldsymbol{\hat{\beta}}_{n}>c_{\rho}) C_{\psi} X_{r}X_{s}\right]
 \right| \\
 &\leq \phi_2 \mathcal{K}_0^2 \left| n^{-1} \sum_{i = 1}^n \left\{\mathds{1}\left(|\boldsymbol{x}_{i}^\top \boldsymbol{\beta}_{0} - c_{\rho}| \leq \mathcal{K}_0 \|\boldsymbol{\hat{\beta}}_n - \boldsymbol{\beta}_0\| \right) + \mathds{1} \left( \boldsymbol{x}_i^\top \boldsymbol{\beta}_0 > c_{\rho}\right)  \right\} -  \mathbb{E}\left[
\mathds{1}(\boldsymbol{X}^\top \boldsymbol{\hat{\beta}}_{n}>c_{\rho})\right] 
 \right| \\
 & \leq \phi_2 \mathcal{K}_0^2 \left| o_P(1) + n^{-1} \sum_{i = 1}^n \mathds{1} \left( \boldsymbol{x}_i^\top \boldsymbol{\beta}_0 > c_{\rho}\right) +  \mathbb{E} \mathds{1}\left(|\boldsymbol{X}^\top \boldsymbol{\beta}_{0} - c_{\rho}| \leq \mathcal{K}_0 \|\boldsymbol{\hat{\beta}}_n - \boldsymbol{\beta}_0\| \right) - \mathbb{E}[
\mathds{1}(\boldsymbol{X}^\top \boldsymbol{\beta}_{0}>c_{\rho})]  
 \right| \\
 & \leq \phi_2 \mathcal{K}_0^2 \left[ o_P(1) + \mathcal{K}_0^2 k_1 \ell + \left|n^{-1} \sum_{i = 1}^n \mathds{1} \left( \boldsymbol{x}_i^\top \boldsymbol{\beta}_0 > c_{\rho}\right) - \mathbb{E}\left[
\mathds{1}(\boldsymbol{X}^\top \boldsymbol{\beta}_{0}>c_{\rho})\right]\right|\right] \tag{E.51} \label{E.51} \\
&\leq \phi_2 \mathcal{K}_0^4 k_1 \ell + o_P(1) , \text{ as } n \to \infty \\
& \xrightarrow{} 0, \text{ as } \ell \to 0, 
 \end{align*}
since $\ell > 0$ is an arbitrarily small number and the WLLN applies to the last term of \eqref{E.51}. Note that the first term of the second line of the above expression is $o_P(1)$ by Assumption~\ref{a:3.6} using the same procedure as in \eqref{E.50}. This establishes that $|S_{2, n}| \xrightarrow{P} 0$.

Thus, by \eqref{Eq:A.45}, we have $\hat{\boldsymbol{\Sigma}}_{2,\boldsymbol{\beta}}^{(r,s)} \xrightarrow{P} \boldsymbol{\Sigma}_{2,\boldsymbol{\beta}_{0}}^{(r,s)}$ as $n \to \infty$. Since $r$ and $s$ are arbitrary row and column indices, respectively, of $\hat{\boldsymbol{\Sigma}}_{2,\boldsymbol{\beta}}$. This implies that $\hat{\boldsymbol{\Sigma}}_{2,\boldsymbol{\beta}} \xrightarrow{P} \boldsymbol{\Sigma}_{2,\boldsymbol{\beta}_{0}}$ as $n \to \infty$. Similarly, we can show that the other blocks of $\boldsymbol{\hat{\Sigma}}$, that is, $\hat{\boldsymbol{\Sigma}}_{2,\boldsymbol{\delta}}$ and $\hat{\boldsymbol{\Sigma}}_{1,\boldsymbol{\delta}}$ are also consistent. We omit the proofs of consistency of $\hat{\boldsymbol{\Sigma}}_{2,\boldsymbol{\delta}}$ and $\hat{\boldsymbol{\Sigma}}_{1,\boldsymbol{\delta}}$ as the proof for $\hat{\boldsymbol{\Sigma}}_{2,\boldsymbol{\delta}}$ follows along the same line with minor notational changes, and the proof for $\hat{\boldsymbol{\Sigma}}_{1,\boldsymbol{\delta}}$ is trivial. Hence, we conclude that $\boldsymbol{\hat{\Sigma}}$ is a consistent estimator of $\boldsymbol{\Sigma}$.  

\noindent\textbf{Consistency of $\hat{\mathbb{D}}:$}
We establish the consistency of $\hat{\mathbb{D}}$ (see \eqref{Eq:3.15}) by showing the consistency of its blocks
$\hat{\mathbb{D}}_{1}$ and $\hat{\mathbb{D}}_{2}$. We begin with $\hat{\mathbb{D}}_{1}$, which we
show is a consistent estimator of $\mathbb{D}_{1}$. To this end, let
$\hat{\mathbb{D}}^{(r,s)}_{1}$ denote the $(r,s)$-th element of $\hat{\mathbb{D}}_{1}$, and let
$\mathbb{D}^{(r,s)}_{1}$ denote the corresponding element of $\mathbb{D}_{1}$. It suffices to show
that $\hat{\mathbb{D}}^{(r,s)}_{1}$ is a consistent estimator of $\mathbb{D}^{(r,s)}_{1}$ for
an arbitrary element $(r,s)$. Throughout, $z_{i,r}$ and $z_{i,s}$ denote the $r$-th and $s$-th elements of the
vector $\boldsymbol{z}_{i}$, and $Z_{r}$ and $Z_{s}$ denote the $r$-th and $s$-th elements of the
random vector $\boldsymbol{Z}$. Adding and subtracting the term $n^{-1}\sum_{i=1}^{n}\vartheta_{i}^{2}z_{i,r}z_{i,s}$ in $\hat{\mathbb{D}}^{(r,s)}_{1}-\mathbb{D}^{(r,s)}_{1}$ and then applying the triangle inequality, we obtain
\begin{equation}\label{Eq:A.51}
\begin{split}
   \left|\hat{\mathbb{D}}^{(r,s)}_{1}-\mathbb{D}^{(r,s)}_{1}\right| &\leq\left|\frac{1}{n}\sum_{i=1}^{n}e_{i}^{2}z_{i,r}z_{i,s}-\frac{1}{n}\sum_{i=1}^{n}\vartheta_{i}^{2}z_{i,r}z_{i,s}\right|+\left|\frac{1}{n}\sum_{i=1}^{n}\vartheta_{i}^{2}z_{i,r}z_{i,s}-\mathbb{E}\left[\vartheta^{2}Z_{r}Z_{s}\right]\right|\\
   &\leq |\mathcal{U}_{1,n}|+|\mathcal{U}_{2,n}|
   \end{split}
\end{equation}
 We now show that $|\mathcal{U}_{1,n}|=o_{P}(1)$ and $|\mathcal{U}_{2,n}|=o_{P}(1)$. We begin with the following.
\begin{equation}\label{Eq:A.52}
\begin{split}
   \left|\mathcal{U}_{1,n}\right|&\leq\frac{1}{n}\sum_{i=1}^{n}|e_{i}^{2}-\vartheta_{i}^{2}||z_{i,r}z_{i,s}|\\
   &\leq 2\;\|\hat{\boldsymbol{\delta}}_{n}-\boldsymbol{\delta}_{0}\|\;\frac{1}{n}\sum_{i=1}^{n}|\vartheta_{i}|\|\boldsymbol{z}_{i}\|^{3}+\|\hat{\boldsymbol{\delta}}_{n}-\boldsymbol{\delta}_{0}\|^{2}\;\frac{1}{n}\sum_{i=1}^{n}\|\boldsymbol{z}_{i}\|^{4},\\
   &\leq 2\;\mathcal{K}_{1}^{3}\|\hat{\boldsymbol{\delta}}_{n}-\boldsymbol{\delta}_{0}\|\;\frac{1}{n}\sum_{i=1}^{n}|\vartheta_{i}|+\|\hat{\boldsymbol{\delta}}_{n}-\boldsymbol{\delta}_{0}\|^{2}\mathcal{K}_{1}^{4},
   \end{split}
\end{equation}
where we use the fact that $e_{i}^{2}-\vartheta_{i}^{2}=-2\vartheta_{i}\boldsymbol{z}_{i}^\top (\hat{\boldsymbol{\delta}}_{n}-\boldsymbol{\delta}_{0})+\big(\boldsymbol{z}_{i}^\top (\hat{\boldsymbol{\delta}}_{n}-\boldsymbol{\delta}_{0})\big)^{2}.$ By the consistency of $\hat{\boldsymbol{\delta}}_{n},$ we have $\|\hat{\boldsymbol{\delta}}_{n}-\boldsymbol{\delta}_{0}\|\xrightarrow{P} 0.$ Under Assumption~\ref{a:3.5} and the moment condition in equation~\ref{Eq:2.3}, the sample averages 
$\frac{1}{n}\sum_{i=1}^{n}|\vartheta_{i}|$
converge in probability to 0 by the weak law of large numbers. Hence, the term defined in Equation~\ref{Eq:A.52} on the right-hand side converges to zero in probability, and therefore $\left|\mathcal{U}_{1,n}\right|=o_{P}(1).$

 Next, consider the term $\mathcal{U}_{2,n}$ in Equation~\ref{Eq:A.51}. Under Assumption~\ref{a:3.5} and the moment condition in Equation~\ref{Eq:2.3}, the weak law of large numbers gives
$$\frac{1}{n}\sum_{i=1}^{n}\vartheta_{i}^{2}z_{i,r}z_{i,s}\xrightarrow{P}\mathbb{E}\left[\vartheta^{2}Z_{r}Z_{s}\right],$$
and therefore $|\mathcal{U}_{2,n}|=o_{P}(1)$.

Thus, by \eqref{Eq:A.51}, we have $\hat{\mathbb{D}}^{(r,s)}_{1} \xrightarrow{P} \mathbb{D}^{(r,s)}_{1}$ as $n \to \infty$. Since $r$ and $s$ are arbitrary row and column indices, respectively, of $\hat{\mathbb{D}}^{(r,s)}_{1}$. This implies that $\hat{\mathbb{D}}^{(r,s)}_{2} \xrightarrow{P} \mathbb{D}^{(r,s)}_{2}$ as $n \to \infty$. Hence, we conclude that $\hat{\mathbb{D}}_{1}$ is a consistent estimator of  $\mathbb{D}_{1}$.
 
 Similarly, for $\hat{\mathbb{D}}_{2}$, let $\hat{\mathbb{D}}^{(r,s)}_{2}$ and $\mathbb{D}^{(r,s)}_{2}$ denote the $(r,s)$-th elements of $\hat{\mathbb{D}}_{2}$ and $\mathbb{D}_{2}$, respectively. It suffices to show that $\hat{\mathbb{D}}^{(r,s)}_{2}\xrightarrow{P}\mathbb{D}^{(r,s)}_{2}$ for every $(r,s)$. To this end, we consider the difference $\hat{\mathbb{D}}^{(r,s)}_{2}-\mathbb{D}^{(r,s)}_{2}$ and add and subtract the term
$\frac{1}{n}\sum_{i=1}^{n}\mathds{1}(\boldsymbol{x}_{i}^\top\hat{\boldsymbol{\beta}}_{n}>c_{\rho})\,\psi^{2}(\hat{\varepsilon})\,x_{i,r}x_{i,s}.$
Applying the triangle inequality, we obtain
\begin{equation}\label{Eq:A.53}
\begin{split}
\left|\hat{\mathbb{D}}^{(r,s)}_{2}-\mathbb{D}^{(r,s)}_{2}\right|&=\left|\frac{1}{n}\sum\limits_{i=1}^{n}\mathds{1}(\hat{\boldsymbol{x}}_{i}^\top \boldsymbol{\hat{\beta}}_{n}>c_{\rho}) \psi^{2}(\hat{\varepsilon}) \hat{x}_{i,r}\hat{x}_{i,s}-\mathbb{E}\left[\mathds{1}(\boldsymbol{X}^\top\boldsymbol{\beta}_{0}>c_{\rho})\psi^{2}(\varepsilon)X_{r}X_{s}\right]\right|\\
&\leq |\mathcal{S}_{1,n}|+|\mathcal{S}_{2,n}|,
\end{split}
\end{equation}
where
\begin{equation}\label{Eq:A.54}
   \mathcal{S}_{1,n}= \frac{1}{n}\sum\limits_{i=1}^{n}\mathds{1}(\hat{\boldsymbol{x}}_{i}^\top \boldsymbol{\hat{\beta}}_{n}>c_{\rho}) \psi^{2}(\hat{\varepsilon}) \hat{x}_{i,r}\hat{x}_{i,s}-\frac{1}{n}\sum\limits_{i=1}^{n}\mathds{1}(\boldsymbol{x}_{i}^\top \boldsymbol{\hat{\beta}}_{n}>c_{\rho}) \psi^{2}(\hat{\varepsilon})x_{i,r}x_{i,s},
\end{equation}
and 
\begin{equation}\label{Eq:A.55}
    \mathcal{S}_{2,n}=\frac{1}{n}\sum\limits_{i=1}^{n}\mathds{1}(\boldsymbol{x}_{i}^\top \boldsymbol{\hat{\beta}}_{n}>c_{\rho}) \psi^{2}(\hat{\varepsilon})x_{i,r}x_{i,s}-\mathbb{E}\left[\mathds{1}(\boldsymbol{X}^\top\boldsymbol{\beta}_{0}>c_{\rho})\psi^{2}(\varepsilon)X_{r}X_{s}\right].
\end{equation}
We now show that $|\mathcal{S}_{1,n}| \xrightarrow{P} 0$ and $|\mathcal{S}_{2,n}| \xrightarrow{P} 0$
as $n \to \infty$. The argument relies on the same quantities and bounds
established earlier. Recall that $D_{i,n}$, $C_{i,n}$ and $D^{x}_{i,n}$ are
defined in Equations~\eqref{R:1}, \eqref{R:2} and \eqref{R:3}, respectively.
As before, $|D_{i,n}| \le \mathds{1}(A_{i,n})$ and
$|C_{i,n}| \le \mathds{1}(A_{i,n})$, where $A_{i,n}$ is the event defined in
Equation~\eqref{Eq:A.48}. Likewise,
$|D^{x}_{i,n}| \le g_{n}(\boldsymbol{x}_i, \boldsymbol{z}_i)$, with
$g_{n}(\boldsymbol{x}_i, \boldsymbol{z}_i)$ and the constants $C_0$,
$\mathcal{K}_0$ and $\mathcal{K}_1$ exactly as defined previously
(see Assumption~\ref{a:3.5}). The derivations of these bounds are given in Results~\ref{Re:1}--\ref{Re:3}.
     
Next, addition and subtraction of $\mathds{1}(\boldsymbol{\hat{x}}_{i}^{\top}\boldsymbol{\hat{\beta}}_{n}>c_{\rho}) x_{i, r} x_{i,s} \psi^{2}(\hat{\varepsilon}_{i})$ and $ \mathds{1}(\boldsymbol{x}_{i}^{\top}\boldsymbol{\beta}_{0}>c_{\rho}) D^x_{i, n} \psi^{2}(\hat{\varepsilon})$ in \eqref{Eq:A.54} yields: 
\begin{align*}
  |\mathcal{S}_{1,n}| &= \left|n^{-1} \sum_{i = 1}^n \psi^{2}(\hat{\varepsilon}_{i}) \left[ D_{i, n}  x_{i, r} x_{i, s} + C_{i, n} D^x_{i, n}+\mathds{1} \left(\boldsymbol{x_i}^\top\boldsymbol{\beta}_0 > c_{\rho} \right) D^x_{i, n}\right]\right|\\
  & \leq n^{-1} \sum_{i = 1}^n |\psi^{2}(\hat{\varepsilon}_{i})| \left[ |D_{i, n}| |x_{i, r} x_{i, s}| + |C_{i, n}| |D^x_{i, n}| + | D^x_{i, n}| \right] \\
  & \leq n^{-1} | \psi^{2}(\hat{\varepsilon})| \sum_{i = 1}^n \left[\mathds{1}(A_{i,n}) \left\{\|\boldsymbol{x}_{i}\|^2 + g_n(\boldsymbol{x}_i, \boldsymbol{z}_i)\right\} + g_n(\boldsymbol{x}_i, \boldsymbol{z}_i) \right]\\
  & \leq | \psi^{2}(\hat{\varepsilon})| \left\{ n^{-1} \mathcal{K}_0^2 \sum_{i = 1}^n \mathds{1}(A_{i,n}) + 2 g_n(\boldsymbol{x}_i, \boldsymbol{z}_i)  \right\} \tag{E.8} \label{E.8} \\
  & \xrightarrow{P} 0, \text{ as } n \to \infty, 
\end{align*} 
 where the last line follows from $\psi^{2}(\hat{\varepsilon})\xrightarrow{P}\psi^{2}(\varepsilon)$ as $n\to\infty$ and using the following two facts. First, $g_{n}(\boldsymbol{x}_i,\boldsymbol{z}_i)=o_{P}(1)$, because $\hat{\boldsymbol{\delta}}_{n}$ is consistent for $\boldsymbol{\delta}_{0}$ (see Lemma~\ref{l:1}). Second, $n^{-1}\sum_{i=1}^{n}\mathds{1}(A_{i,n})=o_{P}(1)$ (see \eqref{Eq:A.49}), which follows from the consistency of $\hat{\boldsymbol{\beta}}_n$ and $\hat{\boldsymbol{\delta}}_n$ together with Markov’s inequality and Assumption~\ref{a:3.6}, applied with $\ell=\min(\ell_1,\ell_2)$ and $\ell_0=\max(\ell_1,\ell_2)$. Thus, $|\mathcal{S}_{1,n}|=o_{P}(1)$. 

Now, we proceed to show that $|\mathcal{S}_{2,n}| = o_P(1)$. We add and subtract $n^{-1} \sum\limits_{i=1}^{n}\mathds{1}(\boldsymbol{x}_{i}^\top \boldsymbol{\hat{\beta}}_{n}>c_{\rho}) \psi^{2}(\varepsilon_{i}) x_{i,r}x_{i,s}$ and $\mathbb{E}\left[
\mathds{1}(\boldsymbol{X}^\top \boldsymbol{\hat{\beta}}_{n}>c_{\rho}) \psi^{2}(\varepsilon) X_{r}X_{s}
\right]$ to $\mathcal{S}_{2,n}$ (see \eqref{Eq:A.55}), and apply the triangle inequality to obtain the following:
\begin{equation} \label{Eq:A.56}
\begin{split}
    |\mathcal{S}_{2,n}| &\leq \left| \psi^{2}(\hat{\varepsilon}) - \psi^{2}(\varepsilon) \right| n^{-1} \sum_{i = 1}^n \mathds{1}(\boldsymbol{x}_{i}^\top \boldsymbol{\hat{\beta}}_{n}>c_{\rho})  \left|x_{i,r}x_{i,s}\right| \\
    &+ \left|\psi^{2}(\varepsilon) \right|\mathbb{E}\left[ \left|X_{r}X_{s}\right|
\left|\mathds{1}(\boldsymbol{X}^\top \boldsymbol{\hat{\beta}}_{n}>c_{\rho}) - \mathds{1}(\boldsymbol{X}^\top\boldsymbol{\beta}_{0}>c_{\rho})\right|
\right] \\
    &+ \left|n^{-1} \sum_{i = 1}^n\mathds{1}(\boldsymbol{x}_{i}^\top \boldsymbol{\hat{\beta}}_{n}>c_{\rho}) \psi^{2}(\varepsilon_{i}) x_{i,r}x_{i,s} -  \mathbb{E}\left[
\mathds{1}(\boldsymbol{X}^\top \boldsymbol{\hat{\beta}}_{n}>c_{\rho}) \psi^{2}(\varepsilon) X_{r}X_{s}\right]
 \right| \\
&= \mathcal{T}_{1,n} + \mathcal{T}_{2,n} + \mathcal{T}_{3,n} \text{ (say) }.
\end{split}
\end{equation}
Now, we show that $\mathcal{T}_{i, n} = o_P(1)$ for $i = 1, 2, 3$. Consider the following:
\begin{align*}
    \mathcal{T}_{1,n}&:= \left| \psi^{2}(\hat{\varepsilon}) - \psi^{2}(\varepsilon) \right| n^{-1} \sum_{i = 1}^n \mathds{1}(\boldsymbol{x}_{i}^\top \boldsymbol{\hat{\beta}}_{n}>c_{\rho})  \left|x_{i,r}x_{i,s}\right| \\
        &\leq \left| \psi^{2}(\hat{\varepsilon}) - \psi^{2}(\varepsilon) \right| n^{-1} \sum_{i = 1}^n \left\{ \left|\mathds{1}(\boldsymbol{x}_{i}^\top \boldsymbol{\hat{\beta}}_{n}>c_{\rho}) - \mathds{1}(\boldsymbol{x}_{i}^\top \boldsymbol{\beta}_{0}>c_{\rho}) \right| +   |\mathds{1}(\boldsymbol{x}_{i}^\top \boldsymbol{\beta}_{0}>c_{\rho})| \right\} \left\|\boldsymbol{x}_i\right\|^2 \\
        & \leq \left| \psi^{2}(\hat{\varepsilon}) - \psi^{2}(\varepsilon) \right| \mathcal{K}_0^2\left[ n^{-1} \sum_{i = 1}^n \mathds{1} \left(\left|\boldsymbol{x}_{i}^\top \boldsymbol{\beta}_{0}>c_{\rho}\right| \leq \left\|\boldsymbol{x}_i\right\| \|\boldsymbol{\hat{\beta}}_n - \boldsymbol{\beta}_0\| \right) + 1\right] \tag{E.52} \label{E.52} \\
        & \xrightarrow{P} 0, \text{ as } n \to \infty, 
\end{align*}
where the last line follows from the consistency of $\psi^{2}(\hat{\varepsilon})$, along with an application (similar to (\eqref{Eq:A.49})) of Markov’s inequality using Assumption~\ref{a:3.6}.  

Next, consider $\mathcal{T}_{2, n}$, we have:
\begin{align*}
   \mathcal{T}_{2, n} &:=  \left| \psi^{2}(\varepsilon)\right|\mathbb{E}\left[ \left|X_{r}X_{s}\right|
\left|\mathds{1}(\boldsymbol{X}^\top \boldsymbol{\hat{\beta}}_{n}>c_{\rho}) - \mathds{1}(\boldsymbol{X}^\top\boldsymbol{\beta}_{0}>c_{\rho}\right|\right] \\
& \leq \phi_{1}^{2}\mathcal{K}_{0}^{2}\mathbb{E}\left[ 
 \mathds{1}\left(\left|c_{\rho}-\boldsymbol{X}^\top\boldsymbol{\beta}_{0}\right|\leq \|\boldsymbol{X}\|\|\boldsymbol{\hat{\beta}}_{n}-\boldsymbol{\beta}_{0}\|\right)\right] \\
 & \leq \phi_{1}^{2}\mathcal{K}_{0}^{2}\mathbb{E}\left[ 
 \mathds{1}\left(\left|c_{\rho}-\boldsymbol{X}^\top\boldsymbol{\beta}_{0}\right|\leq \mathcal{K}_{0}\ell\right)\right]+\mathbb{P}\left[\|\boldsymbol{\hat{\beta}}_n - \boldsymbol{\beta}_0\| > \ell\right] \\
& \leq \phi_{1}^{2}\mathcal{K}_{0}^{2}k_{1}\ell+o(1), \text{ as } n \to \infty, \tag{By Assumption~\ref{a:3.6}} \\
& \xrightarrow{} 0 \text{ as } \ell \to 0,
\end{align*}
where $\phi_1$ is defined in Assumption~\ref{a:3.7}.

Lastly, we show that $\mathcal{T}_{3,n} = o_P(1)$ as follows. Consider the following:
\begin{align*}
   \mathcal{T}_{3,n}&:= \left| n^{-1} \sum_{i = 1}^n\mathds{1}(\boldsymbol{x}_{i}^\top \boldsymbol{\hat{\beta}}_{n}>c_{\rho}) \psi^{2}(\varepsilon) x_{i,r}x_{i,s} -  \mathbb{E}\left[
\mathds{1}(\boldsymbol{X}^\top \boldsymbol{\hat{\beta}}_{n}>c_{\rho}) \psi^{2}(\varepsilon) X_{r}X_{s}\right]
 \right| \\
 &\leq \phi_1 \mathcal{K}_0^2 \left| n^{-1} \sum_{i = 1}^n \left\{\mathds{1}\left(|\boldsymbol{x}_{i}^\top \boldsymbol{\beta}_{0} - c_{\rho}| \leq \mathcal{K}_0 \|\boldsymbol{\hat{\beta}}_n - \boldsymbol{\beta}_0\| \right) + \mathds{1} \left( \boldsymbol{x}_i^\top \boldsymbol{\beta}_0 > c_{\rho}\right)  \right\} -  \mathbb{E}\left[
\mathds{1}(\boldsymbol{X}^\top \boldsymbol{\hat{\beta}}_{n}>c_{\rho})\right]  
 \right| \\
 & \leq \phi_1 \mathcal{K}_0^2 \left| o_P(1) + n^{-1} \sum_{i = 1}^n \mathds{1} \left( \boldsymbol{x}_i^\top \boldsymbol{\beta}_0 > c_{\rho}\right) +  \mathbb{E} \mathds{1}\left(|\boldsymbol{X}^\top \boldsymbol{\beta}_{0} - c_{\rho}| \leq \mathcal{K}_0 \|\boldsymbol{\hat{\beta}}_n - \boldsymbol{\beta}_0\| \right) - \mathbb{E}
\mathds{1}(\boldsymbol{X}^\top \boldsymbol{\beta}_{0}>c_{\rho})  
 \right| \\
 & \leq \phi_2 \mathcal{K}_0^2 \left[ o_P(1) + \mathcal{K}_0^2 k_1 \ell + \left|n^{-1} \sum_{i = 1}^n \mathds{1} \left( \boldsymbol{x}_i^\top \boldsymbol{\beta}_0 > c_{\rho}\right) - \mathbb{E}\left[
\mathds{1}(\boldsymbol{X}^\top \boldsymbol{\beta}_{0}>c_{\rho})\right]\right|\right] \tag{E.53} \label{E.53} \\
&\leq \phi_2 \mathcal{K}_0^4 k_1 \ell + o_P(1) , \text{ as } n \to \infty \\
& \xrightarrow{} 0, \text{ as } \ell \to 0,
 \end{align*}
since $\ell > 0$ is an arbitrarily small number and the WLLN applies to the last term of \eqref{E.53}. Note that the first term of the second line of the above expression is $o_P(1)$ by Assumption~\ref{a:3.6} using the same procedure as in \eqref{Eq:A.49}. This establishes that $|\mathcal{S}_{2, n}| \xrightarrow{P} 0$.

Thus, by \eqref{Eq:A.53}, $\hat{\mathbb{D}}^{(r,s)}_{2} \xrightarrow{P} \mathbb{D}^{(r,s)}_{2}$ as $n \to \infty$. Since $r$ and $s$ are arbitrary row and column indices, this convergence holds for every entry of the matrix, and therefore $\hat{\mathbb{D}}_{2} \xrightarrow{P} \mathbb{D}_{2}$ as $n \to \infty$. That is, $\hat{\mathbb{D}}_{2}$ is a consistent estimator of $\mathbb{D}_{2}$. Since we have already shown that $\hat{\mathbb{D}}_{1}$ is a consistent estimator of $\mathbb{D}_{1}$, it follows that $\hat{\mathbb{D}}$ is a consistent estimator of $\mathbb{D}$. This completes the proof of Theorem~\ref{th3}. 
\end{proof}
\subsection{Auxiliary Results}
\begin{R1} \label{Re:1}
Define
\begin{equation}\label{R:1}
D_{i,n}:=\left|\mathds{1}\big(\hat{\boldsymbol{x}}_i^{\top}\hat{\boldsymbol{\beta}}_{n}>c_{\rho}\big)-\mathds{1}\big(\boldsymbol{x}_i^{\top}\hat{\boldsymbol{\beta}}_{n}>c_{\rho}\big)\right|.
\end{equation}
Under Assumption~\ref{a:3.5} and  $C_0:=\sup_{\boldsymbol\beta\in\mathcal B}\|\boldsymbol\beta\|<\infty$, we have for every $i$ and $n$
$$D_{i,n}\;\le\;\mathds{1}\Big(|\boldsymbol{x}_i^{\top}\boldsymbol{\beta}_0-c_\rho|\le \mathcal{K}_1\|\hat{\boldsymbol\delta}_n-\boldsymbol\delta_0\|\,\|\hat{\boldsymbol\beta}_n-\boldsymbol\beta_0\|+\mathcal{K}_1C_0\|\hat{\boldsymbol\delta}_n-\boldsymbol\delta_0\|+\mathcal{K}_0\|\hat{\boldsymbol\beta}_n-\boldsymbol\beta_0\|\Big).$$
\begin{proof}[\bf{Proof}]
Since the two indicators differ only if $c_{\rho}$ lies between $\boldsymbol{x}_i^{\top}\hat{\boldsymbol{\beta}}_{n}$ and $\hat{\boldsymbol{x}}_i^{\top}\hat{\boldsymbol{\beta}}_{n}$, we have
\begin{align*}
|D_{i,n}|
&\leq \mathds{1}\Big(|\boldsymbol{x}_i^{\top}\hat{\boldsymbol{\beta}}_{n}-c_{\rho}|\leq|\hat{\boldsymbol{x}}_i^{\top}\hat{\boldsymbol{\beta}}_{n}-\boldsymbol{x}_i^{\top}\hat{\boldsymbol{\beta}}_{n}|\Big)\\
&= \mathds{1}\Big(|\boldsymbol{x}_i^{\top}\hat{\boldsymbol{\beta}}_{n}-c_{\rho}|+|\boldsymbol{x}_i^{\top}\hat{\boldsymbol{\beta}}_{n}-\boldsymbol{x}_i^{\top}\boldsymbol{\beta}_{0}|\leq|\hat{\boldsymbol{x}}_i^{\top}\hat{\boldsymbol{\beta}}_{n}-\boldsymbol{x}_i^{\top}\hat{\boldsymbol{\beta}}_{n}|+|\boldsymbol{x}_i^{\top}\hat{\boldsymbol{\beta}}_{n}-\boldsymbol{x}_i^{\top}\boldsymbol{\beta}_{0}|\Big) \tag{L.2} \label{L.2} \\
&\leq \mathds{1}\Big(|\boldsymbol{x}_i^{\top}\boldsymbol{\beta}_{0}-c_{\rho}|\leq|\hat{\boldsymbol{x}}_i^{\top}\hat{\boldsymbol{\beta}}_{n}-\boldsymbol{x}_i^{\top}\hat{\boldsymbol{\beta}}_{n}|+|\boldsymbol{x}_i^{\top}\hat{\boldsymbol{\beta}}_{n}-\boldsymbol{x}_i^{\top}\boldsymbol{\beta}_{0}|\Big) \tag{L.3} \label{L.3}\\
&\leq \mathds{1}\Big(|\boldsymbol{x}_i^{\top}\boldsymbol{\beta}_{0}-c_{\rho}|\leq \|\boldsymbol{z}_i\|\,\|\hat{\boldsymbol{\delta}}_{n}-\boldsymbol{\delta}_{0}\|\,\|\hat{\boldsymbol{\beta}}_{n}\|+\|\boldsymbol{x}_i\|\,\|\hat{\boldsymbol{\beta}}_{n}-\boldsymbol{\beta}_{0}\|\Big)\\
&\leq \mathds{1}\Big(|\boldsymbol{x}_i^{\top}\boldsymbol{\beta}_{0}-c_{\rho}|\leq \|\boldsymbol{z}_i\|\,\|\hat{\boldsymbol{\delta}}_{n}-\boldsymbol{\delta}_{0}\|\|\hat{\boldsymbol{\beta}}_{n}-\boldsymbol{\beta}_{0}\|+\|\boldsymbol{z}_i\|\,\|\hat{\boldsymbol{\delta}}_{n}-\boldsymbol{\delta}_{0}\| \|\boldsymbol{\beta}_{0}\|+\|\boldsymbol{x}_i\|\,\|\hat{\boldsymbol{\beta}}_{n}-\boldsymbol{\beta}_{0}\|\Big),\\
&\leq \mathds{1}\Big(|\boldsymbol{x}_i^{\top}\boldsymbol{\beta}_{0}-c_{\rho}|\leq \mathcal{K}_{1}\,\|\hat{\boldsymbol{\delta}}_{n}-\boldsymbol{\delta}_{0}\|\|\hat{\boldsymbol{\beta}}_{n}-\boldsymbol{\beta}_{0}\|+\mathcal{K}_{1}\,\|\hat{\boldsymbol{\delta}}_{n}-\boldsymbol{\delta}_{0}\| C_{0}+\mathcal{K}_{0}\,\|\hat{\boldsymbol{\beta}}_{n}-\boldsymbol{\beta}_{0}\|\Big),\tag{L.5}\label{AE:1}
\end{align*} 
where the last inequality follows from using Assumption~\ref{a:3.5} and $C_0 := \sup_{\boldsymbol{\beta}\in\mathcal{B}}\|\boldsymbol{\beta}\| < \infty$ by compactness of $\mathcal{B}$. Further, \eqref{L.2} follows from using
the fact that $\mathds{1}(a\leq t)\leq\mathds{1}(b\leq t)$ whenever $b\leq a$, with
$a=|\boldsymbol{x}_i^{\top}\hat{\boldsymbol{\beta}}_{n}-c_{\rho}|+|\boldsymbol{x}_i^{\top}\hat{\boldsymbol{\beta}}_{n}-\boldsymbol{x}_i^{\top}\boldsymbol{\beta}_{0}|$,
$b=|\boldsymbol{x}_i^{\top}\boldsymbol{\beta}_{0}-c_{\rho}|$, and $t$ the right-hand side of the second line. Here $b\leq a$ by the triangle inequality,
$$
|\boldsymbol{x}_i^{\top}\boldsymbol{\beta}_{0}-c_{\rho}|\leq|\boldsymbol{x}_i^{\top}\hat{\boldsymbol{\beta}}_{n}-c_{\rho}|+|\boldsymbol{x}_i^{\top}\hat{\boldsymbol{\beta}}_{n}-\boldsymbol{x}_i^{\top}\boldsymbol{\beta}_{0}|.$$
The last line follows from the Cauchy–Schwarz inequality and $\|\hat{\boldsymbol{x}}_i-\boldsymbol{x}_i\|\leq\|\boldsymbol{z}_i\|\|\hat{\boldsymbol{\delta}}_n-\boldsymbol{\delta}_0\|$:
$$
|\hat{\boldsymbol{x}}_i^{\top}\hat{\boldsymbol{\beta}}_{n}-\boldsymbol{x}_i^{\top}\hat{\boldsymbol{\beta}}_{n}|\leq\|\hat{\boldsymbol{x}}_i-\boldsymbol{x}_i\|\|\hat{\boldsymbol{\beta}}_{n}\|,
\qquad
|\boldsymbol{x}_i^{\top}(\hat{\boldsymbol{\beta}}_{n}-\boldsymbol{\beta}_{0})|\leq\|\boldsymbol{x}_i\|\|\hat{\boldsymbol{\beta}}_{n}-\boldsymbol{\beta}_{0}\|.$$
\end{proof}
\end{R1}
\begin{R1} \label{Re:2}
Define
\begin{equation}\label{R:2}
C_{i,n}:=\left|\mathds{1}\big(\hat{\boldsymbol{x}}_i^{\top}\hat{\boldsymbol{\beta}}_{n}>c_{\rho}\big)-\mathds{1}\big(\boldsymbol{x}_i^{\top}\boldsymbol{\beta}_{0}>c_{\rho}\big)\right|.
\end{equation}
Under Assumption~\ref{a:3.5} and  $C_0:=\sup_{\boldsymbol\beta\in\mathcal B}\|\boldsymbol\beta\|<\infty$, we have for every $i$ and $n$
$$C_{i,n}\;\le\;\mathds{1}\Big(|\boldsymbol{x}_i^{\top}\boldsymbol{\beta}_0-c_\rho|\le \mathcal{K}_1\|\hat{\boldsymbol\delta}_n-\boldsymbol\delta_0\|\,\|\hat{\boldsymbol\beta}_n-\boldsymbol\beta_0\|+\mathcal{K}_1C_0\|\hat{\boldsymbol\delta}_n-\boldsymbol\delta_0\|+\mathcal{K}_0\|\hat{\boldsymbol\beta}_n-\boldsymbol\beta_0\|\Big).$$
\begin{proof}[\textbf{Proof}]
Since the two indicators differ only if $c_{\rho}$ lies between $\boldsymbol{x}_i^{\top}\boldsymbol{\beta}_{0}$ and $\hat{\boldsymbol{x}}_i^{\top}\hat{\boldsymbol{\beta}}_{n}$, we have
\begin{align*}
|C_{i,n}|
&\leq \mathds{1}\Big(|\boldsymbol{x}_i^{\top}\boldsymbol{\beta}_{0}-c_{\rho}|\leq|\hat{\boldsymbol{x}}_i^{\top}\hat{\boldsymbol{\beta}}_{n}-\boldsymbol{x}_i^{\top}\boldsymbol{\beta}_{0}|\Big)\\
&\leq \mathds{1}\Big(|\boldsymbol{x}_i^{\top}\boldsymbol{\beta}_{0}-c_{\rho}|\leq|\hat{\boldsymbol{x}}_i^{\top}\hat{\boldsymbol{\beta}}_{n}-\boldsymbol{x}_i^{\top}\hat{\boldsymbol{\beta}}_{n}|+|\boldsymbol{x}_i^{\top}\hat{\boldsymbol{\beta}}_{n}-\boldsymbol{x}_i^{\top}\boldsymbol{\beta}_{0}|\Big)\\
&\leq \mathds{1}\Big(|\boldsymbol{x}_i^{\top}\boldsymbol{\beta}_{0}-c_{\rho}|\leq \|\boldsymbol{z}_i\|\,\|\hat{\boldsymbol{\delta}}_{n}-\boldsymbol{\delta}_{0}\|\,\|\hat{\boldsymbol{\beta}}_{n}\|+\|\boldsymbol{x}_i\|\,\|\hat{\boldsymbol{\beta}}_{n}-\boldsymbol{\beta}_{0}\|\Big)\\
&\leq \mathds{1}\Big(|\boldsymbol{x}_i^{\top}\boldsymbol{\beta}_{0}-c_{\rho}|\leq \|\boldsymbol{z}_i\|\,\|\hat{\boldsymbol{\delta}}_{n}-\boldsymbol{\delta}_{0}\|\|\hat{\boldsymbol{\beta}}_{n}-\boldsymbol{\beta}_{0}\|+\|\boldsymbol{z}_i\|\,\|\hat{\boldsymbol{\delta}}_{n}-\boldsymbol{\delta}_{0}\| \|\boldsymbol{\beta}_{0}\|+\|\boldsymbol{x}_i\|\,\|\hat{\boldsymbol{\beta}}_{n}-\boldsymbol{\beta}_{0}\|\Big),\\
&\leq\mathds{1}\Big(|\boldsymbol{x}_i^{\top}\boldsymbol{\beta}_{0}-c_{\rho}|\leq \mathcal{K}_{1}\,\|\hat{\boldsymbol{\delta}}_{n}-\boldsymbol{\delta}_{0}\|\|\hat{\boldsymbol{\beta}}_{n}-\boldsymbol{\beta}_{0}\|+\mathcal{K}_{1}\,\|\hat{\boldsymbol{\delta}}_{n}-\boldsymbol{\delta}_{0}\| C_{0}+\mathcal{K}_{0}\,\|\hat{\boldsymbol{\beta}}_{n}-\boldsymbol{\beta}_{0}\|\Big),\tag{AE:2}\label{AE:2}
\end{align*}
where the last inequality follows from Assumption~\ref{a:3.5}
and $C_0 := \sup_{\boldsymbol{\beta}\in\mathcal{B}}\|\boldsymbol{\beta}\| < \infty$ by compactness of $\mathcal{B}$.The first line holds because, if the two indicators differ, then either
$\hat{\boldsymbol{x}}_i^{\top}\hat{\boldsymbol{\beta}}_{n}>c_{\rho}\geq\boldsymbol{x}_i^{\top}\boldsymbol{\beta}_{0}$ or
$\boldsymbol{x}_i^{\top}\boldsymbol{\beta}_{0}>c_{\rho}\geq\hat{\boldsymbol{x}}_i^{\top}\hat{\boldsymbol{\beta}}_{n}$.
In both cases $c_{\rho}$ lies between the two values, so
$$
|\boldsymbol{x}_i^{\top}\boldsymbol{\beta}_{0}-c_{\rho}|\leq|\hat{\boldsymbol{x}}_i^{\top}\hat{\boldsymbol{\beta}}_{n}-\boldsymbol{x}_i^{\top}\boldsymbol{\beta}_{0}|.$$
The second line follows from the triangle inequality,
$$
|\hat{\boldsymbol{x}}_i^{\top}\hat{\boldsymbol{\beta}}_{n}-\boldsymbol{x}_i^{\top}\boldsymbol{\beta}_{0}|\leq|\hat{\boldsymbol{x}}_i^{\top}\hat{\boldsymbol{\beta}}_{n}-\boldsymbol{x}_i^{\top}\hat{\boldsymbol{\beta}}_{n}|+|\boldsymbol{x}_i^{\top}\hat{\boldsymbol{\beta}}_{n}-\boldsymbol{x}_i^{\top}\boldsymbol{\beta}_{0}|,$$
together with the fact that $\mathds{1}(b\leq t)\leq\mathds{1}(b\leq t')$ whenever $t\leq t'$.
The third line follows from the Cauchy--Schwarz inequality and $\|\hat{\boldsymbol{x}}_i-\boldsymbol{x}_i\|\leq\|\boldsymbol{z}_i\|\|\hat{\boldsymbol{\delta}}_n-\boldsymbol{\delta}_0\|$:
$$
|\hat{\boldsymbol{x}}_i^{\top}\hat{\boldsymbol{\beta}}_{n}-\boldsymbol{x}_i^{\top}\hat{\boldsymbol{\beta}}_{n}|\leq\|\hat{\boldsymbol{x}}_i-\boldsymbol{x}_i\|\|\hat{\boldsymbol{\beta}}_{n}\|,
\qquad
|\boldsymbol{x}_i^{\top}(\hat{\boldsymbol{\beta}}_{n}-\boldsymbol{\beta}_{0})|\leq\|\boldsymbol{x}_i\|\|\hat{\boldsymbol{\beta}}_{n}-\boldsymbol{\beta}_{0}\|.$$
\end{proof}
\end{R1}

\begin{R1} \label{Re:3}
Define
\begin{equation}\label{R:3}
D^{x}_{i,n}:=\hat{x}_{i,r}\hat{x}_{i,s}-x_{i,r}x_{i,s}.
\end{equation}
Under Assumption~\ref{a:3.5}, we have for every $i$ and $n$
$$|D^{x}_{i,n}|\leq \mathcal{K}_{1}\,\|\hat{\boldsymbol{\delta}}_{n}-\boldsymbol{\delta}_{0}\|
\Big(\mathcal{K}_{1}\,\|\hat{\boldsymbol{\delta}}_{n}-\boldsymbol{\delta}_{0}\|
+2\mathcal{K}_{0}\Big).$$
\begin{proof}[\textbf{Proof}]
Adding and subtracting \(x_{i,r}\hat{x}_{i,s}\) and applying the triangle inequality gives
\begin{align*}
|D^{x}_{i,n}|
&= \big|(\hat{x}_{i,r}-x_{i,r})\,\hat{x}_{i,s}
+x_{i,r}\,(\hat{x}_{i,s}-x_{i,s})\big|\\
&\leq |\hat{x}_{i,r}-x_{i,r}|\,|\hat{x}_{i,s}|
+|x_{i,r}|\,|\hat{x}_{i,s}-x_{i,s}|\\
&\leq \|\hat{\boldsymbol{x}}_{i}-\boldsymbol{x}_{i}\|
\Big(|\hat{x}_{i,s}|+|x_{i,r}|\Big)\\
&\leq \|\hat{\boldsymbol{x}}_{i}-\boldsymbol{x}_{i}\|
\Big(\|\hat{\boldsymbol{x}}_{i}\|+\|\boldsymbol{x}_{i}\|\Big)\\
&\leq \|\boldsymbol{z}_{i}\|\,\|\hat{\boldsymbol{\delta}}_{n}-\boldsymbol{\delta}_{0}\|
\Big(\|\boldsymbol{z}_{i}\|\,\|\hat{\boldsymbol{\delta}}_{n}-\boldsymbol{\delta}_{0}\|
+2\|\boldsymbol{x}_{i}\|\Big)\\
&\leq \mathcal{K}_{1}\,\|\hat{\boldsymbol{\delta}}_{n}-\boldsymbol{\delta}_{0}\|
\Big(\mathcal{K}_{1}\,\|\hat{\boldsymbol{\delta}}_{n}-\boldsymbol{\delta}_{0}\|
+2\mathcal{K}_{0}\Big)\\
&= \mathcal{K}_{1}^{2}\,\|\hat{\boldsymbol{\delta}}_{n}-\boldsymbol{\delta}_{0}\|^{2}
+2\mathcal{K}_{0}\mathcal{K}_{1}\,\|\hat{\boldsymbol{\delta}}_{n}-\boldsymbol{\delta}_{0}\|.\tag{AE:3}\label{AE:3}
\end{align*}
Here, the inequality
$\|\hat{\boldsymbol{x}}_{i}-\boldsymbol{x}_{i}\|
\leq \|\boldsymbol{z}_{i}\|\,\|\hat{\boldsymbol{\delta}}_{n}-\boldsymbol{\delta}_{0}\|$
together with
$
\|\hat{\boldsymbol{x}}_{i}\|
\leq \|\boldsymbol{x}_{i}\|+\|\hat{\boldsymbol{x}}_{i}-\boldsymbol{x}_{i}\|
\leq \|\boldsymbol{x}_{i}\|
+\|\boldsymbol{z}_{i}\|\,\|\hat{\boldsymbol{\delta}}_{n}-\boldsymbol{\delta}_{0}\|$
yields the bound involving \(\|\boldsymbol{z}_{i}\|\) and \(\|\boldsymbol{x}_{i}\|\).  
Moreover, by Assumption~\ref{a:3.5}, \(\|\boldsymbol{x}_{i}\|\leq\mathcal{K}_{0}\) and
\(\|\boldsymbol{z}_{i}\|\leq\mathcal{K}_{1}\), which gives the final bound.
\end{proof}
\end{R1}
\bibliographystyle{chicago} 
\bibliography{Reference}
\end{document}